\documentclass[structabstract]{aa}
\usepackage{textcomp}
\usepackage{graphicx}
\usepackage{txfonts}
\usepackage{natbib}
\usepackage{longtable}
\usepackage{multirow}
\usepackage{lscape}

\newcommand\nobr{\mbox{-}}
\newcommand\micron{~\hbox{\textmu}m}
\begin{document}
   \title{Warm gas towards young stellar objects in Corona Australis}
   \subtitle{\textit{Herschel}/PACS observations from the DIGIT key programme}
%\titlerunning{\textit{Herschel}/DIGIT observations of warm gas associated with YSOs in CrA}
   \author{Johan~E.~Lindberg
          \inst{1,2}
          \and
          Jes~K.~J{\o}rgensen\inst{2,1}
          \and
          %...
          %\and
          Joel~D.~Green\inst{3}
          \and
          Gregory~J.~Herczeg\inst{4,5}
          \and
          Odysseas~Dionatos\inst{1,2,6}
          \and
          Neal~J.~Evans~II\inst{3}
          \and
          Agata~Karska\inst{5}
          \and
          Susanne~F.~Wampfler\inst{1,2}
          %\and
          %... (other authors and exact order TBD)
          }

   \institute{{Centre for Star and Planet Formation, Natural History Museum of
Denmark, University of Copenhagen, {\O}ster Voldgade 5-7, DK\nobr1350
K{\o}benhavn K, Denmark}\\
              \email{jlindberg@snm.ku.dk}
         	\and
      {Niels Bohr Institute, University of Copenhagen, Juliane Maries Vej 30,
DK\nobr2100 K{\o}benhavn {\O}, Denmark}
			\and
		{The University of Texas at Austin, Department of Astronomy, 2515 Speedway, Stop C1400, Austin, TX 78712-1205, USA}
			\and
		{Kavli Institute for Astronomy and Astrophysics, Peking University, Beijing, 100871, PR China}
			\and
		{Max Planck Institute for Extraterrestrial Physics, Postfach 1312, 85741, Garching, Germany}
			\and
		{University of Vienna, Department of Astronomy, T\"{u}rkenschanzstrasse 17, 1180 Vienna, Austria}\\
         }

   \date{Received July 1, 2013; accepted November 27, 2013}

% \abstract{}{}{}{}{}
% 5 {} token are mandatory

  \abstract
  % context heading (optional)
  % {} leave it empty if necessary
   {The effects of external irradiation on the chemistry and physics in the protostellar envelope around low-mass young stellar objects are poorly understood. The Corona Australis star-forming region contains the R~CrA dark cloud,
comprising several low-mass protostellar cores irradiated by an
intermediate-mass young star.}
  % aims heading (mandatory)
   {We study the effects on the warm gas and dust in a group of low-mass young
stellar objects from the irradiation by the young luminous Herbig~Be star
R~CrA.}
  % methods heading (mandatory)
   {\textit{Herschel}/PACS far-infrared datacubes of two low-mass star-forming
regions in the R~CrA dark cloud are presented. The distribution of CO, OH,
H$_2$O, [\ion{C}{ii}], [\ion {O}{i}], and continuum emission is investigated. We have developed a deconvolution algorithm which we use to deconvolve the maps, separating the point-source emission from the extended
emission. We also construct rotational diagrams of the molecular species.}
  % results heading (mandatory)
   {By deconvolution of the \textit{Herschel} data, we find large-scale (several thousand AU) dust continuum and spectral line emission not associated with the point sources. Similar rotational temperatures are found for the warm CO ($282\pm4$~K), hot CO ($890\pm84$~K), OH ($79\pm4$~K), and H$_2$O ($197\pm7$~K)
emission, respectively, in the point sources and
the extended emission. The rotational temperatures are also similar to what is
found in other more isolated cores. The extended dust continuum emission is found in two ridges similar in extent and
temperature to molecular millimetre emission, indicative of external heating from the
Herbig~Be star R~CrA.}
  % conclusions heading (optional), leave it empty if necessary
   {Our results show that a nearby luminous star does not increase the
molecular excitation temperatures in the warm gas around a young stellar object (YSO). However,
the emission from photodissociation products of H$_2$O, such as OH and O, is
enhanced in the warm gas associated with these protostars and their surroundings compared to similar objects
not suffering from external irradiation.}


   \maketitle
%
%________________________________________________________________

\section{Introduction}

One of the open questions in low-mass star formation is how the irradiation from intermediate-mass stars affects the chemistry,
temperature, and excitation conditions in the warm gas around low-mass young
stellar objects. With the resolution of the
\textit{Herschel Space Observatory}, superior to that of previous far-infrared
telescopes, in combination with deconvolution algorithms, we can now address
this question. Both the spectral line
emission from the gas and the dust continuum emission from the warm regions
peak in the far-infrared (FIR) part of the electromagnetic spectrum. CO, the second most abundant molecule in the interstellar medium (ISM) after H$_2$, has a large number of transitions in this band. In
addition, water and its related species, OH, have their most important
transitions in this band.

%Only a small number of telescopes have allowed studies in the FIR wavelength regime between $\sim50$\micron\ and $\sim200$\micron. With
%space observatories such as IRAS \citep[Infrared Astronomical
%Satellite;][]{neugebauer84}, ISO \citep[Infrared Space
%Observatory;][]{kessler96}, and the \textit{Spitzer Space Telescope} \citep{werner04}, and airborne observatories such as the Kuiper Airborne Observatory (KAO) and the Stratospheric Observatory for Infrared Astronomy \citep[SOFIA;][]{sofia}, this window
%was opened. 
With the advent of the \textit{Herschel Space Observatory}
\citep{pilbratt10}, FIR observations with unprecedented spatial and spectral
resolution have been made available. The \textit{Herschel} observations of
low-mass YSOs reveal numerous lines of CO, H$_2$O, and OH, along with atomic
lines like [\ion{O}{i}] and [\ion{C}{ii}] \citep[e.g.][]{herczeg12,kristensen12,green13}. In most studied sources, the CO
rotational diagrams can be fitted with warm and hot components, with
rotational temperatures of about 300~K and 900~K, respectively \citep{green13,karska13,manoj13}. The OH and H$_2$O emission
is usually characterised by somewhat lower rotational temperatures 
around 100~K \citep{goicoechea12,herczeg12,wampfler13}. Most of these studies of low-mass YSOs have targeted isolated embedded objects. To better understand low-mass star formation in more dynamic environments, studies of small groups of resolvable embedded objects are warranted.

This paper presents PACS \citep[Photodetector Array Camera and Spectrometer;][]{poglitsch10} maps of
the R~Coronae Australis (R~CrA) dark cloud, which harbours one of the closest star-forming regions, located at a distance of 130~pc
\citep{neuhauser08}. The cloud is named for the young star R~CrA, which
has spectral classifications ranging from F5 to B5 (e.g., F5: \citealt{hillenbrand92}; A0: \citealt{manoj06}; B8: \citealt{bibo92}; B5: \citealt{gray06}).
The cloud was mapped in CO by \citet{loren79}, who found an
elongated cloud
extended over about 2 by 0.5~pc and peaking near R~CrA. Higher resolution maps 
of C$^{18}$O with the SEST (Swedish-ESO Submillimetre Telescope) 
revealed several dense molecular clumps with masses 
between $2~M_{\odot}$ and $50~M_{\odot}$ near R~CrA \citep{harju93}.
Surveys of the clumps have revealed a number of embedded protostars,
with IRS7 in the clump to the southeast of R~CrA and IRS5 in the clump
to the west.
\citet{taylor84} report that IRS7 is the most reddened source in the region, 
having a visual extinction of more than 25~mag. 
\citet{brown87} split IRS7 into two sources separated by 14\arcsec, 
using VLA 6~cm observations.
For a more complete description and
references, see \citet{neuhauser08}.

The R~CrA region (including IRS7) has previously been studied at FIR
wavelengths using the ISO telescope \citep{lorenzetti99,giannini99},
detecting lines from CO, OH, \ion{O}{i}, and \ion{C}{ii}. However,
these studies were limited by the ISO angular resolution of 80\arcsec,
which made it impossible to separate the sources in the region.
\citet{sicilia13} present 100\micron\ and 160\micron\ photometry maps
of the CrA region observed with \textit{Herschel}/PACS. The IRS7 clump
is positioned with its centre between the two Herbig~Ae/Be stars R~CrA
and T~CrA, and harbours a handful of Class~0/I Young Stellar Objects
(YSOs).  \citet{nutter05} report detections of four cores within the
IRS7 and IRS5 clumps in SCUBA 450\micron\ and 850\micron\ data:
SMM~1A, with no mid-IR counterpart, proposed to be a pre-stellar core;
SMM~1B, coincident with the mid-IR source IRS7B; SMM~1C, a Class~0
protostar; and SMM~4, coincident with the IRS5
clump. \citet{peterson11} report SMA 1.3~mm point-source continuum
detections at the positions of SMM~1B (IRS7B), SMM~1C, and SMM~4
(IRS5N).  The IRS5 clump contains two protostellar sources, IRS5A and
IRS5N. IRS5 is situated at a slightly greater projected distance from
R~CrA than is IRS7. IRS5A is not detected in SMA 1.3~mm continuum
observations, whereas IRS5N shows significantly fainter continuum
emission than the IRS7 sources \citep[95 mJy in IRS5N, 320 mJy in
IRS7B;][]{peterson11}.

Through single-dish APEX observations, \citet{schoier06} found
elevated H$_2$CO and CH$_3$OH abundances and rotational temperatures
in IRS7A and IRS7B, which were suggested to be caused by increased
internal heating or outflows. \citet{vankempen09a} used mid-$J$ CO
observations to find an EW outflow centred at IRS7A, but also found
that the CO line fluxes in the region are too high to originate from
heating by the embedded protostars, and proposed that the heating
originates from R~CrA.  This was confirmed by \citet{lindberg12}, who
found large-scale ($\sim 10\,000$~AU) H$_2$CO emission heated to
40--60~K by external irradiation from R~CrA. The IRS5 sources are
found to be less affected by the irradiation from R~CrA (the H$_2$CO
rotational temperature in a 29\arcsec\ beam is 47~K for IRS7B
and 28~K for IRS5A; Lindberg et~al., in prep.).

In Sect.~2 we will describe the methods of observations and data
reductions and show the first results. In Sect.~3 follows a discussion
of our deconvolution algorithm and the results thereof. In Sect.~4 we
investigate models of the excitation conditions of the gas. In Sect.~5
we discuss the interpretations of our results, and in Sect.~6 we list
our conclusions.

\section{Observations and data description}

In this section we first describe the observations and methods of data
reduction. A more detailed description can be found in
\citet{green13}. We then discuss the first results of these
reductions.

\subsection{Observation setup and data reduction}

% plot_map.py
\begin{figure*}[!htb]
    \centering
    \includegraphics[width=1.0\linewidth]{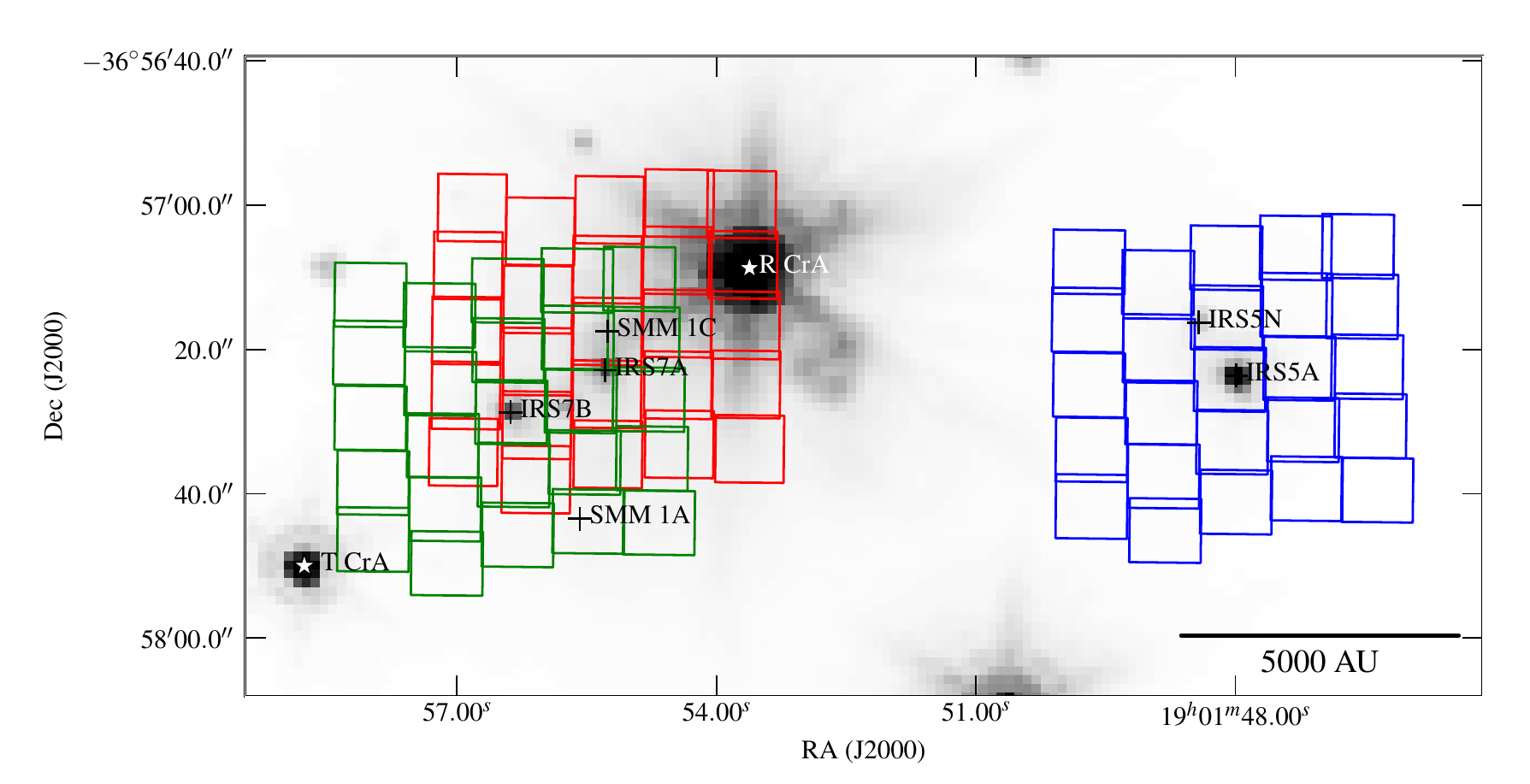} \\
    \caption{Schematic overview of the three PACS footprints, centred at SMM~1C (red),
IRS7B (green), and IRS5A (blue), overlaid on a Spitzer 4.5\micron\ image. The spaxel positions as recorded by PACS on the sky are at the center of each spaxel, and each spaxel measures $9\farcs4\times9\farcs4$ in this representation.
}
    \label{fig:overview}
\end{figure*}

\begin{table*}
\centering
\caption[]{Observational parameters: Positions give the centre coordinates of the
central spaxel. All footprints are $5\times5$ spaxel grids with
approximately 9\farcs25 separation between the spaxel centres.}
\label{tab:obsparam}
\begin{tabular}{l l l l l l}
\noalign{\smallskip}
\hline
\hline
\noalign{\smallskip}
Region & RA & Dec & P.A.\tablefootmark{a} & Observing IDs & Date of observation \\
 & (J2000.0) & (J2000.0) & [\degr] \\
\noalign{\smallskip}
\hline
\noalign{\smallskip}
IRS7A & 19:01:55.3 & $-$36:57:17.0 & $-0.22$ & 1342206990, 1342206989 & 2010-10-23 \\
IRS7B & 19:01:56.4 & $-$36:57:28.3 & \phantom{$-$}2.47 & 1342207807, 1342207808
& 2010-11-02 \\
IRS5A & 19:01:48.1 & $-$36:57:22.7 & \phantom{$-$}2.40 & 1342207806, 1342207805
& 2010-11-01 \\
\noalign{\smallskip}
\hline
\end{tabular}
\tablefoot{
	\tablefoottext{a}{Position angle with respect to an equatorial grid coordinate system.}
     	}
\end{table*}

Integral-field spectroscopy observations in the far infrared (FIR)
regime that cover parts of the R~CrA star-forming region were
performed by the Photodetector Array Camera and Spectrometer
\citep[PACS;][]{poglitsch10} on board the ESA \textit{Herschel Space
  Observatory} \citep{pilbratt10}, with a spectral range from
55~\hbox{\textmu}m to 210~\hbox{\textmu}m. The observations were
carried out in range-spectroscopy mode. The spectral resolution varies
between 0.013~\hbox{\textmu}m and 0.13~\hbox{\textmu}m ($\Delta \varv
\approx 55$--318~km~s$^{-1}$, $\lambda/\Delta\lambda \approx
950$--$5500$).

These observations are part of the ``Dust, Ice and Gas in Time'' (DIGIT; PI:
N. Evans) Open-Time Key Programme, a survey of the change of FIR spectral features
through the evolution of young stellar objects. For details of the DIGIT Key
Programme, see \citet{green13}. The observed PACS fields are shown compared to
the point-source positions and a Spitzer 4.5\micron\ image in
Fig.~\ref{fig:overview}, and the telescope pointings are found in
Table~\ref{tab:obsparam}. The configuration of the $9\farcs4\times9\farcs4$ spaxels shown in Fig.~\ref{fig:overview} is a simplified model: in reality, no spaxels overlap. This is taken into account in the deconvolution method described in Sect.~\ref{sec:deconv_meth}.

The spectral line fluxes in each spaxel were calculated using the HIPE 8.0.2489
reduction of the data, but corrected by the continuum value in the HIPE 6.1
reduction. The HIPE 8.0.2489 reduction was found to provide on average $\sim 40$\% better
signal-to-noise ratio (spanning from almost no improvement up to about a factor 2 at specific wavelengths), whereas the HIPE 6.1 reduction produced more
accurate and reliable spectral energy distributions (SEDs) by $\sim20\%$. These SEDs are consistent with PACS photometry within $\sim10\%$ \citep{green13}. The method of
combining these two reductions is thoroughly described in \citet{green13}. For
the continuum flux densities, the HIPE 6.1 reduction was used.

For each spectral line in each spaxel, a first-degree polynomial
baseline calculated from the surrounding line-free channels was
subtracted from the spectrum. The total line flux was then calculated
by summing the channel flux densities within a typical linewidth
(mostly 0.1--0.5\micron) and multiplying this sum by the channel
width. Of the OH doublets, only four out of nine are resolved. In the
unresolved cases, the combined total flux for both OH lines is
calculated, and half of this value is used for each of the components
in the rotational diagram analysis. The statistical errors are
calculated from the rms noise around each spectral line (any lines
weaker than 3$\sigma$ are ignored) to get the total standard deviation
of the spectral line flux. If nothing else is stated, all errors given
in this paper are at the 1$\sigma$ level. In addition to the
statistical error, a systematic calibration error of 20\% of the flux
(see \citealt{green13}) is used when calculating quantities such as
the rotational temperatures and total number of molecules from the
rotational diagrams.

The two bands below 103~\hbox{\textmu}m are considerably noisier than the bands
above this wavelength, and between approximately 94~\hbox{\textmu}m and
103~\hbox{\textmu}m the noise makes line fluxes very difficult to estimate. The
spectrometer suffers from leakage in the wavelength ranges 70--73\micron,
98--105\micron, and above 190\micron, which produces ghost images of lines from the
next higher grating order. Fluxes at these wavelengths are thus less reliable
than at other wavelengths \citep{herczeg12}.

In addition to the Herschel data, we also used the ISIS spectrograph
on the \textit{William Herschel Telescope} (WHT) on 6 August 2012 to
obtain a low-resolution ($R\sim1\,000$) optical spectrum of R~CrA
covering 3\,000--10\,000~\AA. The spectrum was flux calibrated against
the spectrophotometric standard LTT~7987. The analysis of this
spectrum will be discussed in Sect.~\ref{sec:rcra_spect}.

%plot_spectra.py
%\begin{figure*}[!htb]
%    \centering
%    $\begin{array}{c@{\hspace{0.0cm}}c@{\hspace{0.0cm}}c}
%    \includegraphics{smm1c_spect.png} &
%    \includegraphics{irs7a_spect.png} \\
%    \includegraphics{irs7b_spect.png} &
%    \includegraphics{rcra_spect.png} \\
%    \includegraphics{irs5a_spect.png} &
%    \includegraphics{irs5n_spect.png} \\
%    \end{array}$
%    \caption{Spectra of the spaxels most nearby to each of the six point
%sources. No deconvolution or correction factors have been applied to these
%data. They will thus appear fainter in the red part compared to the true
%spectra.}
%    \label{fig:spectra}
%\end{figure*}

\subsection{First look}

In this section, we provide a first look at the PACS continuum and spectral maps before they are treated by our deconvolution algorithm.

\subsubsection{Continuum maps}

\begin{figure*}[!htb]
    \centering
   	$\begin{array}{c@{\hspace{0.0cm}}c@{\hspace{0.0cm}}c}
	 \includegraphics[width=0.48\linewidth]{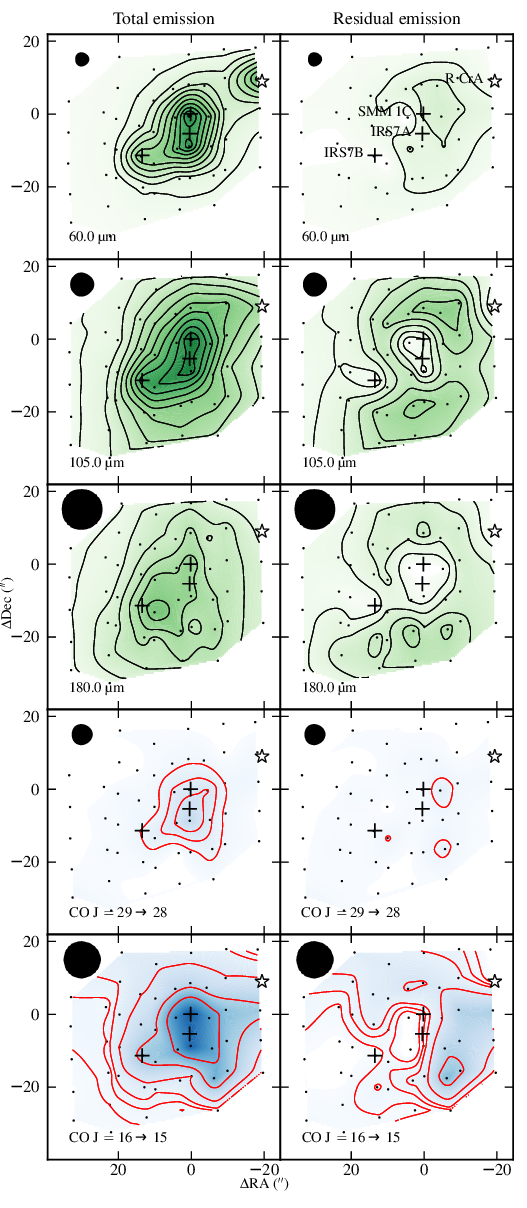} &
	 \includegraphics[width=0.48\linewidth]{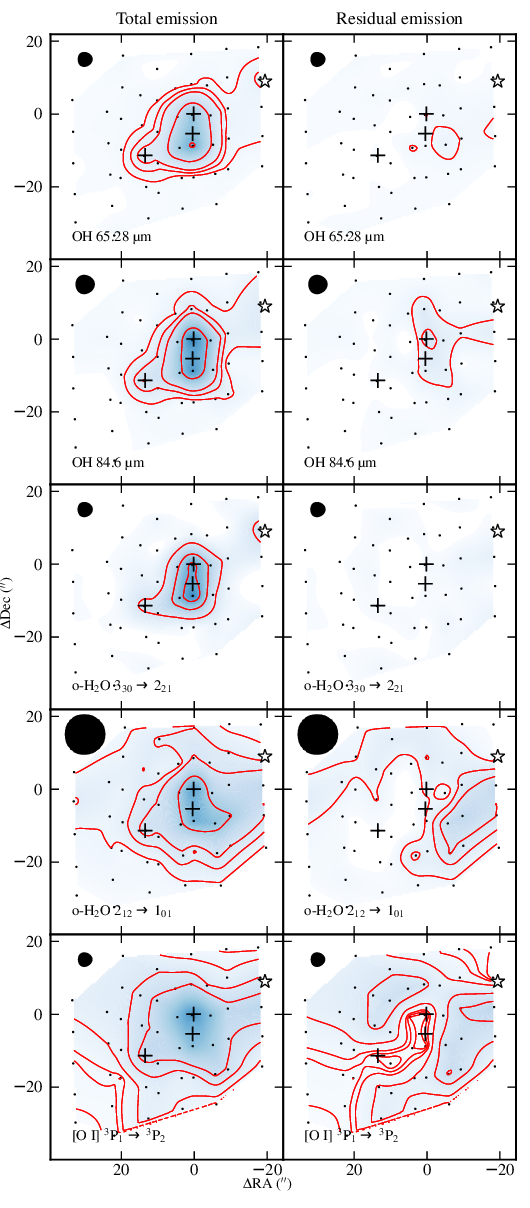} \\
	 \end{array}$
    \caption{IRS7 contour maps of continuum emission at three wavelengths (green with black contours), and some
important spectral lines (the same as in Table~4 of \citealt{green13}, blue with red contours).
The first column in each panel shows the total emission, whereas the second column in each panel shows the residual
emission after deconvolution (emission not associated with the point sources). Such
maps of the whole band of continuum and all spectral lines can be found in
Appendices~\ref{app:contmaps}--\ref{app:linemaps}. The point sources used for the deconvolution are marked with crosses, except for R~CrA, which is marked with a star symbol (the sources are identified in the top row). The dots indicate the PACS spaxel centres. The \textit{Herschel} PSF for each observation is shown in the top left corner. Contour levels are at 10~Jy for the continuum maps, and at $5\sigma$, $10\sigma$, $15\sigma$, $30\sigma$, $60\sigma$, and $90\sigma$ for the line maps. The blue colour maps have the same ranges for each separate molecule, with maximum fluxes of $7\times10^{-16}$~W~m$^{-2}$ for CO, $1\times10^{-15}$~W~m$^{-2}$ for OH, $4\times10^{-16}$~W~m$^{-2}$ for o-H$_2$O, and $1\times10^{-14}$~W~m$^{-2}$ for [\ion{O}{i}].
%$6\times10^{-17}$~W~m$^{-2}$ for the CO and H$_2$O maps, $5\times10^{-17}$~W~m$^{-2}$ for the OH maps, \textbf{and $3\times10^{-16}$~W~m$^{-2}$ for the \ion{O}{i} maps}. The rms level varies between $1\times10^{-17}$~W~m$^{-2}$ and $3\times10^{-17}$~W~m$^{-2}$ across the spectrum.
}
    \label{fig:important_7}
\end{figure*}

\begin{figure*}[!htb]
    \centering
   	$\begin{array}{c@{\hspace{0.0cm}}c@{\hspace{0.0cm}}c}
	 \includegraphics[width=0.48\linewidth]{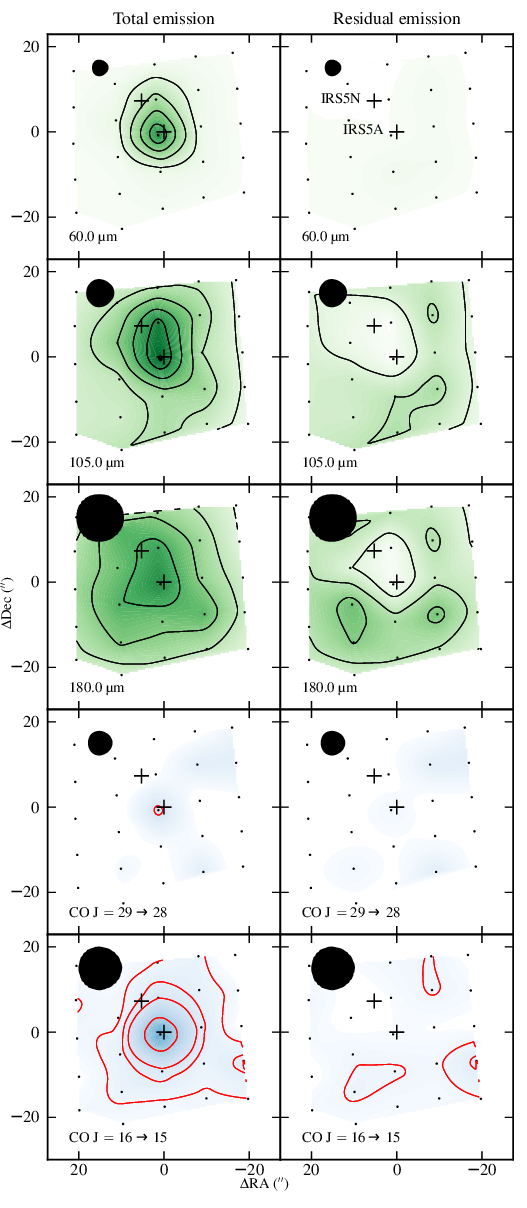} &
	 \includegraphics[width=0.48\linewidth]{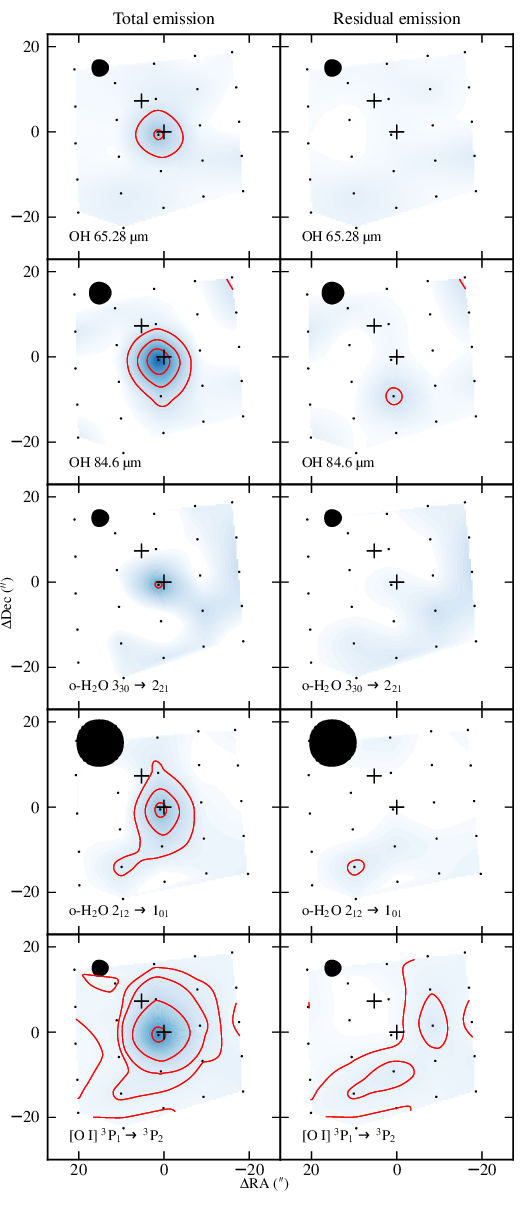} \\
	 \end{array}$
    \caption{As Fig.~\ref{fig:important_7}, but for IRS5. Contour levels are at 3~Jy for the continuum maps, and as in Fig.~\ref{fig:important_7} for the line maps. The blue colour maps have the same ranges for each separate molecule, with maximum fluxes of $1.5\times10^{-16}$~W~m$^{-2}$ for CO, $2\times10^{-16}$~W~m$^{-2}$ for OH, $1\times10^{-16}$~W~m$^{-2}$ for o-H$_2$O, and $2\times10^{-15}$~W~m$^{-2}$ for [\ion{O}{i}].
    %$2\times10^{-17}$~W~m$^{-2}$ for the CO and H$_2$O maps, $1\times10^{-17}$~W~m$^{-2}$ for the OH maps, \textbf{and $2\times10^{-16}$~W~m$^{-2}$ for the \ion{O}{i} maps}. The rms level varies between $1\times10^{-17}$~W~m$^{-2}$ and $6\times10^{-17}$~W~m$^{-2}$ across the spectrum. Dashed contours correspond to negative fluxes.
    }
    \label{fig:important_5}
\end{figure*}

The total emission continuum maps of IRS7 and IRS5 at three representative wavelengths can be found in Figs.~\ref{fig:important_7}--\ref{fig:important_5}, and continuum maps at 20\micron\ steps can be found in Figs.~\ref{fig:contmaps7}--\ref{fig:contmaps5}. These continuum maps show extended emission in the
order of 30--60\arcsec\ in size. The larger sizes are found in the longer
wavelengths (which could be suggested to be attributed only to the larger beam
size at these wavelengths, but the shapes of the emission suggest differently; see also Sect.~\ref{sec:continuum_pomac}). 
The highest continuum flux densities in the strongest illuminated spaxels are found around 85\micron\
in IRS7 and 105\micron\ in IRS5, but the spatially integrated flux density is
strongest around 120\micron\ in IRS7 and 155\micron\ in IRS5 (since the longer
wavelength data have more extended PSFs, which spreads the signal over a larger
solid angle). It is, however, difficult to draw conclusions about the stages
of evolution of the individual sources from the total emission, since it is a
combination of several compact sources and any extended emission. For such a
study, deconvolution of the emission is necessary.

The continuum emission in the IRS7 region is very extended in comparison to
the line emission (see Sect.~\ref{sec:results_line_maps}), 
and has at least some signal across most of the two
PACS fields that cover the region (see
Figs.~\ref{fig:important_7}--\ref{fig:important_5}). Without any deconvolution,
it is impossible to distinguish the point sources detected in mid-IR and mm
data (IRS7A, IRS7B, SMM~1C, and R~CrA).

IRS5 consists of two separate sources, of which the northern source IRS5N is
detected in SMA continuum, whereas the southern IRS5A is
detected in \textit{Spitzer} continuum and line emission. It is difficult to distinguish the two sources in the PACS data.

\subsubsection{Spectral line maps}
\label{sec:results_line_maps}

% plot_map.py
\begin{figure*}[!htb]
    \centering
    \includegraphics[width=1.0\linewidth]{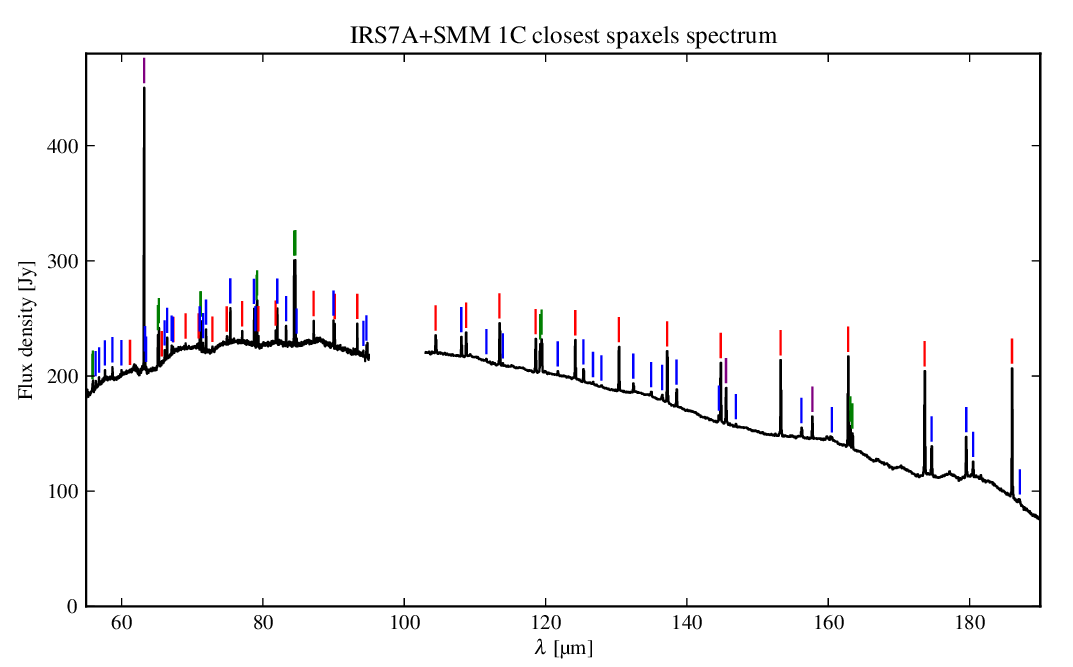} \\
    \caption{Sum of the spectra of the spaxels most nearby to IRS7A and SMM~1C. No deconvolution or correction factors have been applied to the spectrum. It thus appears fainter in the red part compared to the true spectrum. The coloured vertical lines indicate the detected spectral line species: CO (red), OH (green), H$_2$O (blue), and atomic species (purple).
}
    \label{fig:irs7a_spectrum}
\end{figure*}

The sum of the PACS spectra of the
spaxels closest to IRS7A and SMM~1C is shown in Fig.~\ref{fig:irs7a_spectrum}. The detected species are indicated by different colours. The total emission spectral line emission maps of some important lines are found in Figs.~\ref{fig:important_7}--\ref{fig:important_5}, and all spectral line emission maps can be found in
Figs.~\ref{fig:comaps7}--\ref{fig:atomicmaps5}. The CO emission in IRS7 is generally found
to be more extended than the OH and H$_2$O emission. The emission from the
atomic species \ion{C}{ii} and \ion{O}{i} detected in IRS7 is even more
extended. In particular, the [\ion{C}{ii}] line emission peaks off-source in a position east
of IRS7A. Compared to the CO, OH, and H$_2$O emission in the IRS7 cloud, which
is mainly centred on the three point sources, the [\ion{C}{ii}] emission in IRS7
is more extended in the EW-direction. The
[\ion{O}{i}] emission is similar to the CO and OH emission, but apparently much stronger
in SMM~1C than in IRS7A. There is also a relatively strong band of extended
emission in a band W and NE of IRS7A. For IRS5, the spectral line emission seems point-like.

The CO, OH, and H$_2$O emission
detected by PACS in the IRS7 field is centrally peaked on the three protostellar sources in the
IRS7 field (SMM~1C, IRS7A, and IRS7B), with the strongest emission from SMM~1C
and IRS7A. This is consistent with the centrally peaked (but also extended) CO~$J=6\rightarrow5$ and HCO$^+$ emission found by \citet{vankempen09a,vankempen09b}, but different from the very extended H$_2$CO and CH$_3$OH emission detected in SMA
(Submillimeter Array) and APEX (Atacama Pathfinder Experiment) observations
\citep{lindberg12}.
None or very little FIR line emission is seen around SMM~1A, despite the fact
that this region shows very strong H$_2$CO and CH$_3$OH emission in the SMA mm data.

\section{Image analysis}
\label{sec:psfcorr}

As is clearly seen in the total emission continuum and line maps (Figs.~\ref{fig:important_7}--\ref{fig:important_5}), the emission originating from the different sources in the IRS7 and IRS5 fields cannot be easily separated, and a deconvolution method needs to be applied. To test the hypothesis that most of the emission can be accounted for by the previously known point sources, we need to deconvolve the data with the point-spread function (PSF) of the observations.

\subsection{Deconvolution method}
\label{sec:deconv_meth}

The effective diameter of the \textit{Herschel Space Observatory} is 3.28~m.
The \textit{Herschel} PSF is slightly triangular projected on the sky due to the three-point mount of the telescope dish (see Fig.~\ref{fig:pacsbeam}), but it can be roughly
approximated by a Gaussian with a full width at half maximum (FWHM) between 4 and 12\arcsec\ depending on the
wavelength (for our deconvolution algorithm we instead use PSFs constructed from the simulated telescope PSF; see Appendix~\ref{app:pomac}). The PACS spectrometer detector array consists of 25 spaxels, positioned in a square 5$\times$5
pattern. The separations between the spaxel centres are on average
approximately 9\farcs25. This has the effect that emission from a point source will spill over to adjacent
spaxels, particularly for the
longer wavelengths, since the telescope PSF will be larger than the spaxel size. This is normally
easy to correct for when observing a single well-centred point source,
because it is then in principle sufficient to use the central spaxel flux (or
flux density) multiplied by a wavelength-dependent PSF correction
factor\footnote{Refer to the PACS manual v.~2.4 for a detailed description of
the PSF correction factor and the PACS beam:
\url{http://herschel.esac.esa.int/Docs/PACS/html/pacs_om.html}}. However, if
observing off-centred point sources, several point sources in the same field,
or extended emission (or combinations of these), the interpretation of the data
becomes more difficult. We find several point sources in the IRS7 field,
some of them not well-centred on spaxel centres, and also signs of extended
emission in the PACS field, and therefore need another method to separate the
emission from the different possible origins. The IRS7
field was observed in two partially overlapping PACS pointings, which creates
an almost Nyquist-sampled map for all but the shortest wavelengths in the overlap region.

To be able to distinguish point-source emission from more extended emission,
the signal must be deconvolved from the PSF. However, the spatial resolution of
the signal is limited by the design of the PACS instrument, only providing 25
data points for this signal per pointing. We have developed a method that still
can provide deconvolution of point sources from the PACS data, called
POMAC\footnote{Poor Man's CLEAN.}. The method is based on the CLEAN algorithm
\citep{hogbom74}, often used to deconvolve undersampled maps
in radio interferometry. Due to the low number of data points, our method relies on \textit{a priori} knowledge about
the point-source positions. In this case, we employ ALMA data from \citet{lindberg13_alma}, and SMA and \textit{Spitzer} data from \citet{peterson11} to establish the positions of the YSOs with an accuracy of $\lesssim1$\arcsec.  
Based on the larger sample of more isolated sources \citep{green13}, we
expect that a major part of the emission seen with \textit{Herschel}, both continuum and line,
will originate from the point sources seen at mid-infrared and submillimetre
wavelengths, and we use the deconvolution to test this hypothesis. The CLEAN algorithm is used
with the modification that it is only allowed to identify these pre-defined
point sources as sources of emission. The algorithm then iterates over the PACS data with customary break
criteria (such as avoiding subtraction below the noise floor in any spaxel).
After this, the residual map can be studied to identify previously unknown
point sources as well as extended emission. Repeating the process after adding
new point sources will eventually leave all extended emission in the residual map
(however, still convolved with the PSF). The algorithm was tested on the PACS spectrometer continuum of the disc source HD~100546 \citep{sturm10}, which is not expected to differ significantly from a point source in the continuum. The deconvolution produced results within errors of those obtained using the PACS PSF correction factor across the whole PACS band, and no significant residuals were noted. A more detailed explanation of the
POMAC algorithm, including a description of how the telescope PSFs were generated, can be found
in Appendix~\ref{app:pomac}.

\subsubsection{Definition of point sources}

The sources treated in this paper have been studied in several previous papers,
giving them many different names. To facilitate comparison with other work, a
list of the most common names for these point sources found in the literature
can be found in Table~\ref{tab:pointsource}. We will maintain the name usage 
of \citet{lindberg12}.

\begin{table}[!htb]
\centering
\caption[]{Continuum point sources in mid-IR and/or (sub)mm data.}
\label{tab:pointsource}
\begin{tabular}{l l l l l l}
\noalign{\smallskip}
\hline
\hline
\noalign{\smallskip}
Name & RA & Dec & Other names \\
& (J2000.0) & (J2000.0) & \\
\noalign{\smallskip}
\hline
\noalign{\smallskip}
IRS7A\tablefootmark{a} & 19:01:55.33 & $-$36:57:22.4 & IRS7W, IRS7 \\
SMM~1C\tablefootmark{a} & 19:01:55.31 & $-$36:57:17.0 & SMA~2, Brown 9 \\
IRS7B\tablefootmark{a} & 19:01:56.42 & $-$36:57:28.4 & IRS7E, SMM~1B, \\
                                & & & SMA~1 \\
CXO 34\tablefootmark{a} & 19:01:55.78 & $-$36:57:27.9 & FP-34 \\
R~CrA\tablefootmark{b} & 19:01:53.67 & $-$36:57:08.0 & & \\
IRS5A\tablefootmark{b} & 19:01:48.03 & $-$36:57:22.2 & CrA-19, IRS5ab \\
IRS5N\tablefootmark{c} & 19:01:48.47 & $-$36:57:14.9 & CrA-20, SMM~4, \\
                                & & & FP-25 \\
\noalign{\smallskip}
\hline
\end{tabular}
\tablefoot{
	\tablefoottext{a}{Coordinates from ALMA 0.8~mm continuum map \citep{lindberg13_alma}.}
	\tablefoottext{b}{Coordinates from \textit{Spitzer} 4.5\micron\ image data \citep{peterson11}.}
	\tablefoottext{c}{Coordinates from SMA 1.3~mm maps \citep{peterson11}.}
     	}
\end{table}

To match
the line and continuum emission with compact objects we rely
on \textit{a priori} position data from other observations with better
resolution, thus obtained in other wavebands. To cover all sources that could
be visible in the FIR, we used both longer (submillimetre) and shorter
(mid-infrared) wavelength observations to identify possible point sources.
\citet{peterson11} used SMA 226~GHz continuum observations to identify four
continuum peaks in the IRS7 and IRS5 regions (IRS7B, SMA2 / SMM~1C, IRS5N, and
R~CrA); and Spitzer 4.5\micron\ observations to find five continuum
peaks in the same fields (R~CrA, CrA-19 / IRS5A, IRS7B, IRS7A, and CXO~34). We identified four of these sources (IRS7B, SMM~1C, IRS7A, and CXO~34) in ALMA Cycle~0 observations of the 342~GHz continuum centred at IRS7B \citep{lindberg13_alma}. The ALMA coordinates are of superior accuracy, and will be used for these sources. The
coordinates for these sources can be found in Table~\ref{tab:pointsource}.

From these point-source positions, we selected the six most prominent sources
(only excluding CXO~34, which is almost an order of magnitude weaker at
4.5~\hbox{\textmu}m than the second weakest infrared source) to use for the
deconvolution of both the continuum and the line emission. The excluded source
CXO~34 is not only very weak, but also situated between the much stronger
sources IRS7B and IRS7A, and any far-infrared emission originating from CXO~34 would not be possible to
disentangle from the emission of the surrounding sources.

IRS7A and SMM~1C have an angular separation of only 5\arcsec, i.e. not
spatially resolved by the PACS array.  We attempted to separate the
emission from IRS7A and SMM~1C, but we found that the results were
unreliable, also for the shorter wavelengths.  Therefore, we have
decided to treat IRS7A and SMM~1C together, summing the fluxes given
for the two sources. This is not ideal, however, since they are
suggested to be of different types (Class~0 and Class~I,
respectively). Since the line and continuum emission in this region is
extended in the north-south direction it is likely that both sources
contribute to this emission.

\subsection{Results of the continuum deconvolution}
\label{sec:continuum_pomac}

\begin{figure*}[!htb]
    \centering
    \includegraphics[width=0.48\linewidth]{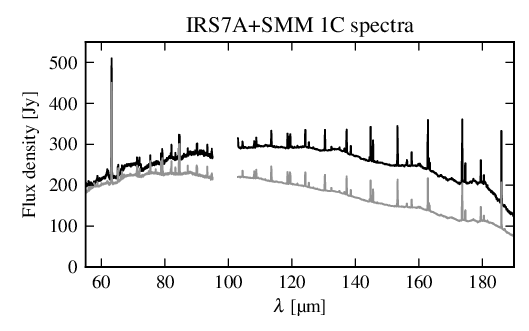} \\
    $\begin{array}{c@{\hspace{0.0cm}}c@{\hspace{0.0cm}}c}
    \includegraphics[width=0.48\linewidth]{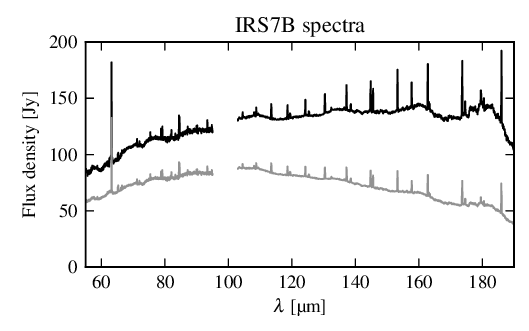} &
    \includegraphics[width=0.48\linewidth]{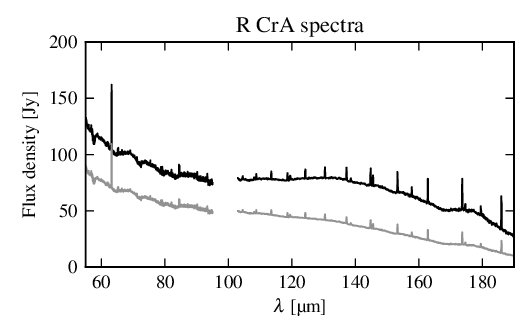} \\
    \includegraphics[width=0.48\linewidth]{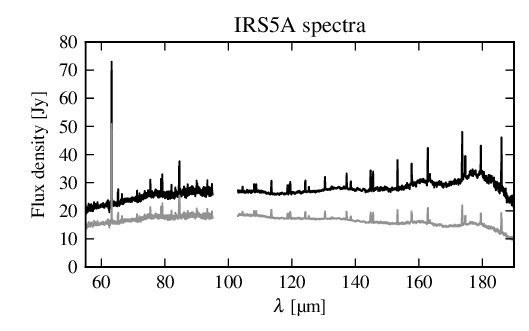} &
    \includegraphics[width=0.48\linewidth]{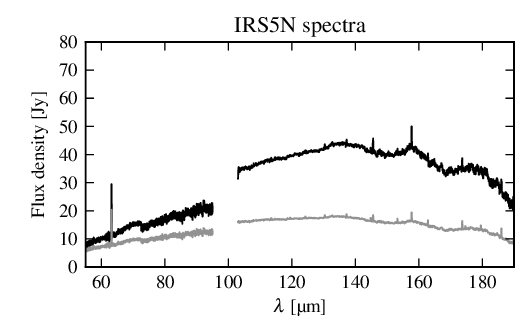} \\
    \end{array}$
    \caption{Spectra of the deconvolved point-source emission (black) and the non-deconvolved data (grey). Note that since
the flux density in the line channels are sums of the line and continuum flux densities,
these will not be properly cleaned (line strengths are under-estimated in these
spectra, due to the more extended continuum). The deconvolved spectra shown here are used for a
sanity check on the continuum deconvolution (see
Sect.~\ref{sec:continuum_pomac}), but not for extracting line fluxes (see
Sect.~\ref{sec:line_pomac} for a description of the method used for the line
flux deconvolution).}
    \label{fig:spectra_pomac}
\end{figure*}

Deconvolved and non-deconvolved spectra of the point sources are found in
Fig.~\ref{fig:spectra_pomac}. When comparing the non-deconvolved and deconvolved spectra,
we find that the spectral lines are stronger compared to the continuum in the
deconvolved data (especially in the IRS7 sources), indicating that the spectral
line emission is less extended than the continuum emission, a pattern which
is also seen in the larger DIGIT embedded objects sample \citep[hereby
referred to as ``the DIGIT sample'';][]{green13}, although data of
more crowded regions \citep[e.g. Serpens;][]{dionatos13} or strong
outflow sources \citep[e.g. L1448-MM;][]{lee13} show spatial
extent. As in the other sources in the DIGIT sample, the emission from
[\ion{O}{i}] is more extended than that from the other species. We also find that
IRS5N only shows a few spectral lines, but has a continuum as strong as that of IRS5A. The bumps in the spectra around 180\micron\ are caused by an
instrument leakage effect.

%plot_spectra.py

The continuum deconvolution is performed on averaged line-free spectral boxes
every 5\micron. As a sanity check, these results are compared to the continuum
level found in the channel-by-channel deconvolution
(see above and Fig.~\ref{fig:spectra_pomac}), and the two methods are found to be in good agreement.

The total emission maps as well as the deconvolution residual maps at three
different wavelengths can be found in the first six panels of
Figs.~\ref{fig:important_7}--\ref{fig:important_5}. Continuum total emission
maps and residual maps in 20\micron\ steps can be found in
Figs.~\ref{fig:contmaps7}--\ref{fig:contmaps5}.

\begin{figure}[!htb]
	\centering
    \includegraphics[width=1.0\linewidth]{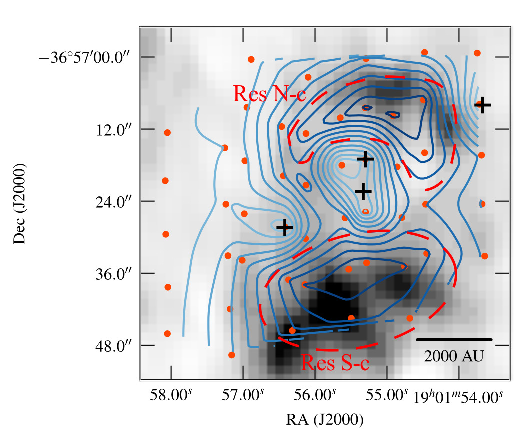}
    \caption{The 110\micron\ continuum residual map (coloured contours) of IRS7
    overplotted on the H$_2$CO $3_{03}\rightarrow2_{02}$ emission (greyscale) from
    SMA+APEX observations \citep{lindberg12}. The orange dots show the PACS spaxel
    centres. Contour levels are at 5~Jy intervals and go from light blue to dark blue in colour as the level of emission increases (i.e., the residual emission is at a minimum near the centre of the figure). The red dashed lines illustrate our definition of the two extended (residual) ridge regions (Res~N\nobr c and Res~S\nobr c) in IRS7
continuum. The crosses show the point sources used for the deconvolution.}
    \label{fig:cont_extreg}
\end{figure}

Running the POMAC deconvolution algorithm on the continuum data leaves strong residual emission
in a shape that remains constant across most of the PACS spectral band, except for the shortest wavelengths.
We find only a little continuum residual emission in the
centre of the IRS7 field, but strong residual emission is found in
two ridges north and south of the YSOs. This could be an effect of the stop criteria used by the POMAC algorithm, which assume that all emission on the point-source positions shall be attributed to the point sources, and should thus not leave considerable residuals on-source. We did several POMAC runs with different stop criteria. None of them could reproduce a smooth continuum residual, but either showed the ridges or left significant emission on the point source positions. The continuum residual at 110\micron\ is shown in Fig.~\ref{fig:cont_extreg}, which is used to define the residual regions Res~N\nobr c and Res~S\nobr c (c for continuum). These regions are chosen to coincide with the spaxels with a residual continuum spectral flux density of at least 30~Jy at 110\micron. Also shown are the ridges of molecular gas detected in H$_2$CO and CH$_3$OH
\citep{lindberg12}, which bear resemblance in shape
and position to the PACS continuum residuals. Res~S\nobr c coincides partially with the pre-stellar core SMM~1A detected in SCUBA submm observations \citep{nutter05}. These similarities indicate that the deconvolution algorithm and the used stop criteria produce accurate results.
Also in the IRS5 field there is some residual continuum emission, but only at
wavelengths longer than 100\micron. 

% extregs.py

\subsection{Results of the spectral line deconvolution}
\label{sec:line_pomac}

For the deconvolution of the spectral lines, the flux of each spectral line in
each spaxel is first measured after subtraction of linear baselines, producing total line flux maps. These total line flux maps are then the input data to the
POMAC algorithm. The line strengths are somewhat higher (approximately 20\%)
compared to those found if extracting line strengths from the
continuum-subtracted channel-by-channel-deconvolved spectra
(Fig.~\ref{fig:spectra_pomac}). This is due to the stop criteria, making the
algorithm reach the noise floor earlier in the channel-by-channel data than in the
line flux data, since the S/N level is lower in the individual channels than for the total line fluxes.

The line flux maps as well as the deconvolution residual maps of seven important
spectral lines can be found in the 14 last panels of
Figs.~\ref{fig:important_7}--\ref{fig:important_5}. All total line intensity
maps and residual maps can be found in
Figs.~\ref{fig:comaps7}--\ref{fig:atomicmaps5} in Appendix~\ref{app:linemaps}. The extracted point source line fluxes are listed in Table~\ref{tab:herschel_lineparams}. Note that, as in the case of the continuum deconvolution, the POMAC algorithm will attribute all on-source emission to the point sources, and not leave residuals at the point-source positions. One could also assume that the point sources are sitting on a plateau of extended emission, but attempts to model that situation with different stop criteria have been unsuccessful. In any case, since the residual emission is much fainter than the point-source emission, such a solution would not change the results of excitation analysis of the point sources dramatically. It is difficult to give an exact estimate on this contribution, since the residual emission is primarily found west of the IRS7 point sources, and a smooth distribution of the extended emission around these sources is thus not possible. For CO, the molecule with the most prominent extended emission, the point sources line fluxes would be $\lesssim25\%$ lower assuming a flat distribution of the extended emission as strong as the residual west of the point sources, but it shall be noted that this is a worst-case scenario. Our estimates show that the errors on the rotational temperatures (Sect.~\ref{sec:rotdiag_analysis}) would increase by $\sim60\%$ assuming this scenario. The reported extended emission should, on the other hand, be seen as a lower limit on the amount of extended gas.

Some spectral lines suffer heavily from line blending, and will not be
considered in the further analysis. These are listed in
Table~\ref{tab:blended}. In addition, only spectral lines between 55\micron\
and 100\micron, and 103\micron\ and 195\micron\ are considered, due to leakage
and/or a high noise level in the outer parts of the bands. The instrument suffers from leakage also in the ranges 70--73\micron\ and 98--105\micron, so
line strengths in these bands are less reliable than those in other bands.

\begin{table*}
\centering
\caption[]{Spectral lines in the observed spectral band which are not considered in the analysis since they suffer from line
blending.}
\label{tab:blended}
\begin{tabular}{l l l l l}
\noalign{\smallskip}
\hline
\hline
\noalign{\smallskip}
Species & Transition & Wavelength & Blend & Transition \\
& & [\hbox{\textmu}m] & & \\
\noalign{\smallskip}
\hline
\noalign{\smallskip}
o-H$_2$O & $6_{25}\rightarrow5_{14}$ & \phantom{0}65.2 & OH & $^2\Pi_{3/2}(J=9/2-\rightarrow7/2+)$ \\
CO & $J=35\rightarrow34$ & \phantom{0}74.9 & o-H$_2$O &
$7_{25}\rightarrow6_{34}$ \\
o-H$_2$O & $7_{25}\rightarrow6_{34}$ & \phantom{0}74.9 & CO &
$J=35\rightarrow34$ \\
CO & $J=31\rightarrow30$ & \phantom{0}84.4 & OH &
$^2\Pi_{3/2}(J=7/2+\rightarrow5/2-)$ \\
OH & $^2\Pi_{3/2}(J=7/2+\rightarrow5/2-)$ & \phantom{0}84.4 & CO & $J=31\rightarrow30$ \\
o-H$_2$O & $6_{25}\rightarrow6_{16}$ & \phantom{0}94.6 & o-H$_2$O &
$4_{41}\rightarrow4_{32}$ \\
o-H$_2$O & $4_{41}\rightarrow4_{32}$ & \phantom{0}94.7 & o-H$_2$O &
$6_{25}\rightarrow6_{16}$\\
CO & $J=23\rightarrow22$ & 113.4 & o-H$_2$O & $4_{14}\rightarrow3_{03}$ \\
o-H$_2$O & $4_{14}\rightarrow3_{03}$ & 113.5 & CO & $J=23\rightarrow22$ \\
p-H$_2$O & $3_{22}\rightarrow3_{13}$ & 156.2 & o-H$_2$O &
$5_{23}\rightarrow4_{32}$ \\
o-H$_2$O & $5_{23}\rightarrow4_{32}$ & 156.3 & p-H$_2$O &
$3_{22}\rightarrow3_{13}$ \\
\noalign{\smallskip}
\hline
\end{tabular}

\end{table*}

\subsubsection{CO emission patterns}

We find CO emission associated with all the pre-defined point sources in the deconvolution, although only very faint emission is found to originate from IRS5N.

Studying the residual maps (Figs.~\ref{fig:cont_extreg} and \ref{fig:comaps7}--\ref{fig:comaps5}), considerable extended CO emission is found
southwest, east, north, and west of the point sources in the IRS7 field for the
lower-\textit{J} CO lines ($J \lesssim 25$). For the higher-\textit{J} CO
lines, the S/N is too low to find more than traces of such emission, but this
will be further discussed in Sect.~\ref{sec:co_extended_ex}. In the IRS5 field,
the CO emission is completely point-like, leaving no residuals after the
deconvolution.

The relatively strong extended emission in the SW part could
 also be explained by
a CO point source in a position not associated with any YSO. No corresponding
point source has been identified in the SMA data, in the \textit{Spitzer} data,
or in any source catalogue, but we cannot rule out that this is a very faintly
emitting YSO. However, we do not consider this to be a point source, and the
excitation conditions of this emission will be treated together with that of
the other extended CO emission in Sect.~\ref{sec:co_extended_ex}. This peak
bears resemblance to an outflow front, but it does not align well with the EW
outflow found by \citet{vankempen09a}. The origin is thus uncertain.

The CO line emission in IRS5 is well-centred on the spaxel corresponding to
IRS5A (see CO emission maps and CO residual maps in Figs.~\ref{fig:important_5} and \ref{fig:comaps5}),
and when running the POMAC code on this data assuming IRS5A as the only point
source, only marginal residuals are found. It is thus reasonable to believe
that IRS5N produces only a very low amount of CO emission.

\subsubsection{OH and H$_2$O emission patterns}

The OH and H$_2$O (p-H$_2$O and o-H$_2$O) line emission will be treated in the
same section due to their similar emission pattern. IRS7A+SMM~1C are strong emitters
of lines from these molecules, whereas IRS7B is much weaker in the OH and H$_2$O transitions. The OH, p-H$_2$O, and o-H$_2$O line emission maps and residual
maps are found in Figs.~\ref{fig:ohmaps7}--\ref{fig:oh2omaps5}. There is less
extended emission in the OH and H$_2$O line data compared to the CO data, in
particular, there is no considerable extended emission southwest of IRS7. There
is, however, some extended OH and H$_2$O emission west of IRS7.

As in the case of CO, IRS5A seems to be the dominant emitter of OH and H$_2$O in
the IRS5 field, IRS5N not contributing any significant emission.

\subsubsection{Atomic line emission patterns}
Compared to the CO, OH, and H$_2$O emission in the IRS7 cloud, which is mainly
centred on the three point sources, the [\ion{C}{ii}] emission is more extended
in the EW-direction and peaks in different positions from the other spectral lines in the IRS7 field. However, the [\ion{C}{ii}] PACS data of protostellar sources often suffer from emission in the
off-positions. We have investigated the signal using the two different nod
positions, and the general structure is similar but not identical in these two data sets, which suggests that the detected [\ion{C}{ii}] morphology may partly be an observational effect. The [\ion{C}{ii}] data could thus be unreliable, and will not be further discussed.

The [\ion{O}{i}] emission peaks on the point sources, but also shows strong
extended emission in the whole IRS7 field, with residual peaks similar to those
of OH and H$_2$O, indicating that they trace the same extended gas. This is a
good indicator of large-scale PDR (photo-dissociation region) activity or alternatively outflow-associated shocks, since H$_2$O can be
photo-dissociated into OH and O \citep{hollenbach97}. The [\ion{O}{i}] data do not suffer from off-position emission to a significant level. The line ratio between the 145\micron\ and 63\micron\ [\ion{O}{i}] lines varies between 0.06 and 0.11 for the point sources, which is also indicative of a strong radiation field (i.e. PDR activity; \citealt{kaufman99}).

Around IRS5, however, the [\ion{C}{ii}] and [\ion{O}{i}] emission can be
explained by two point sources centred at IRS5A and IRS5N.

\subsection{Comparing line and continuum emission}

Using our deconvolution algorithm POMAC, we find that the FIR line emission
mostly originates from the (sub)mm/mid-IR continuum point sources (but there is also
some CO and OH line emission from residual regions: Res~SW\nobr l, Res~E\nobr l, Res~N\nobr l, and Res~W\nobr l), whereas the FIR continuum shows a much more extended shape (see Fig.~\ref{fig:cont_extreg}). After
using the deconvolution algorithm we find that most of the continuum emission
not associated with point sources can be found in two ridges extending in the
east-west direction, positioned north and south of the YSO point sources. These
ridges coincide with molecular (H$_2$CO and CH$_3$OH) emission detected
in millimetre data, proposed to be heated by external irradiation from the Herbig~Be star R~CrA \citep{lindberg12}.

% extregs.py
\begin{figure}[!htb]
	\centering
    \includegraphics[width=1.0\linewidth]{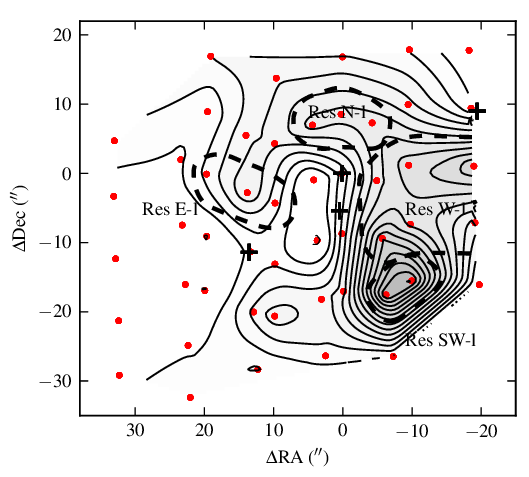}
    \caption{The dashed lines illustrate our definition of the four extended (residual) regions Res~SW\nobr l, Res~E\nobr l, Res~N\nobr l, and Res~W\nobr l in IRS7 CO and OH. The contour map shows the CO $J=19\rightarrow18$ residual map. The red dots show
the PACS spaxel centres. Contours are $3\sigma = 1.55 \times 10^{-17}$~W~m$^{-2}$.}
    \label{fig:co_extreg}
\end{figure}

The northern ridge (Res~N\nobr c and Res~N\nobr l) coincides not only with the northern
H$_2$CO ridge observed by \citet{lindberg12}, but also to some extent with
HCO$^+$ $J=3\rightarrow2$ emission \citep{groppi07} and X-ray emission
\citep{forbrich07}. Another possible explanation for the 
physical conditions found in the region could thus be that the gas is dominated
by X-ray irradiation.

\section{Analysis}

\subsection{Analysis of the spectral energy distributions}
\label{sec:sed_analysis}

\begin{figure*}[!htb]
    \centering
    $\begin{array}{c@{\hspace{0.0cm}}c@{\hspace{0.0cm}}c}
    \includegraphics[width=0.48\linewidth]{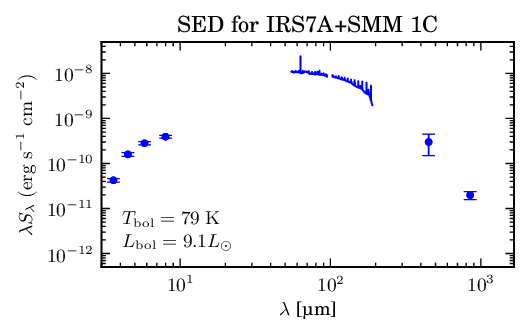} &
    \includegraphics[width=0.48\linewidth]{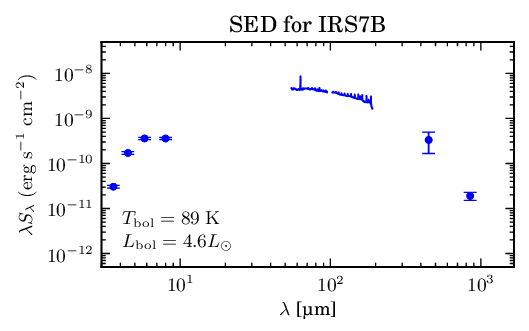} \\
    \includegraphics[width=0.48\linewidth]{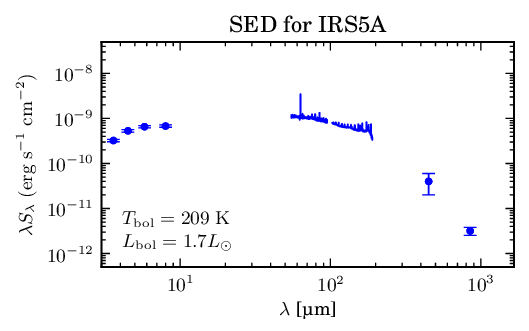} &
    \includegraphics[width=0.48\linewidth]{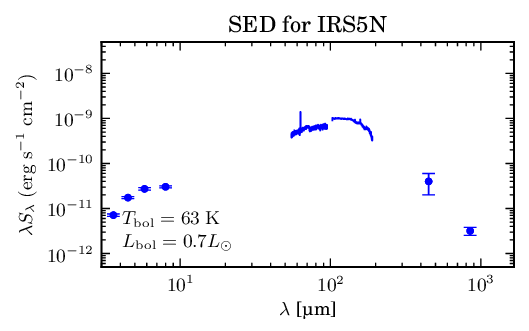} \\
    \end{array}$
    \caption{Spectral energy distributions of the deconvolved point sources. The continuum data points come from \textit{Spitzer} and SCUBA data. The spectra are deconvolved PACS spectra. Refer to Table~\ref{tab:sedfluxes} for details and references.}
    \label{fig:seds}
\end{figure*}

The spectral energy distributions (SEDs) of the low-mass point sources, 
can be found in Fig.~\ref{fig:seds}, where continuum data points from 
\textit{Spitzer} and SCUBA
measurements as well as the deconvolved \textit{Herschel}/PACS spectra have been included.
Bolometric luminosities are calculated from integration of a
first degree spline fit to the SED data points (including line-free points across the \textit{Herschel} spectrum) and bolometric temperatures are calculated from the mean frequencies of these spline fits \citep[see][]{myers98}. The results are shown in Table~\ref{tab:sed}. For the sources with strong
submm emission, the SCUBA data have been linearly extrapolated to allow for
integration of the SEDs up to 1.3~mm \citep[cf.][]{jorgensen09}. The SED of R~CrA is treated in Sect.~\ref{sec:rcra_spect}.

The luminosities of the observed low-mass sources (IRS7A+SMM~1C, IRS7B, IRS5A, and IRS5N) are
all in the order $1~L_{\odot}$--$10~L_{\odot}$. All these sources fulfil the $L_{\mathrm{bol}}/L_{\mathrm{submm}} < 200$ criterion for Class~0 sources \citep{andre93}.
On the other hand, the low-mass sources all have
bolometric temperatures that fall in the Class~I range, except IRS5N, which is
a Class~0 source. However, as discussed previously, IRS7A+SMM~1C is a binary unresolved by
\textit{Herschel}, where the components have very different mid-IR and submm
spectral energy distributions. It is thus likely that the bolometric temperature of SMM~1C
is lower, and that of IRS7A is higher.

Differences between infrared and submillimetre continuum emission are also found
between IRS5A and IRS5N in the IRS5 field. It has been suggested that
IRS5A, being a binary ($\sim 100$~AU), has all the dust in the disc
cleared away \citep{jensen96,peterson11}, explaining why it is not
 detected by SMA mm observations.
According to \citet{peterson11}, both IRS5A and IRS5N are
Class~I sources or younger, with IRS5N having a steeper mid-IR spectral slope
$\alpha$ than IRS5A. The observations of IRS5 can be used as a comparison for the IRS7 sources, since they should be less affected by the irradiation from R~CrA. IRS5A has a mid-IR luminosity a few times higher than that of IRS7A and IRS7B; however, it shows only moderate line emission in the FIR (\textit{Herschel}) data.

\begin{table}[!tb]
\centering
\caption[]{Results of SED fits.}
\label{tab:sed}
\begin{tabular}{l l l l}
\noalign{\smallskip}
\hline
\hline
\noalign{\smallskip}
Source & $T_{\mathrm{bol}}$ & $L_{\mathrm{bol}}$ & $L_{\mathrm{bol}}/L_{\mathrm{submm}}$\\
& [K] & $[L_{\odot}]$ \\
\noalign{\smallskip}
\hline
\noalign{\smallskip}
IRS7A+SMM~1C & \phantom{0}79 & \phantom{0}9.1 & \phantom{0}99 \\
IRS7B & \phantom{0}89 & \phantom{0}4.6 & \phantom{0}48 \\
R~CrA & 889 & 53.4 & \phantom{0}... \\
IRS5A & 209 & \phantom{0}1.7 & 135 \\
IRS5N & \phantom{0}63 & \phantom{0}0.7 & \phantom{0}55 \\
\noalign{\smallskip}
\hline
\end{tabular}
\end{table}

\begin{figure}[!tb]
	\centering
    \includegraphics[width=1.0\linewidth]{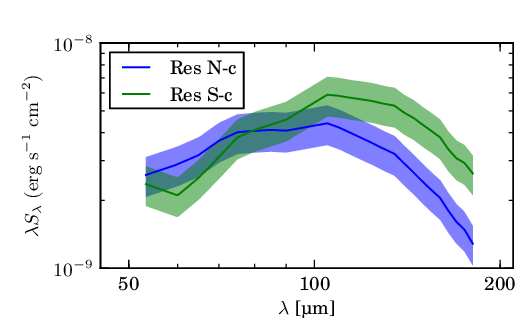}
    \caption{Spectral energy distributions of the continuum residual regions Res~N\nobr c and Res~S\nobr c within the PACS band. The shading shows the 20\% calibration uncertainty of the \textit{Herschel}/PACS spectrometer.}
    \label{fig:sed_res}
\end{figure}

The spectral energy distributions of the residual structures
Res~N\nobr c and Res~S\nobr c within the PACS band (Fig.~\ref{fig:sed_res}) are found to peak around the same wavelength as typical embedded YSOs do (such as the Class~0 source IRS5N). 
The SED peak around 100\micron\ corresponds to a black-body dust temperature of $\sim40$~K.

\subsubsection{Spectral classification and SED of R~CrA}
\label{sec:rcra_spect}

The WHT optical spectrum of R~CrA is consistent with a photosphere of a star with spectral type B3--A0, by comparison to the \citet{pickles98} compilation of photospheric templates. The higher Balmer lines are seen in absorption and are used for the spectral comparison. The large range in acceptable spectral types is caused by the possibility of emission and absorption affecting the Balmer line equivalent widths.  
Strong emission is detected in H$\alpha$ and H$\beta$. P-Cygni and inverse P-Cygni absorption are detected in some lines by \citet{brown13}. Both emission and absorption may affect the equivalent widths in the photospheric lines. A high resolution spectrum would be required to improve the spectral type.  

This spectral type is consistent with the most reliable literature spectral types of B5 \citep{gray06} and B8 \citep{bibo92}, which were also obtained using blue spectra. Other spectral types range from A0--F5 \citep[e.g.][]{joy45,greenstein47,hillenbrand92,vieira03} but are typically based on red spectra, which are much less sensitive to the spectral type of hot stars.  The spectrum cannot be well fit with the median interstellar extinction law, using a total-to-selective extinction ratio $R_{\mathrm{V}}=3.1$. For the \citet{weingartner01} extinction law with $R_{\mathrm{V}}=5.5$, the $A_{\mathrm{V}}=4.5$ for an A0 spectral type and 5.3 for a B3 spectral type. A higher $A_{\mathrm{V}}$ may be obtained with a higher $R_{\mathrm{V}}$ \citep{manoj06}, however such a high $R_{\mathrm{V}}$ is not necessary to explain the shape of the optical spectrum.

The V magnitude at the time of our observation was $\sim 13.2$~mag., as measured in our spectrum. The parameters B6 spectral type, $A_{\mathrm{V}}=5.0$, and $d=130$~pc lead to a luminosity of $22~L_{\odot}$ -- much smaller than that inferred from the total SED and much smaller than that expected for a young B star. The V-band magnitude is variable by $\sim 3$~magnitudes \citep{bibo92}. At its brightest, the star could be $350~L_{\odot}$, assuming no change in the measured extinction. Alternatively, the measured luminosity may be much lower than the real luminosity if the star is seen edge-on, as found in the Robitaille models (see below).

\begin{figure}[!tb]
	\centering
    \includegraphics[width=1.0\linewidth]{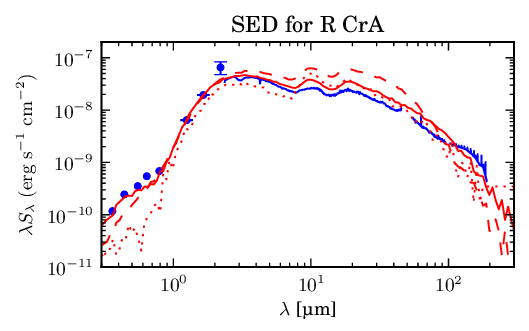}
    \caption{The R~CrA SED (blue) with optical (SAAO) and 2MASS data points, and ISO SWS and \textit{Herschel}/PACS spectra. Note that this SED has a different $x$-axis than the SEDs in Fig.~\ref{fig:seds}.
    Overplotted are the best, fifth best, and tenth best fits from the Robitaille model (red solid, dashed, and dotted).
The best fit corresponds to Robitaille model ID 3011150, with a disc observed at an inclination
of $\sim81\degr$, an $A_{\mathrm{V}} = 3.27$, the distance 130~pc, and
apertures similar to those of the instruments used for the actual observations. The sawtooth pattern at long wavelengths originates from the model being not very accurate at these wavelengths. The ten best fits all correspond to sources with luminosities of at least $480~L_{\odot}$ and high inclinations ($>80\degr$). For the PACS fit, the continuum flux densities at 70\micron, 100\micron, and 160\micron\ have been used.
}
    \label{fig:robitaille}
\end{figure}

The SED of R~CrA including SAAO, 2MASS, ISO SWS, and \textit{Herschel}/PACS data points, is shown in Fig.~\ref{fig:robitaille} (blue data points and spectra). R~CrA is found to have a bolometric
temperature of $\sim900$~K, making it a Class~II YSO; and a luminosity of
$53~L_{\odot}$, significantly lower than the previous value $99~L_{\odot}$--$166~L_{\odot}$ \citep[][the lower value is from integration of the SED and the higher value is from a model of the extinction-free SED]{bibo92}. This discrepancy largely owes to the fact that \citet{bibo92} used KAO data for the FIR data points, and the large KAO beam included most of the R~CrA cloud. As a result, the KAO 100\micron\ flux is 7~times higher than our deconvolved value for R~CrA measured with PACS. Another contribution to the large spread in the spectral classification data in the literature (F5 to B5) could be variability of the source \citep[see e.g.][]{herbst99}.

We here attempt to constrain the physical properties of R~CrA by the use of another method: using a database of SED
models of YSOs \citep{robitaille06} and an
online\footnote{\url{http://caravan.astro.wisc.edu/protostars/}} fitting tool
\citep{robitaille07}, we find that the observed SED can best be explained by a source with stellar mass $M\approx6~M_{\odot}$, a stellar temperature corresponding to a B3 star, a total stellar luminosity
$L\approx900~L_{\odot}$, a nearly edge-on disc (inclination $80\degr$), and an extinction of $A_{\mathrm{V}} \approx 3.3$. The fit is found in Fig.~\ref{fig:robitaille}, together with the fifth and tenth best fits from this model. The ten best fits all correspond to sources with $L_{\mathrm{bol}}\gtrsim480~L_{\odot}$ and nearly edge-on discs.
We also investigate the model SED of a star with the same properties as the best fit, but with a face-on disc. It is found to have flux densities
more than an order of magnitude higher in the UV/optical and a few times
higher in the infrared/submm than the edge-on counterpart. Thus, assuming that R~CrA behaves like this model star, it will heat some
parts of the surrounding regions much more efficiently than other parts,
perhaps giving rise to the ridge-like structures of heated gas and dust. Differences in the density distribution could also contribute to the uneven temperatures in the region. Since the SED database is not exhaustive, and due to the large number of free parameters, the use of this model and the resulting interpretation could be unreliable. \citet{lindberg12} estimate a minimum luminosity for R~CrA of $100L_{\odot}$ in order to heat the molecular gas to the measured temperatures.

\subsection{Rotational diagram analysis}
\label{sec:rotdiag_analysis}

If the line emission is optically thin, the flux of a spectral line can be converted
into a population of molecules in the upper state of the rotational transition it represents.
Rotational temperatures and the total number of emitting molecules 
can be estimated by fitting a line to a plot of upper-state population
versus upper-state energy \citep{goldsmith99,green13}. 
To evaluate the line flux of each spectral line for each YSO, one
could either use the line flux in the spaxel closest to the YSO (with a
wavelength-dependent correction factor\footnote{Refer to the PACS manual v.~2.4, Figure~4.5: \url{http://herschel.esac.esa.int/Docs/PACS/html/pacs_om.html}} applied), or the POMAC method described
in Sect.~\ref{sec:psfcorr} and Appendix~\ref{app:pomac}. When studying an isolated point source, the
difference in the results of these two methods should be small, at least if the
amount of extended emission is reasonably low. However, in a field with several
YSOs (like IRS7), the emission from the YSOs would spill over into each other's
spaxels, so that a strong emitter could influence the measured flux in a weaker
nearby source. This spill-over contribution is minimised when the POMAC method
is used. In this section, all rotational diagrams are calculated with fluxes
estimated from the POMAC algorithm.

\subsubsection{CO -- point-source emission}
\label{sec:co_point_rot}
In the non-deconvolved YSO spaxels, up to 28 CO lines are detected, from the
$J=13\rightarrow12$ line at 200\micron\ to the $J=40\rightarrow39$ line
at 66\micron. However, some CO lines, including the $J=13\rightarrow12$ line, lie in
the leakage spectral region \citep{green13}, and others are blended with other spectral lines (see Table~\ref{tab:blended}). These lines will not be used in the
rotational diagrams. We also detect five $^{13}$CO lines ($J=14\rightarrow13$ to $J=21\rightarrow20$; three of the eight lines in this range are blended with stronger spectral lines).

\begin{figure*}[!htb]
    \centering
    $\begin{array}{c@{\hspace{0.0cm}}c@{\hspace{0.0cm}}c}
    \includegraphics[width=0.48\linewidth]{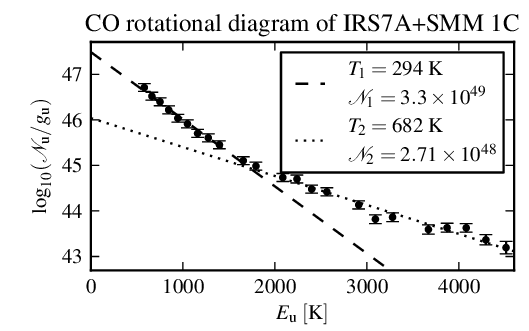} &
    \includegraphics[width=0.48\linewidth]{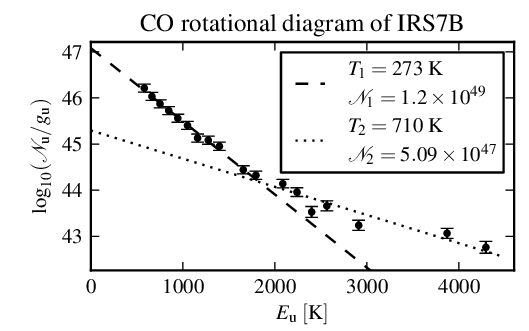} \\
    \includegraphics[width=0.48\linewidth]{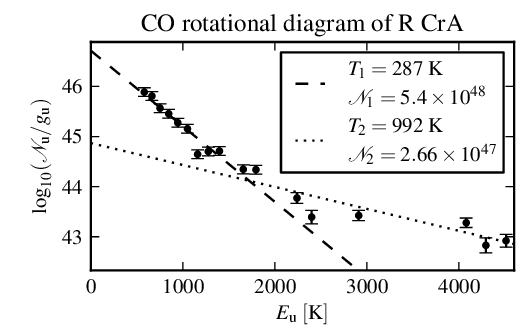} &
    \includegraphics[width=0.48\linewidth]{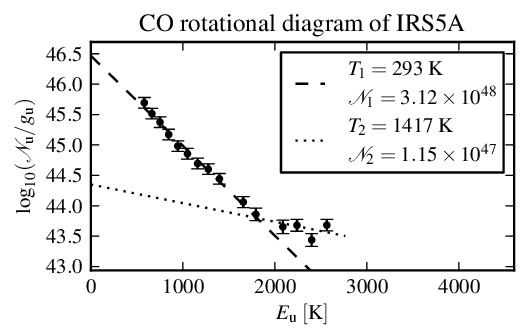} \\
    \includegraphics[width=0.48\linewidth]{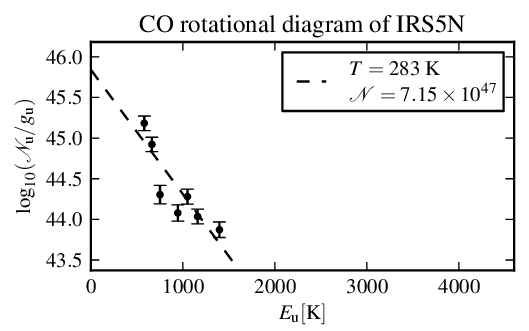} \\
    \end{array}$
    \caption{CO rotational diagrams of the point sources in the deconvolved PACS data. Some high-$J$ data points are missing in some sources since many of these lines lie around the detection limit and the noise level is fluctuating. $\mathcal{N}$ is the total number of CO molecules in each source given the rotational fit.}
    \label{fig:co_ex}
\end{figure*}

The rotational temperature found by this method will correspond to the kinetic
temperature if the cloud is homogeneous and in LTE, and the spectral lines are optically thin. However, in the
case of CO, two rotational temperatures fit the data much better than
does a single rotational temperature 
(see \citealp{green13} for a
justification of the use of a two-component fit; and also \citealp{manoj13}; \citealp{karska13}; \citealp{dionatos13}).
In the case of CrA, the warm component has a temperature of approximately 300~K for all point
sources, whereas the hot component has a larger spread around 650--1400~K
(mainly due to a lower S/N), which both are in the same order as in many other embedded YSOs \citep[see e.g.][]{green13,manoj13,herczeg12}. 

In Table~\ref{tab:rottemp_co}, the
calculated CO rotational temperatures and total number of molecules are given along with the same properties of other similar sources in the literature. 
The errors on the temperatures and numbers of molecules are
calculated assuming a systematic error of 20\% on the line fluxes -- even though
the line fluxes are from the same spectrum, the systematic errors are assumed
to be independent.
The CO rotational diagrams are found in Fig.~\ref{fig:co_ex}.

The CO rotational temperatures towards IRS5A established from the POMAC fluxes are found to be within errors of those derived from
just using the central spaxel flux and the PSF correction factor, as expected for this relatively isolated point source. IRS5N has five times as faint CO emission as IRS5A, but has a similar CO rotational temperature in the warm component. No hot component is detected in IRS5N (if present at a ratio relative to the warm component
seen in other sources, it would be below the detection limit).

\begin{table*}
\centering
\caption[]{CO rotational temperatures and total number of molecules.}
\label{tab:rottemp_co}
\begin{tabular}{l l l l l} % l l l l}
\noalign{\smallskip}
\hline
\hline
\noalign{\smallskip}
YSO & $T_{\mathrm{warm}}$\tablefootmark{a} &
$\mathcal{N}_{\mathrm{warm}}$\tablefootmark{a} &
$T_{\mathrm{hot}}$\tablefootmark{a} &
$\mathcal{N}_{\mathrm{hot}}$\tablefootmark{a} \\
& [K] & [$10^{48}$] & [K] & [$10^{48}$] \\
\noalign{\smallskip}
\hline
\noalign{\smallskip}
IRS7A+SMM~1C\tablefootmark{b} & $294\pm16$ & $33.0\pm3.0$ &
$\phantom{0}682\pm\phantom{0}34$ & $2.7\phantom{0}\pm0.3$ \\
IRS7B & $273\pm14$ & $12.0\pm1.1$ & $\phantom{0}710\pm\phantom{0}54$ &
$0.51\pm0.07$ \\
R CrA & $287\pm16$ & $\phantom{0}5.4\pm0.5$ & $\phantom{0}992\pm\phantom{0}91$ &
$0.27\pm0.03$ \\
IRS5A & $293\pm17$ & $\phantom{0}3.1\pm0.3$ & $1417\pm780$ &
$0.12\pm0.04$ \\
IRS5N & $283\pm24$ & $\phantom{0}0.7\pm0.1$ & \phantom{0}... & ... \\
\noalign{\smallskip}
\hline
\noalign{\smallskip}
Res~SW\nobr l & $285^{+15}_{-33}$ & $\phantom{0}5.0\pm0.5$ &
$\phantom{0}653^{+68}_{-69}$ & $0.35\pm0.06$ \\
Res~E\nobr l & $287^{+16}_{-33}$ & $\phantom{0}1.7\pm0.2$ &
$1015\pm198$ & $0.09\pm0.02$ \\
Res~N\nobr l & $281^{+15}_{-32}$ & $\phantom{0}3.3\pm0.3$ &
$\phantom{0}898\pm334$ & $0.11\pm0.06$ \\
Res~W\nobr l & $253^{+12}_{-26}$ & $13.0\pm1.2$ &
$\phantom{0}751^{+93}_{-94}$ & $0.40\pm0.07$ \\
\noalign{\smallskip}
\hline
\noalign{\smallskip}
CrA point-source average & $286\pm\phantom{0}3$\tablefootmark{c} & $10.8\pm5.2$\tablefootmark{c} & $\phantom{0}950\pm148$\tablefootmark{c}
& $0.90\pm0.52$\tablefootmark{c} \\
CrA extended average & $277\pm\phantom{0}7$\tablefootmark{c} & $\phantom{0}5.8\pm2.2$\tablefootmark{c} &
$\phantom{0}829\pm\phantom{0}69$\tablefootmark{c} & $0.24\pm0.07$\tablefootmark{c} \\
\noalign{\smallskip}
\hline
\noalign{\smallskip}
NGC 1333 IRAS 4B\tablefootmark{d} & 280 & 40 & \phantom{0}880 & 3 \\
Serpens SMM1\tablefootmark{e} & $337\pm40$ & \phantom{0}... &
$\phantom{0}622\pm\phantom{0}30$ & ... \\
Serpens SMM3/4 average\tablefootmark{f} & $260\pm10$\tablefootmark{c} & $49\phantom{.0}\pm6$\tablefootmark{c} & $\phantom{0}800\pm\phantom{0}60$\tablefootmark{c} & $2.2\phantom{0}\pm0.6$\tablefootmark{c} \\
\noalign{\smallskip}
\hline
\noalign{\smallskip}
DIGIT average\tablefootmark{g} & $355\pm\phantom{0}3$\tablefootmark{c} &
$\phantom{0}5.2\pm0.4$\tablefootmark{c} & $\phantom{0}814\pm\phantom{0}29$\tablefootmark{c} & $1.63\pm0.20$\tablefootmark{c} \\
HOPS average\tablefootmark{h} & $288\pm14$\tablefootmark{c} & \phantom{0}... & $\phantom{0}735\pm\phantom{0}37$\tablefootmark{c} & ... \\
\noalign{\smallskip}
\hline
\end{tabular}
\tablefoot{
     	\tablefoottext{a}{The methods used for calculating the error estimates are discussed in Sects.~\ref{sec:co_point_rot}--\ref{sec:co_extended_ex}.}
     	\tablefoottext{b}{The separation between these sources is too small to
allow for treating their line fluxes independently using POMAC.}
     	\tablefoottext{c}{For the sample averages, standard deviations of the mean (the standard deviation of the sample divided by the square root of the sample size) of the significant fits are given.}
     	\tablefoottext{d}{From \citet{herczeg12}.}
     	\tablefoottext{e}{From \citet{goicoechea12}.}
     	\tablefoottext{f}{Average of SMM3b, SMM3c, SMM3r, and SMM4, from \citet{dionatos13}, assuming a distance of 415~pc.}
     	\tablefoottext{g}{From \citet{green13}, including all DIGIT embedded sources except the CrA and Serpens SMM3/4 sources. The Serpens region is, like the CrA region, suffering from confusion and is treated in a separate paper \citep{dionatos13}.}
     	\tablefoottext{h}{From \citet{manoj13}.}
     	}
\end{table*}

\begin{figure}[!htb]
    \centering
    \includegraphics[width=1.0\linewidth]{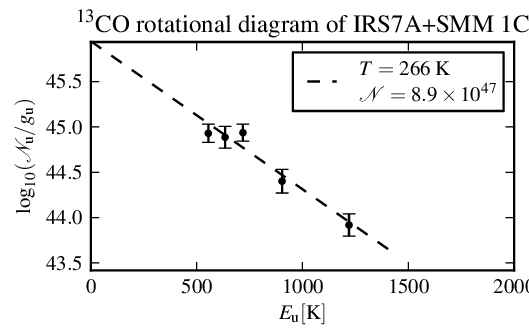} \\
    \caption{$^{13}$CO rotational diagram of IRS7A+SMM~1C from the deconvolved PACS data. $\mathcal{N}$ is the total number of $^{13}$CO molecules in each source given the rotational fit.}
    \label{fig:13co_ex}
\end{figure}

We also perform a rotational diagram fit for the $^{13}$CO data in IRS7A+SMM~1C (see Fig.~\ref{fig:13co_ex}). From the five detected lines, we get a rotational temperature of $266\pm35$~K, which is consistent with the $^{12}$CO temperature $294\pm16$~K. The number of molecules is found to be $(8.90\pm1.58)\times10^{47}$, but if the temperature is constrained to the $^{12}$CO value the number of molecules becomes slightly lower, $7.47\times10^{47}$. With the $^{13}$CO rotational temperature constrained to the $^{12}$CO value, we find the $^{12}$CO/$^{13}$CO abundance ratio to be $44\pm9$, corresponding to an optical depth of $0.56\pm0.24$ assuming the local ISM $^{12}$C/$^{13}$C value of $77\pm7$ \citep{wilson94}. We adopt a CO line width of $\sim7.5$~km~s$^{-1}$ from \textit{Herschel} HIFI observations of the $^{12}$CO $J=16\rightarrow15$ line (Kristensen et al. in prep.). This value is comparable to the $^{12}$CO $J=7\rightarrow6$ quiescent component line width \citep[8~km~s$^{-1}$;][]{vankempen09a}. We use RADEX \citep{vandertak07}, a
non-LTE radiative transfer code for calculations of line strengths in
isothermal homogeneous interstellar clouds, to find that this marginally optically thick result is consistent with a $^{12}$CO column density of $\sim10^{18}$~cm$^{-2}$, corresponding to a size of the emitting region in the order of a few arcseconds ($\sim500$~AU).

We have also investigated whether the data can be fitted
with a single kinetic temperature component of much higher temperature and lower
density, as suggested by \citet{neufeld12}.
RADEX calculations assuming the same line width as above (7.5~km~s$^{-1}$) and the column density derived above ($\sim10^{18}$~cm$^{-2}$) give a single kinetic
temperature component with a best fit at
$\sim5\,000$~K ($>2500$~K with a $1\sigma$ certainty) for an H$_2$ density $(2.5\pm0.5)\times10^4$~cm$^{-3}$ for IRS7A+SMM~1C. For the best solution, the lowest-$J$ ($J\lesssim17$) lines are marginally optically thick ($\tau \sim 1$), while the lines with $J\gtrsim 22$ have optical depths $\tau \ll 1$. The reduced $\chi^2$ value for the best fit is 1.1.

\subsubsection{CO -- extended emission}
\label{sec:co_extended_ex}
To investigate if the extended CO emission in the IRS7 region shows any variation in temperature, it was grouped into four areas: southwest,
east, northwest, and west of the YSOs (see Fig.~\ref{fig:co_extreg}). 
We call the four residual
emission regions Res~SW\nobr l, Res~E\nobr l, Res~N\nobr l, and Res~W\nobr l (l for spectral line). Res~N\nobr l overlaps with the central part of the continuum residual region Res~N\nobr c, and Res~SW\nobr l overlaps with the western part of Res~S\nobr c (see Fig.~\ref{fig:cont_extreg}).
By computing the residual spectral line emission in these regions
we can produce
rotational diagrams for this extended emission.

\begin{figure*}[!htb]
    \centering
    $\begin{array}{c@{\hspace{0.0cm}}c@{\hspace{0.0cm}}c}
    \includegraphics[width=0.48\linewidth]{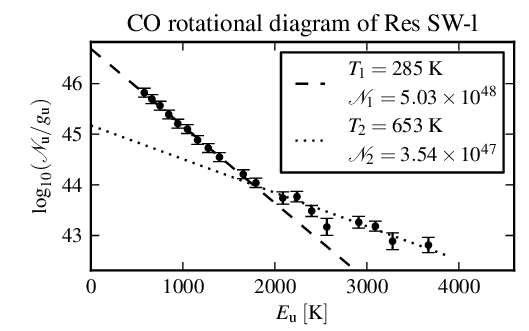} &
    \includegraphics[width=0.48\linewidth]{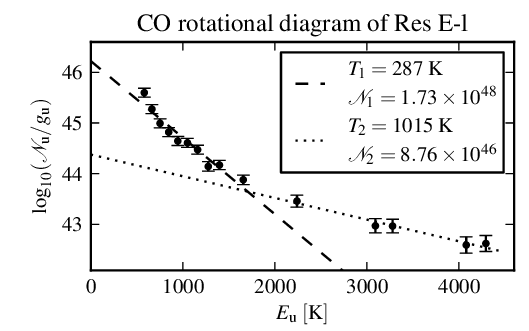} \\
    \includegraphics[width=0.48\linewidth]{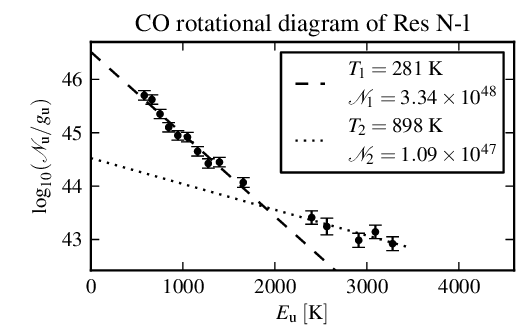} &
    \includegraphics[width=0.48\linewidth]{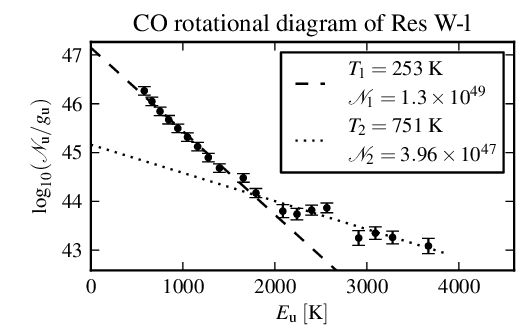} \\
    \end{array}$
    \caption{CO rotational diagrams for the residual emission of the extended regions Res~SW\nobr l, Res~E\nobr l,
Res~N\nobr l, and Res~W\nobr l. $\mathcal{N}$ is the total number of CO molecules in each source given the rotational fit.}
    \label{fig:co_ex_res}
\end{figure*}

The resulting rotational diagrams are found in Fig.~\ref{fig:co_ex_res};
as for the point-source emission, a warm and a hot
component with rotational temperatures around 300~K and 900~K, respectively, are found. The exact properties of the fits are included in
Table~\ref{tab:rottemp_co}. The rotational temperatures of the residual emission are within the errors of the point-source rotational temperatures.

On one hand, the extended emission is not deconvolved, and one can thus argue
that a PSF correction factor needs to be applied to this data. On the other
hand, since these fluxes are sums of emission from several (2--6) spaxels (see Fig.~\ref{fig:co_extreg}), they are
clearly in less need of PSF correction than point-source data. However, to take
this issue into account, we calculate the rotational temperatures of the extended
emission also using the PACS standard PSF correction factor from the PACS
manual, and use the result when establishing the lower boundary of the error estimate of these temperatures.

\subsubsection{OH and H$_2$O}

The excitation conditions of the related species, OH and H$_2$O, are
treated together in this section.

For the unresolved OH doublets, the sum of both lines is measured and divided
by 2. In the rotational diagram fits we exclude the same lines as \citet{wampfler13}. These
excluded lines are the 119\micron\ doublet (which in similar sources is found to be an optically thick transition), the
84.4\micron\ line (CO blend), and the 98\micron\ and 55\micron\ doublets (lines
in leakage regions). They are plotted with open circles in the rotational diagrams.

\begin{figure*}[!htb]
    \centering
    $\begin{array}{c@{\hspace{0.0cm}}c@{\hspace{0.0cm}}c}
    \includegraphics[width=0.48\linewidth]{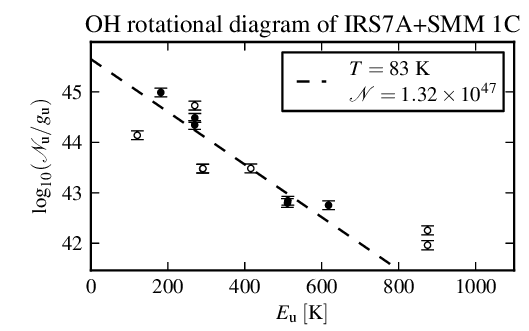} &
    \includegraphics[width=0.48\linewidth]{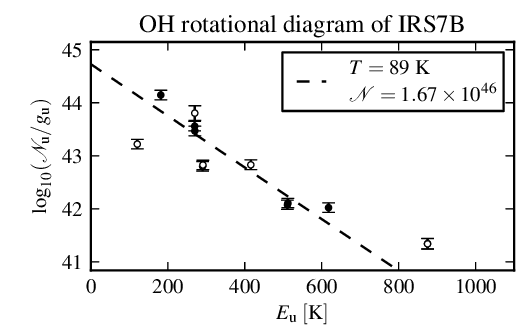} \\
    \includegraphics[width=0.48\linewidth]{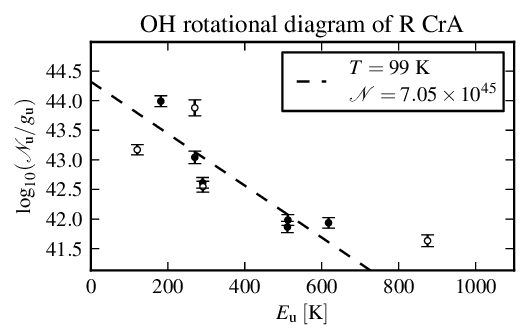} &
    \includegraphics[width=0.48\linewidth]{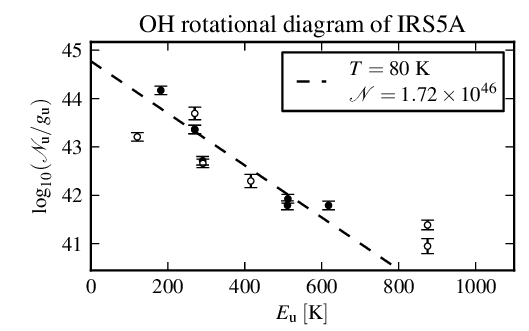} \\
    \end{array}$
    \caption{OH rotational diagrams of the point sources in the deconvolved PACS data. The data points plotted as open circles were not included in the fit due to optical thickness, blends, or leakage. $\mathcal{N}$ is the total number of OH molecules in each source given the rotational fit.}
    \label{fig:oh_ex}
\end{figure*}

\begin{figure*}[!htb]
    \centering
    $\begin{array}{c@{\hspace{0.0cm}}c@{\hspace{0.0cm}}c}
    \includegraphics[width=0.48\linewidth]{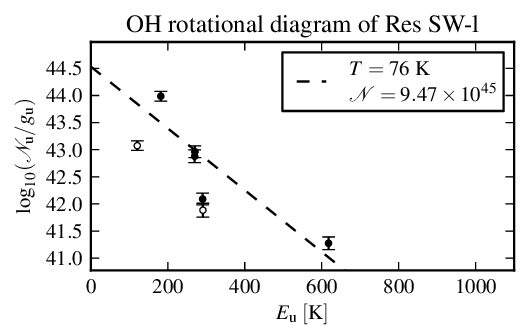} &
    \includegraphics[width=0.48\linewidth]{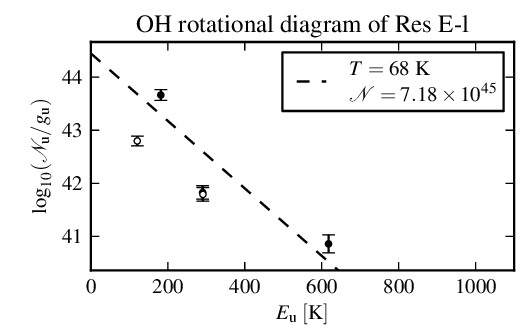} \\
    \includegraphics[width=0.48\linewidth]{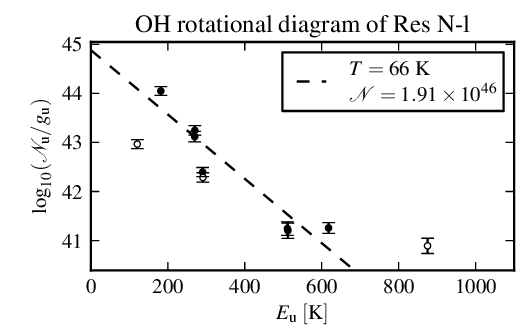} &
    \includegraphics[width=0.48\linewidth]{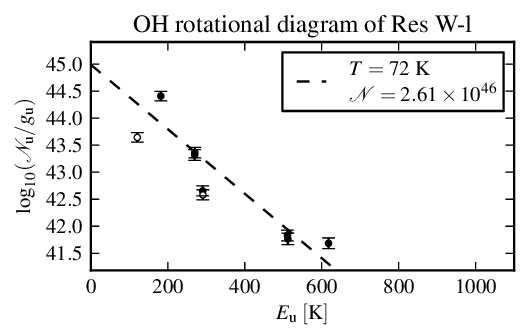} \\
    \end{array}$
    \caption{OH rotational diagrams for the residual emission of the extended regions Res~SW\nobr l,
Res~N\nobr l, Res~E\nobr l, and Res~W\nobr l. The data points plotted as open circles were not included in the fit due to optical thickness, blends, or leakage. $\mathcal{N}$ is the total number of OH molecules in each source given the rotational fit.}
    \label{fig:oh_ex_res}
\end{figure*}

The OH rotational diagrams of the point sources are found in
Fig.~\ref{fig:oh_ex}, and those of the extended emission in
Fig.~\ref{fig:oh_ex_res}. The derived parameters are listed in
Table~\ref{tab:rottemp_oh}, where they are also compared to some other embedded
sources in the literature. The CrA sources have fairly uniform excitation temperatures,
and do not differ significantly from the other sources in the literature. We also produce OH rotational diagrams of the four extended regions (Fig.~\ref{fig:oh_ex_res}). The OH temperatures in the
extended emission are similar to the point sources within $3\sigma$.
%[NJE is the difference statistically significant?]
% JEL: The average point source temperature is 87\pm10, the average extended temperature is 72\pm5, so the averages are not within each other's errors. !!!

\begin{table}
\centering
\caption[]{OH rotational temperatures and total number of molecules.}
\label{tab:rottemp_oh}
\begin{tabular}{l l l}
\noalign{\smallskip}
\hline
\hline
\noalign{\smallskip}
YSO & $T$\tablefootmark{a} & $\mathcal{N}$\tablefootmark{a} \\
& [K] & [$10^{45}$] \\
\noalign{\smallskip}
\hline
\noalign{\smallskip}
IRS7A+SMM~1C & $83\pm3$ & $132\phantom{.0}\pm12$ \\
IRS7B & $89\pm4$ & $\phantom{0}16.7\pm\phantom{0}1.5$ \\
R~CrA & $99\pm5$ & $\phantom{00}7.1\pm\phantom{0}0.7$ \\
IRS5A & $80\pm3$ & $\phantom{0}17.2\pm\phantom{0}1.5$ \\
\noalign{\smallskip}
\hline
\noalign{\smallskip}
Res SW-l & $76\pm4$ & $\phantom{00}9.5\pm\phantom{0}1.1$
\\
Res E-l & $68\pm4$ & $\phantom{00}7.2\pm\phantom{0}1.2$ \\
Res N-l & $66\pm2$ & $\phantom{0}19.1\pm\phantom{0}2.1$
\\
Res W-l & $72\pm2$ & $\phantom{0}26.1\pm\phantom{0}2.5$
\\
\noalign{\smallskip}
\hline
\noalign{\smallskip}
CrA point-source average & $88\pm4$\tablefootmark{b} & $\phantom{0}43.3\pm25.7$\tablefootmark{b} \\
CrA extended average & $71\pm2$\tablefootmark{b} & $\phantom{0}15.5\pm\phantom{0}3.8$\tablefootmark{b}
\\
\noalign{\smallskip}
\hline
\noalign{\smallskip}
NGC 1333 IRAS 4B\tablefootmark{c} & $60$ & 130 \\
Serpens SMM1\tablefootmark{d} & $72\pm8$ & \phantom{0}... \\
Serpens SMM3/4 average\tablefootmark{e} & $88\pm2$\tablefootmark{b} & $\phantom{0}26\phantom{.0}\pm\phantom{0}3$\tablefootmark{b} \\
\noalign{\smallskip}
\hline
\noalign{\smallskip}
DIGIT\tablefootmark{f} & $83\pm3$\tablefootmark{b} & $\phantom{0}24\phantom{.0}\pm\phantom{0}3$\tablefootmark{b} \\
\noalign{\smallskip}
\hline
\end{tabular}
\tablefoot{
		\tablefoottext{a}{The methods used for calculating the error estimates are the same as for the CO rotational diagrams, see Sects.~\ref{sec:co_point_rot}--\ref{sec:co_extended_ex}.} 
     	\tablefoottext{b}{For the sample averages, standard deviations of the mean of the significant fits are given (see Table~\ref{tab:rottemp_co}).}
     	\tablefoottext{c}{From \citet{herczeg12}. Only the cool component based chiefly on PACS data is given here.}
     	\tablefoottext{d}{From \citet{wampfler13}.}	 \\
     	\tablefoottext{e}{Average of SMM3b and SMM4, from \citet{dionatos13}, assuming a distance of 415~pc. This value is calculated excluding the same lines that have been excluded in this paper, and does not agree with the values given in \citet{dionatos13} for this reason.}
     	\tablefoottext{f}{Average of the DIGIT sample, from
     	\citet{green13}, not including the CrA and Serpens sources. This value is calculated excluding the same lines that have been excluded in this paper, and for this reason does not agree with the value given in \citet{green13} (183~K).}}
\end{table}

\begin{figure*}[!htb]
    \centering
    $\begin{array}{c@{\hspace{0.0cm}}c@{\hspace{0.0cm}}c}
    \includegraphics[width=0.48\linewidth]{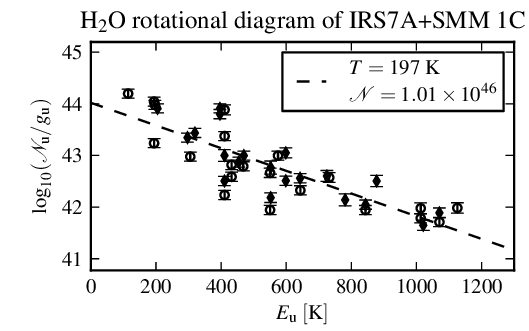} &
    \includegraphics[width=0.48\linewidth]{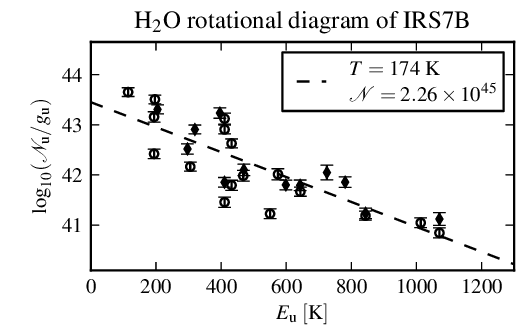} \\
    \includegraphics[width=0.48\linewidth]{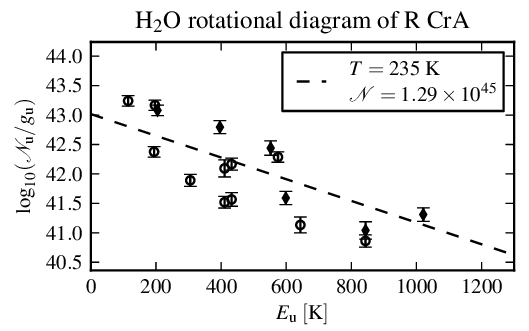} &
    \includegraphics[width=0.48\linewidth]{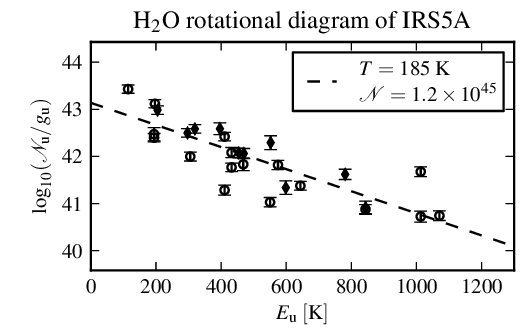} \\
    \includegraphics[width=0.48\linewidth]{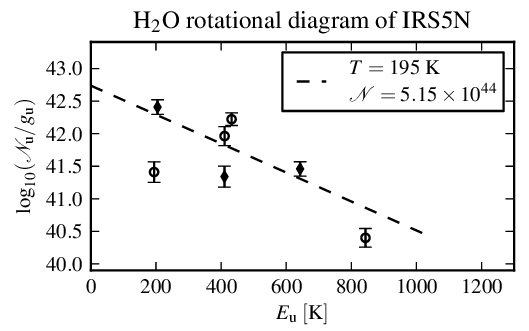} \\
    \end{array}$
    \caption{H$_2$O rotational diagrams of the point sources in the deconvolved PACS data. Ortho lines are marked with open circles and para lines with filled diamonds. $\mathcal{N}$ is the total number of H$_2$O molecules in each source given the rotational fit.}
    \label{fig:h2o_ex}
\end{figure*}

\begin{figure*}[!htb]
    \centering
    $\begin{array}{c@{\hspace{0.0cm}}c@{\hspace{0.0cm}}c}
    \includegraphics[width=0.48\linewidth]{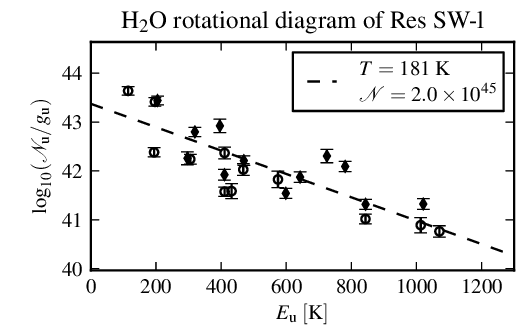} &
    \includegraphics[width=0.48\linewidth]{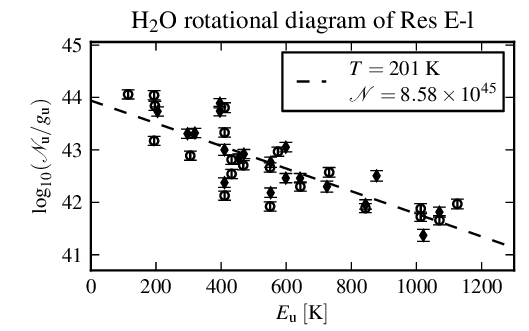} \\
    \includegraphics[width=0.48\linewidth]{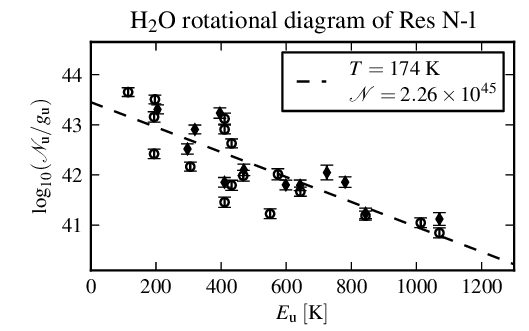} &
    \includegraphics[width=0.48\linewidth]{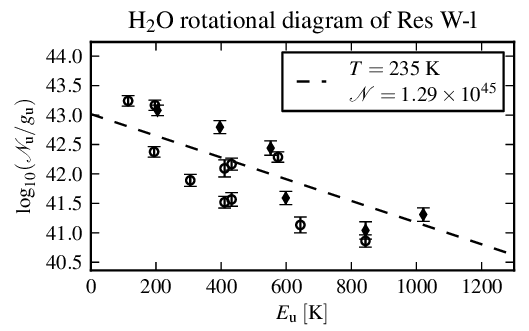} \\
    \end{array}$
    \caption{H$_2$O rotational diagrams for the residual emission of the extended regions Res~SW\nobr l, Res~E\nobr l, Res~N\nobr l, and Res~W\nobr l. Ortho lines are marked with open circles and para lines with filled diamonds. $\mathcal{N}$ is the total number of H$_2$O molecules in each source given the rotational fit.}
    \label{fig:h2o_ex_res}
\end{figure*}

For H$_2$O, we assume an ortho-to-para ratio of 3 \citep{herczeg12}. This
assumption is accounted for in the rotational diagrams (Figs.~\ref{fig:h2o_ex}--\ref{fig:h2o_ex_res}).
The H$_2$O rotational diagrams show quite large spreads, which are mainly due
to subthermal excitation effects and optical depth effects on some of the lines \citep{herczeg12}. The apparent shift between ortho and para lines might be caused by either of these effects, or by an ortho-to-para ratio lower than 3, but without more elaborate radiative transfer models it is impossible to distinguish between these scenarios. The derived rotational temperatures and total numbers of molecules are shown in Table~\ref{tab:rottemp_h2o}.
As in the case of OH, non-LTE radiative transfer models can be used to resolve the optical depth effects \citep{herczeg12}.
Another method would be to exclude
spectral lines suspected to be optically thick. The H$_2$O excitation temperatures are higher than the OH
temperatures. The higher H$_2$O temperature
in R~CrA than in the other sources is likely caused by a smaller number of detected lines, which enhances the
temperature since the stronger lines are optically thick. The observed H$_2$O
temperatures do not vary significantly from the DIGIT average.

As in the CO and OH cases, the H$_2$O rotational temperatures are similar in the extended emission and in the point sources.

\begin{table}
\centering
\caption[]{H$_2$O rotational temperatures and total number of molecules.}
\label{tab:rottemp_h2o}
\begin{tabular}{l l l}
\noalign{\smallskip}
\hline
\hline
\noalign{\smallskip}
YSO & $T$\tablefootmark{a} & $\mathcal{N}$\tablefootmark{a} \\
& [K] & [$10^{45}$] \\
\noalign{\smallskip}
\hline
\noalign{\smallskip}
IRS7A+SMM~1C & $197\pm\phantom{0}4$ & $\phantom{0}10.1\pm0.3$ \\
IRS7B & $174\pm\phantom{0}4$ & $\phantom{00}2.3\pm0.1$ \\
R~CrA & $235\pm\phantom{0}9$ & $\phantom{00}1.3\pm0.1$ \\
IRS5A & $185\pm\phantom{0}5$ & $\phantom{00}1.2\pm0.1$ \\
IRS5N & $195\pm20$ & $\phantom{00}0.5\pm0.1$ \\
\noalign{\smallskip}
\hline
\noalign{\smallskip}
Res~SW\nobr l & $181\pm\phantom{0}5$ & $\phantom{00}2.0\pm0.1$ \\
Res~E\nobr l & $201\pm\phantom{0}4$ & $\phantom{00}8.6\pm0.3$ \\
Res~N\nobr l & $174\pm\phantom{0}4$ & $\phantom{00}2.3\pm0.1$ \\
Res~W\nobr l & $235\pm\phantom{0}9$ & $\phantom{00}1.3\pm0.1$ \\
\noalign{\smallskip}
\hline
\noalign{\smallskip}
CrA point-source average & $197\pm\phantom{0}9$\tablefootmark{b} & $\phantom{00}3.1\pm1.6$\tablefootmark{b} \\
CrA extended average & $198\pm12$\tablefootmark{b} & $\phantom{00}3.5\pm1.5$\tablefootmark{b} \\
\noalign{\smallskip}
\hline
\noalign{\smallskip}
NGC 1333 IRAS 4B\tablefootmark{c} & 110/220 & 100 \\
Serpens SMM1\tablefootmark{d} & $136\pm27$ & \phantom{0}... \\
Serpens SMM3/4 average\tablefootmark{e} & $105\pm\phantom{0}6$\tablefootmark{b} & $\phantom{0}25\phantom{.0}\pm3$\tablefootmark{b} \\
\noalign{\smallskip}
\hline
\noalign{\smallskip}
DIGIT\tablefootmark{f} & $194\pm20$\tablefootmark{b} & $\phantom{00}7.7\pm2.6$\tablefootmark{b} \\
\noalign{\smallskip}
\hline
\end{tabular}
\tablefoot{
		\tablefoottext{a}{The methods used for calculating the error estimates are the same as for the CO rotational diagrams, see Sects.~\ref{sec:co_point_rot}--\ref{sec:co_extended_ex}.}
     	\tablefoottext{b}{For the sample averages, standard deviations of the mean of the significant fits are given (see Table~\ref{tab:rottemp_co}).}
		\tablefoottext{c}{From \citet{herczeg12}. Two separate cool and warm
components.}
		\tablefoottext{d}{From \citet{goicoechea12}.}
		\tablefoottext{e}{Average of SMM3b, SMM3c, SMM3r, and SMM4, from \citet{dionatos13}, assuming a distance of 415~pc.}
     	\tablefoottext{f}{Average of the DIGIT sample, from
\citet{green13}, not including the CrA and Serpens sources.} \\
     	}
\end{table}

\subsubsection{Comparison of the rotational diagrams}

The rotational temperatures of the three molecules studied in the FIR data (CO,
OH, and H$_2$O) are all different (see Tables~\ref{tab:rottemp_co}--\ref{tab:rottemp_h2o}). There are, however, relatively small spreads among the warm CO and OH temperatures, respectively. The warm CO temperature average is significantly lower in the CrA point-source sample than in the DIGIT sample \citep{green13}, but in agreement with the HOPS (\textit{Herschel} Orion Protostar Survey) sample \citep{manoj13}; whereas the hot CO is in agreement between the CrA and DIGIT samples, but higher than in the HOPS sample. The OH and H$_2$O temperatures are similar between the CrA and DIGIT samples, where the DIGIT sample average for OH has been recalculated for consistency using only the OH lines in our fits, and thus does not agree with the average value given by \citet{green13}.

Comparing the number of molecules per source between the CrA and DIGIT samples \citep{green13} shows that the average number of CO molecules is larger in the CrA sample, the average number of OH molecules is larger (but within errors) in the CrA sample, and the average number of H$_2$O molecules is lower in the CrA sample. Comparing these results could however be biased, since it was not possible to construct OH and H$_2$O rotational diagrams for all sources in the DIGIT sample where CO rotational diagrams could be made. Instead, calculations of line ratios in the whole DIGIT sample will be a better tracer of any difference in abundance ratios (see Sect.~\ref{sec:lineratios}).

\section{Discussion}
\label{sec:discussion}

\subsection{Survey of source properties}

In Table~\ref{tab:yso_prop}, some important properties of the studied point sources and extended line emission regions
are tabulated along with the DIGIT sample averages. The properties of the sources in the CrA sample are found to be fairly typical for low-mass embedded protostars. Since IRS7A and SMM~1C cannot be separated in the PACS data it is difficult to draw any conclusions about the classes of the separate sources. However, their different appearance in continuum data of other bands (IRS7A is detected in mid-IR but not mm; SMM~1C is detected in mm but not mid-IR) and their combined $T_{\mathrm{bol}} = 80$~K point towards SMM~1C being a Class~0 source. IRS5N is definitely a Class~0 object, but the other low-mass sources in the sample cannot consistently be assigned to Class~0 or Class~I.

The SEDs of the extended continuum emission (in Res~N\nobr c and Res~S\nobr c; see Fig.~\ref{fig:sed_res}) are similar to
that of the Class~0 source IRS5N, which indicates that this gas has similar
temperatures to those of very young protostellar cores. The dust black-body
temperature of the two continuum ridges is found to be 40--50~K, which is
consistent with the H$_2$CO temperature, estimated to be 40--60~K in
\citet{lindberg12}. \citet{lindberg12} showed that these temperatures cannot
be caused by radiation from the low-mass protostars, but can instead be
explained by external irradiation from R~CrA.

\begin{table*}
\centering
\caption[]{Properties of the studied point sources.}
\label{tab:yso_prop}
\begin{tabular}{l l l l l l l l l c c l}
\noalign{\smallskip}
\hline
\hline
\noalign{\smallskip}
YSO & \multicolumn{2}{c}{CO\tablefootmark{a}} & \multicolumn{2}{c}{OH} &
\multicolumn{2}{c}{H$_2$O} & & & Mid-IR/FIR/mm & FIR/mm & Class\\
& $T_{\mathrm{rot}}$ & $\mathcal{N}$ & $T_{\mathrm{rot}}$ & $\mathcal{N}$ &
$T_{\mathrm{rot}}$ & $\mathcal{N}$ & $T_{\mathrm{bol}}$ & $L_{\mathrm{bol}}$ &
continuum & lines \\
& [K] & [10$^{48}$] & [K] & [10$^{45}$] & [K] & [10$^{45}$] & [K] & [$L_{\sun}$] & detected & detected \\
\noalign{\smallskip}
\hline
\noalign{\smallskip}
IRS7A & 294\tablefootmark{b}/\phantom{0}682\tablefootmark{b} & 35.7\tablefootmark{b} &
83\tablefootmark{b} & 132\tablefootmark{b} & 197\tablefootmark{b}
& 10.1\tablefootmark{b} & \phantom{0}79\tablefootmark{b} &
\phantom{0}9.1\tablefootmark{b} & yes/yes\tablefootmark{b}/no\tablefootmark{c} &
yes\tablefootmark{b}/yes & 0/I \\
SMM~1C & -- & -- &
-- & -- & --
& -- & -- &
-- & no/yes\tablefootmark{b}/yes &
yes\tablefootmark{b}/yes & 0/I \\
IRS7B & 273\phantom{0}/\phantom{0}710 & 12.5 & 89 & \phantom{0}17 & 174 & \phantom{0}2.3 &
\phantom{0}89 & \phantom{0}4.6 & yes/yes/yes & yes/yes & 0/I \\
R CrA  & 287\phantom{0}/\phantom{0}992 & \phantom{0}5.7 & 99 & \phantom{00}7.1 & 235 & \phantom{0}1.3 & 889 & 53 &
yes/yes/faint & yes/yes & II/III \\
IRS5A & 293\phantom{0}/1417 & \phantom{0}3.2 & 80 & \phantom{0}17 & 185 & \phantom{0}1.2 & 209 &
\phantom{0}1.7 & yes/yes/yes & yes/... & 0/I \\
IRS5N & 283\phantom{0}/... & \phantom{0}0.7 & ... & ... & 195 & \phantom{0}0.5 & \phantom{0}63 & \phantom{0}0.7 &
no/yes/yes & no/... & 0 \\
\noalign{\smallskip}
\hline
\noalign{\smallskip}
Res~SW-l & 285\phantom{0}/\phantom{0}653 & \phantom{0}5.0 & 76 & \phantom{00}9.5 & 181 & \phantom{0}2.0 & ... & ... & ... & ... & ... \\
Res~E-l & 287\phantom{0}/1015 & \phantom{0}1.7 & 68 & \phantom{00}7.2 & 201 & \phantom{0}8.6 & ... & ... & ... & ... & ... \\
Res~N-l & 281\phantom{0}/\phantom{0}898 & \phantom{0}3.3 & 66 & \phantom{0}19 & 174 & \phantom{0}2.3 & ... & ... & ... & ... & ... \\
Res~W-l & 253\phantom{0}/\phantom{0}751 & 13.0 & 72 & \phantom{0}26 & 235 & \phantom{0}1.3 & ... & ... & ... & ... & ... \\
\noalign{\smallskip}
\hline
\noalign{\smallskip}
DIGIT\tablefootmark{d} & 355\phantom{0}/\phantom{0}814 & \phantom{0}7.0 & 83 & \phantom{0}24 & 194 & \phantom{0}7.7 &
167 & \phantom{0}6 & ... & ... & ... \\
\noalign{\smallskip}
\hline
\end{tabular}
\tablefoot{
		\tablefoottext{a}{The temperatures are for the warm and hot components, respectively. The
total number of molecules is the sum of the warm and hot component.}
		\tablefoottext{b}{Combined value for IRS7A and SMM~1C.}
		\tablefoottext{c}{Detected in ALMA 0.8~mm continuum \citep{lindberg13_alma}.}
     	\tablefoottext{d}{Average of the DIGIT sample, from
\citet{green13}, not including the CrA and Serpens sources. The OH rotational diagrams have been recalculated excluding the same lines that are excluded in this paper.} \\
     	}
\end{table*}

\subsection{Excitation conditions}

As pointed out earlier, the CO excitation temperatures in the point sources are in good agreement with those found in larger samples of low-mass embedded objects \citep{green13,karska13,manoj13}. One possible explanation for the excitation conditions being similar towards the externally irradiated protostars and in sources not subject to external irradiation is that the
irradiation from R~CrA is not substantial enough to dramatically change the
properties of the high-temperature gas. The excitation conditions of the extended molecular line emission found across the IRS7 field are consistent with those of the compact objects, and do not change significantly across the IRS7 field. The appearance of such extended emission is, however, unusual. The excitation diagrams could be explained
by a single-temperature non-LTE fit
\citep{neufeld12}, assuming a low gas density ($n\sim10^4$~cm$^{-3}$), and that
the gas is collisionally excited to high temperatures ($T\sim5\,000$~K). However, the observation of strong and extended H$_2$CO and CH$_3$OH emission in the field makes this low-density scenario unlikely. Furthermore, H$_2$ densities $\lesssim10^6$~cm$^{-3}$ do not fit with the H$_2$CO optical depth derived by \citet{lindberg12}. Another possibility is that the molecular gas towards the point sources and the extended gas are excited by different excitation mechanisms.

\subsection{Comparing FIR and mm spectral line data}
\label{sec:fir_mm_compare}

Strong CO $J=6\rightarrow5$ and $J=7\rightarrow6$ emission found on an east-west line centred at IRS7A \citet{vankempen09a} are consistent with the residual CO regions Res~E\nobr l and Res~W\nobr l, although the FIR data is more dominated by the point-source emission. The major difference between the morphology of the H$_2$CO and CH$_3$OH mm line data \citep{lindberg12} and the FIR
(\textit{Herschel}) line data is that most of the FIR line emission is
well-aligned with the mm and mid-IR continuum point sources, while the mm lines
appear in more extended structures, which are not centred on these point
sources. However, the residual PACS continuum emission after deconvolution of the
point-source emission (corresponding to extended dust continuum emission) shows shapes very similar to the high-temperature H$_2$CO ridges
observed in the mm (see Fig.~\ref{fig:cont_extreg}). The H$_2$CO rotational temperatures measured in the
SMA/APEX mm data range from 30 to 100~K, but non-LTE modelling shows that the
physical temperatures probably are in the order of 40--60~K \citep{lindberg12}. The PACS SEDs of these
ridges show black-body temperatures ($\sim40$--$50$~K; see Fig.~\ref{fig:sed_res}) consistent with the
H$_2$CO temperatures measured. The H$_2$CO
ridges are observed on relatively large scales ($\sim8\,000$~AU), and are not associated with the point sources.

The POMAC algorithm shows that most
of the FIR molecular line emission originates from the mid-IR/(sub)mm point sources
IRS7A, SMM~1C, IRS7B, R~CrA, and IRS5A. This emission is similar in excitation
to what is found towards sources that are not subject to external irradation. However, through the deconvolution we also find extended line emission in the IRS7 region. Interestingly, the CO, OH, and H$_2$O excitation conditions of the extended emission resemble those near the protostars (see Tables~\ref{tab:rottemp_co}--\ref{tab:rottemp_h2o}), and suggest high temperatures also on these relatively large scales ($>1\,000$~AU from the protostars). Such extended line emission, in particular the hot CO and OH emission, is unusual around low-mass embedded objects \citep[see e.g.][]{vankempen10,green13,karska13}.

Still, the H$_2$CO emission detected by SMA/APEX is even more extended, and not
associated with the point sources. Regardless of the lower spatial resolution of the
PACS data, it is certain that the FIR and mm line emission have different
origins.

\subsection{Water and oxygen chemistry}
\label{sec:lineratios}

Assuming that the region around R~CrA exhibits PDR-like conditions, the OH abundance should be
enhanced with respect to the H$_2$O abundance \citep[see e.g.][]{walsh13}. We
thus want to establish whether the OH/H$_2$O ratio is enhanced in these sources
compared to other sources in similar stages of evolution. A first-order
comparison can be made by analysing the ratios of certain OH and H$_2$O
spectral lines in the DIGIT sample of embedded protostars \citep{green13}. So far, no systematic study of OH and H$_2$O excitation diagrams and abundances in low-mass embedded objects is available. To avoid biases introduced by different amounts of detected lines and different methods of extracting the abundances, we instead compare the ratios of individual spectral line luminosities.

\begin{figure*}[!htb]
    \centering
    $\begin{array}{c@{\hspace{0.0cm}}c@{\hspace{0.0cm}}c}
    \includegraphics[width=0.48\linewidth]{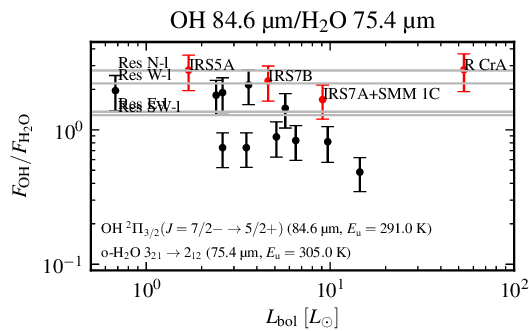} &
    \includegraphics[width=0.48\linewidth]{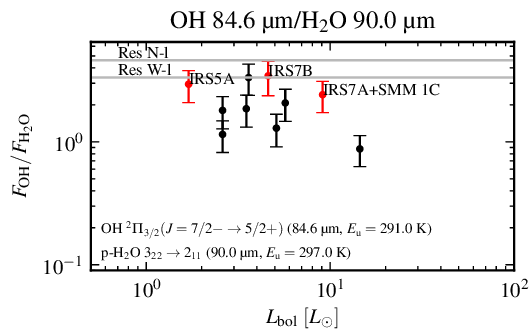} \\
    \includegraphics[width=0.48\linewidth]{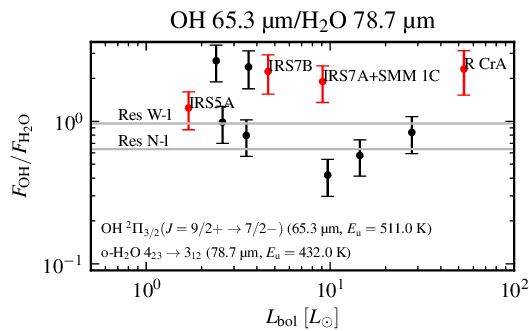} &
    \includegraphics[width=0.48\linewidth]{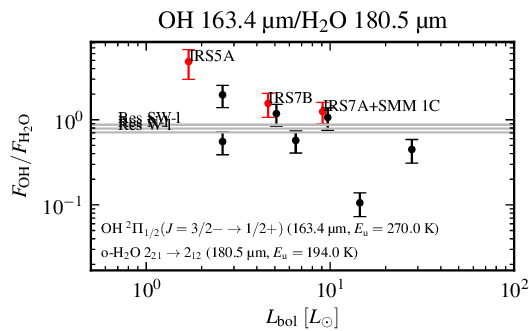} \\
    \includegraphics[width=0.48\linewidth]{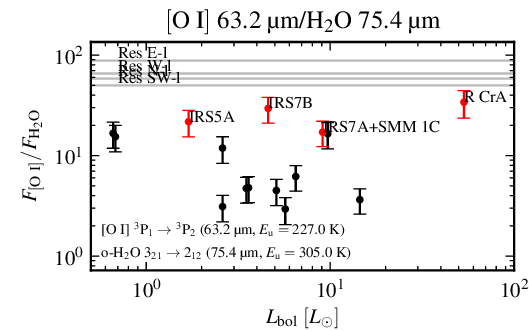} &
    \includegraphics[width=0.48\linewidth]{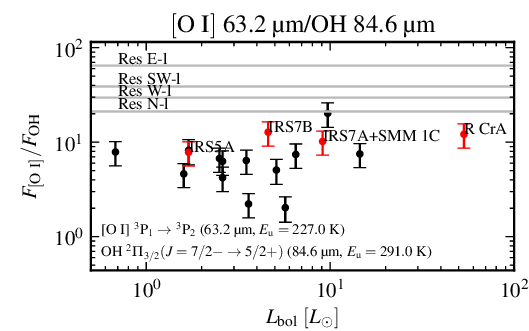} \\
    \includegraphics[width=0.48\linewidth]{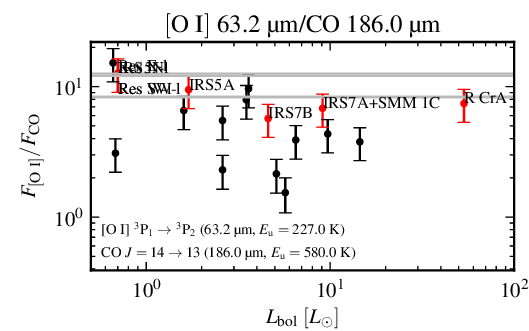} &
    \includegraphics[width=0.48\linewidth]{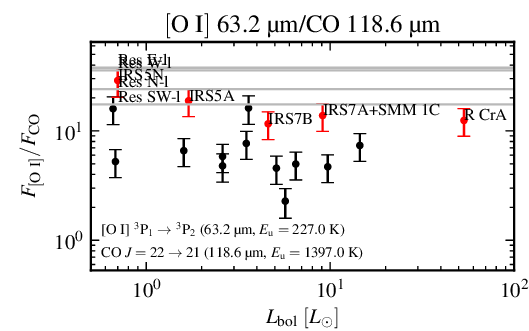} \\
    \end{array}$
    \caption{Ratios of [\ion{O}{i}], OH, H$_2$O, and CO lines plotted versus the bolometric luminosities of the sources. See details in the figures. The CrA
    point sources are marked with red, the black data points are other DIGIT embedded sources \citep{green13}. The CrA extended emission regions are marked with grey lines. Only data points where both lines are significantly detected are included.}
    \label{fig:lineratios}
\end{figure*}

We need to compare OH and H$_2$O lines suspected to be optically thin, that are detected in many sources, that have
small PSFs (short wavelengths) or at least similar PSFs (similar wavelengths), 
and that have similar upper-level energies to remove any
bias. Thus, in the first four panels of Fig.~\ref{fig:lineratios}, we plot four different line ratios of three OH lines and three H$_2$O lines. The wavelengths differ by less than 20\micron\ and the upper level energies by less than 80~K for each of the ratios. Details on the transitions are found in the figure.
We
find that all the CrA sources have an OH/H$_2$O line ratio higher than most
other DIGIT embedded objects, which is indicative of PDR activity \citep{hollenbach97}. Also the extended regions have relatively high OH/H$_2$O ratios. We have also compared
the OH (c.f. Fig.~9 in \citealp{wampfler13}), H$_2$O, and CO
($J=16\rightarrow15$; cf. Fig.~22 in \citealp{green13}) line luminosities to
the bolometric luminosities of the DIGIT and WISH embedded sources, and found
that the sources in CrA fall within the scatter around the linear correlation
for all three lines, although on the higher end in the OH and CO case. We find that the enhanced OH/H$_2$O ratio mainly is due to an increase of the OH flux rather than a decrease of the H$_2$O flux. 

From dissociation of OH, O should be a major destruction product, so we would
also expect an enhancement of the [\ion{O}{i}] line strengths compared to H$_2$O and OH
in a PDR \citep{hollenbach97}. In the third row of diagrams in Fig.~\ref{fig:lineratios}, we compare the line ratios of the [\ion{O}{i}] 63.2\micron\ line to one H$_2$O line and one OH line with similar wavelengths and upper-level energies. We see a very strong enhancement of [\ion{O}{i}], in particular in the extended regions, supporting the hypothesis of a PDR induced by external irradiation.

In the last row of Fig.~\ref{fig:lineratios}, the [\ion{O}{i}] 63.2\micron\ line flux is compared to two CO lines. In the comparison of [\ion{O}{i}] and CO it is more difficult to accommodate our ambition of using similar wavelengths and upper-level energies -- we could have used the [\ion{O}{i}] 145\micron\ line, but this was not desirable for S/N reasons. We find that the [\ion{O}{i}] flux is enhanced in the CrA sources also with respect to the CO flux.

The higher [\ion{O}{i}]/OH, [\ion{O}{i}]/H$_2$O, and OH/H$_2$O line ratios could also indicate a later stage of evolution \citep[cf. Class~II sources in][]{podio12}, but the low bolometric temperatures and $L_{\mathrm{bol}}/L_{\mathrm{submm}}$ ratios of the sources indicate that they are Class~0/I sources, and the (in many cases)
higher line ratios found in the extended emission compared to the point sources indicate that the heating is external in its origin.

\section{Conclusions}

We study \textit{Herschel}/PACS line and continuum maps of the low-mass
star-forming region R~CrA subject to strong irradiation from the nearby
Herbig~Be star R~CrA. In addition, we deconvolve the maps to study the
point-source and extended contributions to the emission. Our main results are
the following:

\begin{enumerate}
   
\item FIR continuum emission is found not only at the (sub)mm and mid-IR
continuum point sources, but also (somewhat fainter) in two ridges north and
south of the IRS7 protostars. These correlate in position with H$_2$CO and CH$_3$OH mm emission, and the
continuum emission peaks give temperatures (40--50~K) similar to the rotational
temperature of the H$_2$CO emission \citep{lindberg12}, both suggesting that
the extended FIR continuum emission traces the dust associated with the externally irradiated material.

\item The rotational temperatures of the warm CO component ($286\pm3$~K), the hot CO component ($950\pm148$~K), OH ($88\pm4$~K), and H$_2$O ($197\pm9$~K) measured towards the continuum point sources are consistent with or lower than those found in larger samples of similar sources, suggesting that the excitation conditions of the dense gas
close to the protostars are not affected by the external irradiation. A $^{13}$CO rotational diagram suggests that the mid-$J$ $^{12}$CO lines are marginally optically thick ($\tau\sim0.6$).

\item CO, OH, and H$_2$O emission not associated with any of the previously known continuum point sources
is detected, and shows excitation conditions similar to the gas near the
protostars ($277\pm7$~K for the warm CO component, $829\pm69$~K for the hot CO component, $71\pm2$~K for OH, and $198\pm12$ for H$_2$O). The warm gas thus exists on much larger scales than can be
explained by heating from the low-mass YSOs. One possible explanation is that
this emission traces radiatively excited low-density gas, but detections of high density tracers such as H$_2$CO and CH$_3$OH challenge this hypothesis. The extent of the FIR molecular emission is larger than previously seen in any low-mass protostellar sources.

\item When comparing the IRS7 and IRS5 fields -- the former with a smaller angular separation from the irradiating Herbig~Be star R~CrA than the latter -- we find that the two fields have similar average rotational temperatures of the warm CO component (285~K and 288~K, respectively), OH (90~K and 80~K, respectively), and H$_2$O (200~K and 190~K, respectively). However, more extended emission (both line and continuum) is seen in the IRS7 field than in the IRS5 field. The higher level of irradiation from R~CrA in IRS7 than in IRS5 does thus not significantly affect the rotational temperatures, but the possible link between extended emission and external irradiation needs to be investigated further.
      
\item The OH/H$_2$O, [\ion{O}{i}]/H$_2$O, [\ion{O}{i}]/OH, and [\ion{O}{i}]/CO line ratios are enhanced in the CrA point sources and extended gas
compared to other embedded objects, which is similar to what has previously been seen in PDRs \citep{hollenbach97}. Typically, these line ratios are enhanced by a factor of 1.5--4.0 in the CrA sources.

\end{enumerate}

To further study the origin of the excitation conditions in protostellar cores
and their surroundings, we propose similar investigations of any extended
emission in PACS observations of embedded objects, including other sources
in regions of potential strong external irradiation. This could for instance include the Orion sources discussed by
\citet{manoj13}, and the isolated sources in the DIGIT sample \citep{green13}.

\begin{acknowledgements}

This research was supported by a grant from the Instrument Center for Danish
Astrophysics (IDA) and a Lundbeck Foundation Group Leader Fellowship to JKJ.
Research at Centre for Star and Planet Formation is funded by the Danish
National Research Foundation and the University of Copenhagen's programme of
excellence. Support for this work, part of the \textit{Herschel} Open Time Key Project
Program, was provided by NASA through an award issued by the Jet Propulsion
Laboratory, California Institute of Technology. The \textit{William Herschel Telescope} is operated on the island of La Palma by the Isaac Newton Group in the Spanish Observatorio del Roque de los Muchachos of the Instituto de Astrof{\'i}sica de Canarias. We thank Nienke van der Marel for helping to obtain the optical spectrum of R~CrA. The authors would also like to thank the anonymous referee, Lars Kristensen, and Tim van Kempen for their useful suggestions and comments which improved the quality of the paper.

\end{acknowledgements}

\bibliographystyle{aa} % style aa.bst
\bibliography{herschel_rcra_arxiv}

\clearpage
\onecolumn
\begin{landscape}
\begin{longtable}{l l l l l l l l l}
\caption[]{Identified spectral lines in the \textit{Herschel}/PACS data and deconvolved
line fluxes in the continuum point sources. All errors are 1$\sigma$ of the
rms. For undetected lines, the $3\sigma$ upper limit lies between
$60\times10^{-18}$~W~m$^{-2}$ and $10\times10^{-18}$~W~m$^{-2}$ from 55\micron\ to
195\micron. Lines in the leakage regions are not tabulated.}\\
\hline
\hline
\noalign{\smallskip}
Species & Transition & Wavelength\tablefootmark{a} & IRS7A+SMM~1C & IRS7B & R
CrA & IRS5A & IRS5N \\
&  & [\hbox{\textmu}m] & [$10^{-18}$~W~m$^{-2}$] & [$10^{-18}$~W~m$^{-2}$] &
[$10^{-18}$~W~m$^{-2}$] & [$10^{-18}$~W~m$^{-2}$] & [$10^{-18}$~W~m$^{-2}$] \\
\noalign{\smallskip}
\hline
\noalign{\smallskip}
OH & $^2\Pi_{1/2}(J=9/2+\rightarrow7/2-)$ & \phantom{0}55.891 & $\phantom{00}352\pm24$ & $\phantom{00}85\pm\phantom{0}9$ & \phantom{00}... & $\phantom{00}34\pm10$ & \phantom{0}... \\
OH & $^2\Pi_{1/2}(J=9/2-\rightarrow7/2+)$ & \phantom{0}55.950 & $\phantom{00}686\pm24$ & $\phantom{00}83\pm\phantom{0}9$ & $\phantom{0}164\pm17$ & $\phantom{00}93\pm10$ & $\phantom{0}36\pm\phantom{0}9$ \\
p-H$_2$O & $4_{31}\rightarrow3_{22}$ & \phantom{0}56.325 & $\phantom{00}347\pm29$ & \phantom{00}... & \phantom{00}... & \phantom{00}... & \phantom{0}... \\
o-H$_2$O & $9_{09}\rightarrow8_{18}$ & \phantom{0}56.816 & $\phantom{00}460\pm30$ & $\phantom{00}97\pm20$ & $\phantom{0}372\pm39$ & $\phantom{00}30\pm\phantom{0}9$ & \phantom{0}... \\
p-H$_2$O & $4_{22}\rightarrow3_{13}$ & \phantom{0}57.637 & $\phantom{00}431\pm39$ & \phantom{00}... & \phantom{00}... & $\phantom{00}69\pm\phantom{0}9$ & \phantom{0}... \\
o-H$_2$O & $4_{32}\rightarrow3_{21}$ & \phantom{0}58.699 & $\phantom{00}547\pm27$ & $\phantom{0}105\pm10$ & \phantom{00}... & $\phantom{00}66\pm\phantom{0}8$ & \phantom{0}... \\
p-H$_2$O & $7_{26}\rightarrow6_{15}$ & \phantom{0}59.987 & $\phantom{00}146\pm17$ & \phantom{00}... & $\phantom{00}67\pm12$ & \phantom{00}... & \phantom{0}... \\
\ion{O}{i} & $^3$P$_1\rightarrow$~$^3$P$_2$ & \phantom{0}63.184 & $14735\pm46$ & $3914\pm26$ & $2399\pm26$ & $1961\pm19$ & $809\pm16$ \\
o-H$_2$O & $8_{18}\rightarrow7_{07}$ & \phantom{0}63.324 & $\phantom{00}712\pm25$ & $\phantom{00}97\pm11$ & \phantom{00}... & $\phantom{00}77\pm\phantom{0}9$ & \phantom{0}... \\
p-H$_2$O & $8_{08}\rightarrow7_{17}$ & \phantom{0}63.458 & $\phantom{00}354\pm31$ & $\phantom{00}61\pm13$ & \phantom{00}... & \phantom{00}... & \phantom{0}... \\
OH & $^2\Pi_{3/2}(J=9/2-\rightarrow7/2+)$ & \phantom{0}65.132 & $\phantom{0}1341\pm30$ & $\phantom{0}247\pm\phantom{0}9$ & $\phantom{0}185\pm13$ & $\phantom{0}163\pm\phantom{0}9$ & \phantom{0}... \\
o-H$_2$O & $6_{25}\rightarrow5_{14}$ & \phantom{0}65.166 & \multicolumn{5}{l}{Blend with OH $^2\Pi_{3/2}(J=9/2-\rightarrow7/2+)$} \\
OH & $^2\Pi_{3/2}(J=9/2+\rightarrow7/2-)$ & \phantom{0}65.279 & $\phantom{0}1197\pm28$ & $\phantom{0}227\pm\phantom{0}8$ & $\phantom{0}140\pm12$ & $\phantom{0}118\pm\phantom{0}7$ & $\phantom{0}16\pm\phantom{0}3$ \\
CO & $J=40\rightarrow39$ & \phantom{0}65.686 & $\phantom{000}88\pm21$ & \phantom{00}... & $\phantom{00}47\pm10$ & \phantom{00}... & \phantom{0}... \\
o-H$_2$O & $7_{16}\rightarrow6_{25}$ & \phantom{0}66.093 & $\phantom{00}387\pm20$ & $\phantom{00}71\pm\phantom{0}7$ & \phantom{00}... & $\phantom{00}34\pm\phantom{0}6$ & \phantom{0}... \\
o-H$_2$O & $3_{30}\rightarrow2_{21}$ & \phantom{0}66.438 & $\phantom{00}662\pm20$ & $\phantom{0}110\pm14$ & $\phantom{0}128\pm13$ & $\phantom{00}74\pm10$ & \phantom{0}... \\
p-H$_2$O & $3_{31}\rightarrow2_{20}$ & \phantom{0}67.089 & $\phantom{00}401\pm24$ & $\phantom{00}90\pm\phantom{0}8$ & \phantom{00}... & \phantom{00}... & $\phantom{0}27\pm\phantom{0}9$ \\
o-H$_2$O & $3_{30}\rightarrow3_{03}$ & \phantom{0}67.269 & $\phantom{00}200\pm19$ & $\phantom{00}35\pm\phantom{0}5$ & \phantom{00}... & \phantom{00}... & \phantom{0}... \\
CO & $J=39\rightarrow38$ & \phantom{0}67.336 & $\phantom{00}117\pm19$ & $\phantom{00}29\pm\phantom{0}6$ & $\phantom{00}34\pm\phantom{0}9$ & \phantom{00}... & \phantom{0}... \\
CO & $J=38\rightarrow37$ & \phantom{0}69.074 & $\phantom{00}192\pm17$ & \phantom{00}... & $\phantom{00}86\pm\phantom{0}7$ & \phantom{00}... & \phantom{0}... \\
CO & $J=37\rightarrow36$ & \phantom{0}70.907 & $\phantom{00}173\pm18$ & $\phantom{00}47\pm\phantom{0}7$ & \phantom{00}... & \phantom{00}... & \phantom{0}... \\
p-H$_2$O & $5_{24}\rightarrow4_{13}$ & \phantom{0}71.067 & $\phantom{00}329\pm12$ & $\phantom{00}64\pm\phantom{0}6$ & $\phantom{00}40\pm\phantom{0}6$ & $\phantom{00}22\pm\phantom{0}6$ & \phantom{0}... \\
OH & $^2\Pi_{1/2}(J=7/2-\rightarrow5/2+)$ & \phantom{0}71.171 & $\phantom{0}1268\pm23$ & $\phantom{0}235\pm\phantom{0}7$ & $\phantom{0}193\pm\phantom{0}9$ & $\phantom{0}138\pm\phantom{0}7$ & \phantom{0}... \\
OH & $^2\Pi_{1/2}(J=7/2+\rightarrow5/2-)$ & \phantom{0}71.215 & \multicolumn{5}{l}{Blend with OH $^2\Pi_{1/2}(J=7/2-\rightarrow5/2+)$} \\
p-H$_2$O & $7_{17}\rightarrow6_{06}$ & \phantom{0}71.540 & $\phantom{00}271\pm13$ & $\phantom{00}42\pm\phantom{0}6$ & $\phantom{00}27\pm\phantom{0}7$ & $\phantom{00}20\pm\phantom{0}5$ & \phantom{0}... \\
o-H$_2$O & $7_{07}\rightarrow6_{16}$ & \phantom{0}71.947 & $\phantom{00}624\pm13$ & $\phantom{0}110\pm\phantom{0}6$ & $\phantom{00}52\pm\phantom{0}7$ & $\phantom{00}54\pm\phantom{0}5$ & $\phantom{0}18\pm\phantom{0}5$ \\
CO & $J=36\rightarrow35$ & \phantom{0}72.843 & $\phantom{00}140\pm17$ & \phantom{00}... & \phantom{00}... & \phantom{00}... & \phantom{0}... \\
CO & $J=35\rightarrow34$ & \phantom{0}74.890 & $\phantom{00}362\pm19$ & $\phantom{00}74\pm\phantom{0}7$ & \phantom{00}... & \phantom{00}... & \phantom{0}... \\
o-H$_2$O & $7_{25}\rightarrow6_{34}$ & \phantom{0}74.945 & \multicolumn{5}{l}{Blend with CO $J=35\rightarrow34$} \\
o-H$_2$O & $3_{21}\rightarrow2_{12}$ & \phantom{0}75.381 & $\phantom{00}861\pm22$ & $\phantom{0}133\pm\phantom{0}7$ & $\phantom{00}71\pm\phantom{0}8$ & $\phantom{00}90\pm\phantom{0}7$ & \phantom{0}... \\
CO & $J=34\rightarrow33$ & \phantom{0}77.059 & $\phantom{00}204\pm19$ & \phantom{00}... & \phantom{00}... & \phantom{00}... & \phantom{0}... \\
o-H$_2$O & $4_{23}\rightarrow3_{12}$ & \phantom{0}78.742 & $\phantom{00}630\pm27$ & $\phantom{0}102\pm12$ & $\phantom{00}60\pm11$ & $\phantom{00}95\pm\phantom{0}6$ & \phantom{0}... \\
p-H$_2$O & $6_{15}\rightarrow5_{24}$ & \phantom{0}78.928 & $\phantom{00}101\pm18$ & $\phantom{00}53\pm\phantom{0}7$ & \phantom{00}... & $\phantom{00}30\pm\phantom{0}5$ & \phantom{0}... \\
OH & $^2\Pi_{1/2}(J=1/2-)\rightarrow$~$^2\Pi_{3/2}(J=3/2+)$ & \phantom{0}79.116 & $\phantom{0}1741\pm28$ & $\phantom{0}250\pm14$ & $\phantom{0}175\pm13$ & $\phantom{0}264\pm\phantom{0}8$ & \phantom{0}... \\
OH & $^2\Pi_{1/2}(J=1/2+)\rightarrow$~$^2\Pi_{3/2}(J=3/2-)$ & \phantom{0}79.179 & \multicolumn{5}{l}{Blend with OH $^2\Pi_{1/2}(J=1/2-)\rightarrow$~$^2\Pi_{3/2}(J=3/2+)$} \\
CO & $J=33\rightarrow32$ & \phantom{0}79.360 & $\phantom{00}162\pm15$ & \phantom{00}... & \phantom{00}... & \phantom{00}... & \phantom{0}... \\
CO & $J=32\rightarrow31$ & \phantom{0}81.806 & $\phantom{00}290\pm15$ & $\phantom{00}37\pm\phantom{0}6$ & $\phantom{00}57\pm\phantom{0}7$ & \phantom{00}... & \phantom{0}... \\
o-H$_2$O & $6_{16}\rightarrow5_{05}$ & \phantom{0}82.031 & $\phantom{00}740\pm22$ & $\phantom{0}162\pm\phantom{0}9$ & $\phantom{00}48\pm11$ & $\phantom{00}83\pm\phantom{0}6$ & \phantom{0}... \\
p-H$_2$O & $6_{06}\rightarrow5_{15}$ & \phantom{0}83.284 & $\phantom{00}397\pm21$ & $\phantom{00}68\pm10$ & \phantom{00}... & \phantom{00}... & $\phantom{0}31\pm\phantom{0}5$ \\
CO & $J=31\rightarrow30$ & \phantom{0}84.411 & \multicolumn{5}{l}{Blend with OH $^2\Pi_{3/2}(J=7/2+\rightarrow5/2-)$} \\
%OH & $^2\Pi_{3/2}(J=7/2+\rightarrow5/2-)$ & \phantom{0}84.420 & $\phantom{0}1433\pm21$ & $\phantom{0}308\pm13$ & $\phantom{0}182\pm12$ \\
OH & $^2\Pi_{3/2}(J=7/2+\rightarrow5/2-)$ & \phantom{0}84.420 & $\phantom{0}1474\pm18$ & $\phantom{0}327\pm14$ & $\phantom{0}172\pm11$ & $\phantom{0}223\pm\phantom{0}7$ & \phantom{0}... \\
OH & $^2\Pi_{3/2}(J=7/2-\rightarrow5/2+)$ & \phantom{0}84.597 & $\phantom{0}1444\pm20$ & $\phantom{0}306\pm14$ & $\phantom{0}197\pm11$ & $\phantom{0}250\pm\phantom{0}8$ & \phantom{0}... \\
o-H$_2$O & $7_{16}\rightarrow7_{07}$ & \phantom{0}84.767 & $\phantom{00}107\pm15$ & \phantom{00}... & \phantom{00}... & $\phantom{00}53\pm\phantom{0}7$ & \phantom{0}... \\
CO & $J=30\rightarrow29$ & \phantom{0}87.190 & $\phantom{00}421\pm17$ & $\phantom{00}73\pm10$ & \phantom{00}... & $\phantom{00}77\pm\phantom{0}6$ & \phantom{0}... \\
p-H$_2$O & $3_{22}\rightarrow2_{11}$ & \phantom{0}89.988 & $\phantom{00}597\pm18$ & $\phantom{00}89\pm11$ & \phantom{00}... & $\phantom{00}85\pm\phantom{0}5$ & \phantom{0}... \\
CO & $J=29\rightarrow28$ & \phantom{0}90.163 & $\phantom{00}413\pm14$ & $\phantom{00}46\pm\phantom{0}8$ & $\phantom{00}34\pm\phantom{0}8$ & $\phantom{00}37\pm\phantom{0}5$ & \phantom{0}... \\
CO & $J=28\rightarrow27$ & \phantom{0}93.349 & $\phantom{00}587\pm15$ & $\phantom{0}106\pm\phantom{0}9$ & $\phantom{00}69\pm10$ & $\phantom{00}56\pm\phantom{0}6$ & \phantom{0}... \\
p-H$_2$O & $5_{42}\rightarrow5_{33}$ & \phantom{0}94.210 & $\phantom{000}87\pm13$ & \phantom{00}... & \phantom{00}... & \phantom{00}... & \phantom{0}... \\
o-H$_2$O & $6_{25}\rightarrow6_{16}$ & \phantom{0}94.644 & $\phantom{00}351\pm18$ & $\phantom{00}46\pm\phantom{0}7$ & $\phantom{00}50\pm\phantom{0}9$ & $\phantom{00}35\pm\phantom{0}7$ & \phantom{0}... \\
o-H$_2$O & $4_{41}\rightarrow4_{32}$ & \phantom{0}94.705 & \multicolumn{5}{l}{Blend with o-H$_2$O $6_{25}\rightarrow6_{16}$} \\
p-H$_2$O & $5_{15}\rightarrow4_{04}$ & \phantom{0}95.627 & $\phantom{00}501\pm14$ & $\phantom{00}64\pm11$ & \phantom{00}... & $\phantom{00}58\pm\phantom{0}8$ & \phantom{0}... \\
OH & $^2\Pi_{3/2}(J=3/2+)\rightarrow$~$^2\Pi_{1/2}(J=5/2-)$ & \phantom{0}96.271 & $\phantom{00}405\pm17$ & $\phantom{00}48\pm12$ & $\phantom{00}57\pm14$ & $\phantom{00}37\pm\phantom{0}8$ & \phantom{0}... \\
OH & $^2\Pi_{3/2}(J=3/2-)\rightarrow$~$^2\Pi_{1/2}(J=5/2+)$ & \phantom{0}96.363 & \multicolumn{5}{l}{Blend with OH $^2\Pi_{3/2}(J=3/2+)\rightarrow$~$^2\Pi_{1/2}(J=5/2-)$} \\
CO & $J=27\rightarrow26$ & \phantom{0}96.773 & $\phantom{00}534\pm19$ & $\phantom{0}136\pm\phantom{0}9$ & \phantom{00}... & $\phantom{00}44\pm\phantom{0}7$ & \phantom{0}... \\
OH & $^2\Pi_{1/2}(J=5/2+\rightarrow3/2-)$ & \phantom{0}98.737 & $\phantom{0}1282\pm30$ & $\phantom{0}285\pm19$ & \phantom{00}... & $\phantom{00}83\pm21$ & \phantom{0}... \\
OH & $^2\Pi_{1/2}(J=5/2-\rightarrow3/2+)$ & \phantom{0}98.764 & \multicolumn{5}{l}{Blend with OH $^2\Pi_{1/2}(J=5/2+\rightarrow3/2-)$} \\
o-H$_2$O & $5_{05}\rightarrow4_{14}$ & \phantom{0}99.493 & $\phantom{00}783\pm34$ & $\phantom{0}123\pm18$ & \phantom{00}... & $\phantom{00}87\pm20$ & \phantom{0}... \\
CO & $J=25\rightarrow24$ & 104.445 & $\phantom{00}665\pm21$ & $\phantom{0}145\pm\phantom{0}7$ & $\phantom{0}150\pm\phantom{0}4$ & $\phantom{00}50\pm\phantom{0}6$ & \phantom{0}... \\
o-H$_2$O & $2_{21}\rightarrow1_{10}$ & 108.073 & $\phantom{00}604\pm10$ & $\phantom{00}92\pm\phantom{0}8$ & $\phantom{00}83\pm\phantom{0}4$ & $\phantom{00}87\pm\phantom{0}3$ & $\phantom{00}9\pm\phantom{0}3$ \\
CO & $J=24\rightarrow23$ & 108.763 & $\phantom{00}720\pm11$ & $\phantom{0}157\pm\phantom{0}6$ & $\phantom{0}126\pm\phantom{0}5$ & $\phantom{00}65\pm\phantom{0}4$ & \phantom{0}... \\
p-H$_2$O & $5_{24}\rightarrow5_{15}$ & 111.628 & $\phantom{00}131\pm12$ & \phantom{00}... & \phantom{00}... & \phantom{00}... & \phantom{0}... \\
CO & $J=23\rightarrow22$ & 113.458 & $\phantom{0}1636\pm13$ & $\phantom{0}489\pm\phantom{0}8$ & $\phantom{0}231\pm\phantom{0}3$ & $\phantom{0}181\pm\phantom{0}3$ & $\phantom{0}21\pm\phantom{0}3$ \\
o-H$_2$O & $4_{14}\rightarrow3_{03}$ & 113.537 & \multicolumn{5}{l}{Blend with CO $J=23\rightarrow22$} \\
p-H$_2$O & $5_{33}\rightarrow5_{24}$ & 113.948 & $\phantom{000}63\pm\phantom{0}9$ & $\phantom{00}18\pm\phantom{0}5$ & \phantom{00}... & \phantom{00}... & \phantom{0}... \\
CO & $J=22\rightarrow21$ & 118.581 & $\phantom{0}1066\pm11$ & $\phantom{0}337\pm\phantom{0}9$ & $\phantom{0}193\pm\phantom{0}3$ & $\phantom{0}104\pm\phantom{0}2$ & $\phantom{0}28\pm\phantom{0}3$ \\
OH & $^2\Pi_{3/2}(J=5/2-\rightarrow3/2+)$ & 119.234 & $\phantom{0}1886\pm17$ & $\phantom{0}227\pm10$ & $\phantom{0}202\pm\phantom{0}6$ & $\phantom{0}221\pm\phantom{0}4$ & $\phantom{0}26\pm\phantom{0}4$ \\
OH & $^2\Pi_{3/2}(J=5/2+\rightarrow3/2-)$ & 119.441 & \multicolumn{5}{l}{Blend with OH $^2\Pi_{3/2}(J=5/2-\rightarrow3/2+)$} \\
o-H$_2$O & $4_{32}\rightarrow4_{23}$ & 121.722 & $\phantom{00}123\pm\phantom{0}7$ & \phantom{00}... & \phantom{00}... & \phantom{00}... & \phantom{0}... \\
CO & $J=21\rightarrow20$ & 124.193 & $\phantom{0}1215\pm10$ & $\phantom{0}364\pm\phantom{0}6$ & $\phantom{0}151\pm\phantom{0}5$ & $\phantom{0}120\pm\phantom{0}2$ & \phantom{0}... \\
p-H$_2$O & $4_{04}\rightarrow3_{13}$ & 125.354 & $\phantom{00}332\pm\phantom{0}8$ & $\phantom{00}98\pm\phantom{0}5$ & \phantom{00}... & $\phantom{00}47\pm\phantom{0}2$ & \phantom{0}... \\
p-H$_2$O & $3_{31}\rightarrow3_{22}$ & 126.714 & $\phantom{000}43\pm\phantom{0}7$ & \phantom{00}... & \phantom{00}... & \phantom{00}... & \phantom{0}... \\
o-H$_2$O & $7_{25}\rightarrow7_{16}$ & 127.884 & $\phantom{000}45\pm\phantom{0}5$ & \phantom{00}... & \phantom{00}... & \phantom{00}... & \phantom{0}... \\
$^{13}$CO & $J=21\rightarrow20$ & 129.891 & $\phantom{000}21\pm\phantom{0}4$ & \phantom{00}... & \phantom{00}... & \phantom{00}... & \phantom{0}... \\
CO & $J=20\rightarrow19$ & 130.369 & $\phantom{0}1203\pm\phantom{0}8$ & $\phantom{0}322\pm\phantom{0}3$ & $\phantom{0}105\pm\phantom{0}2$ & $\phantom{0}118\pm\phantom{0}2$ & $\phantom{0}26\pm\phantom{0}2$ \\
o-H$_2$O & $4_{23}\rightarrow4_{14}$ & 132.408 & $\phantom{00}106\pm\phantom{0}6$ & $\phantom{00}69\pm\phantom{0}3$ & $\phantom{00}24\pm\phantom{0}3$ & $\phantom{00}20\pm\phantom{0}3$ & $\phantom{0}27\pm\phantom{0}3$ \\
o-H$_2$O & $5_{14}\rightarrow5_{05}$ & 134.935 & $\phantom{00}181\pm\phantom{0}5$ & $\phantom{00}19\pm\phantom{0}3$ & $\phantom{00}35\pm\phantom{0}1$ & $\phantom{00}12\pm\phantom{0}1$ & \phantom{0}... \\
o-H$_2$O & $3_{30}\rightarrow3_{21}$ & 136.496 & $\phantom{00}237\pm\phantom{0}7$ & $\phantom{00}81\pm\phantom{0}4$ & $\phantom{00}12\pm\phantom{0}3$ & $\phantom{00}26\pm\phantom{0}2$ & $\phantom{00}9\pm\phantom{0}2$ \\
CO & $J=19\rightarrow18$ & 137.196 & $\phantom{0}1523\pm\phantom{0}8$ & $\phantom{0}471\pm\phantom{0}5$ & $\phantom{0}265\pm\phantom{0}4$ & $\phantom{0}133\pm\phantom{0}2$ & $\phantom{0}35\pm\phantom{0}3$ \\
p-H$_2$O & $3_{13}\rightarrow2_{02}$ & 138.528 & $\phantom{00}507\pm\phantom{0}7$ & $\phantom{0}126\pm\phantom{0}6$ & $\phantom{00}75\pm\phantom{0}4$ & $\phantom{00}60\pm\phantom{0}3$ & $\phantom{0}16\pm\phantom{0}3$ \\
p-H$_2$O & $4_{13}\rightarrow3_{22}$ & 144.518 & $\phantom{00}166\pm\phantom{0}6$ & $\phantom{00}35\pm\phantom{0}4$ & \phantom{00}... & $\phantom{000}8\pm\phantom{0}2$ & \phantom{0}... \\
CO & $J=18\rightarrow17$ & 144.784 & $\phantom{0}1555\pm\phantom{0}6$ & $\phantom{0}520\pm\phantom{0}5$ & $\phantom{0}270\pm\phantom{0}4$ & $\phantom{0}137\pm\phantom{0}2$ & $\phantom{0}17\pm\phantom{0}2$ \\
\ion{O}{i} & $^3$P$_0\rightarrow$~$^3$P$_1$ & 145.525 & $\phantom{0}1029\pm\phantom{0}7$ & $\phantom{0}425\pm\phantom{0}5$ & $\phantom{0}177\pm\phantom{0}3$ & $\phantom{0}125\pm\phantom{0}2$ & $\phantom{0}73\pm\phantom{0}3$ \\
p-H$_2$O & $4_{31}\rightarrow4_{22}$ & 146.923 & $\phantom{000}29\pm\phantom{0}6$ & \phantom{00}... & $\phantom{00}14\pm\phantom{0}3$ & $\phantom{00}10\pm\phantom{0}3$ & \phantom{0}... \\
$^{13}$CO & $J=18\rightarrow17$ & 151.431 & $\phantom{000}30\pm\phantom{0}7$ & $\phantom{00}11\pm\phantom{0}4$ & \phantom{00}... & \phantom{00}... & \phantom{0}... \\
CO & $J=17\rightarrow16$ & 153.267 & $\phantom{0}1800\pm\phantom{0}9$ & $\phantom{0}578\pm\phantom{0}4$ & $\phantom{0}309\pm\phantom{0}4$ & $\phantom{0}161\pm\phantom{0}3$ & \phantom{0}... \\
p-H$_2$O & $3_{22}\rightarrow3_{13}$ & 156.194 & $\phantom{00}291\pm\phantom{0}7$ & $\phantom{00}24\pm\phantom{0}3$ & $\phantom{00}20\pm\phantom{0}4$ & $\phantom{00}17\pm\phantom{0}2$ & $\phantom{00}8\pm\phantom{0}2$ \\
o-H$_2$O & $5_{23}\rightarrow4_{32}$ & 156.265 & \multicolumn{5}{l}{Blend with p-H$_2$O $3_{22}\rightarrow3_{13}$} \\
\ion{C}{ii} & $^2$P$_{\frac{3}{2}}\rightarrow$~$^2$P$_{\frac{1}{2}}$ & 157.741 & $\phantom{00}528\pm\phantom{0}7$ & $\phantom{0}379\pm\phantom{0}5$ & $\phantom{0}137\pm\phantom{0}3$ & $\phantom{0}140\pm\phantom{0}3$ & $123\pm\phantom{0}2$ \\
o-H$_2$O & $5_{32}\rightarrow5_{23}$ & 160.510 & $\phantom{000}62\pm\phantom{0}5$ & \phantom{00}... & \phantom{00}... & \phantom{00}... & \phantom{0}... \\
CO & $J=16\rightarrow15$ & 162.812 & $\phantom{0}2015\pm\phantom{0}8$ & $\phantom{0}601\pm\phantom{0}4$ & $\phantom{0}297\pm\phantom{0}2$ & $\phantom{0}191\pm\phantom{0}2$ & $\phantom{0}16\pm\phantom{0}3$ \\
OH & $^2\Pi_{1/2}(J=3/2+\rightarrow1/2-)$ & 163.124 & $\phantom{00}477\pm\phantom{0}5$ & $\phantom{00}46\pm\phantom{0}3$ & $\phantom{00}17\pm\phantom{0}2$ & $\phantom{00}36\pm\phantom{0}1$ & \phantom{0}... \\
OH & $^2\Pi_{1/2}(J=3/2-\rightarrow1/2+)$ & 163.396 & $\phantom{00}343\pm\phantom{0}5$ & $\phantom{00}56\pm\phantom{0}3$ & \phantom{00}... & $\phantom{00}36\pm\phantom{0}1$ & \phantom{0}... \\
$^{13}$CO & $J=16\rightarrow15$ & 170.290 & $\phantom{000}58\pm\phantom{0}5$ & \phantom{00}... & \phantom{00}... & \phantom{00}... & \phantom{0}... \\
CO & $J=15\rightarrow14$ & 173.631 & $\phantom{0}1952\pm10$ & $\phantom{0}636\pm\phantom{0}5$ & $\phantom{0}379\pm\phantom{0}4$ & $\phantom{0}193\pm\phantom{0}3$ & $\phantom{0}49\pm\phantom{0}2$ \\
o-H$_2$O & $3_{03}\rightarrow2_{12}$ & 174.626 & $\phantom{00}568\pm\phantom{0}6$ & $\phantom{0}192\pm\phantom{0}4$ & $\phantom{00}88\pm\phantom{0}2$ & $\phantom{00}78\pm\phantom{0}3$ & \phantom{0}... \\
o-H$_2$O & $2_{12}\rightarrow1_{01}$ & 179.527 & $\phantom{00}724\pm\phantom{0}8$ & $\phantom{0}206\pm\phantom{0}5$ & $\phantom{00}80\pm\phantom{0}2$ & $\phantom{0}123\pm\phantom{0}3$ & \phantom{0}... \\
o-H$_2$O & $2_{21}\rightarrow2_{12}$ & 180.488 & $\phantom{00}275\pm\phantom{0}5$ & $\phantom{00}36\pm\phantom{0}5$ & \phantom{00}... & $\phantom{000}7\pm\phantom{0}2$ & \phantom{0}... \\
$^{13}$CO & $J=15\rightarrow14$ & 181.608 & $\phantom{000}38\pm\phantom{0}7$ & $\phantom{00}27\pm\phantom{0}4$ & \phantom{00}... & \phantom{00}... & \phantom{0}... \\
CO & $J=14\rightarrow13$ & 185.999 & $\phantom{0}2160\pm\phantom{0}8$ & $\phantom{0}687\pm\phantom{0}5$ & $\phantom{0}323\pm\phantom{0}3$ & $\phantom{0}207\pm\phantom{0}5$ & $\phantom{0}64\pm\phantom{0}3$ \\
p-H$_2$O & $4_{13}\rightarrow4_{04}$ & 187.111 & $\phantom{00}111\pm\phantom{0}6$ & \phantom{00}... & $\phantom{00}11\pm\phantom{0}2$ & \phantom{00}... & \phantom{0}... \\
$^{13}$CO & $J=14\rightarrow13$ & 194.546 & $\phantom{000}30\pm\phantom{0}3$ & $\phantom{00}27\pm\phantom{0}2$ & \phantom{00}... & \phantom{00}... & \phantom{0}... \\
\label{tab:herschel_lineparams}
\end{longtable}
\tablefoot{
     	\tablefoottext{a}{Wavelengths are from the LAMDA database \citep{lamda}.}
\\
     	}
\end{landscape}
%}

\begin{appendix}

\section{Continuum maps}
\label{app:contmaps}

\begin{figure*}[!htb]
	\centering
	$\begin{array}{c@{\hspace{0.0cm}}c@{\hspace{0.0cm}}c}
    \includegraphics[width=0.48\linewidth]{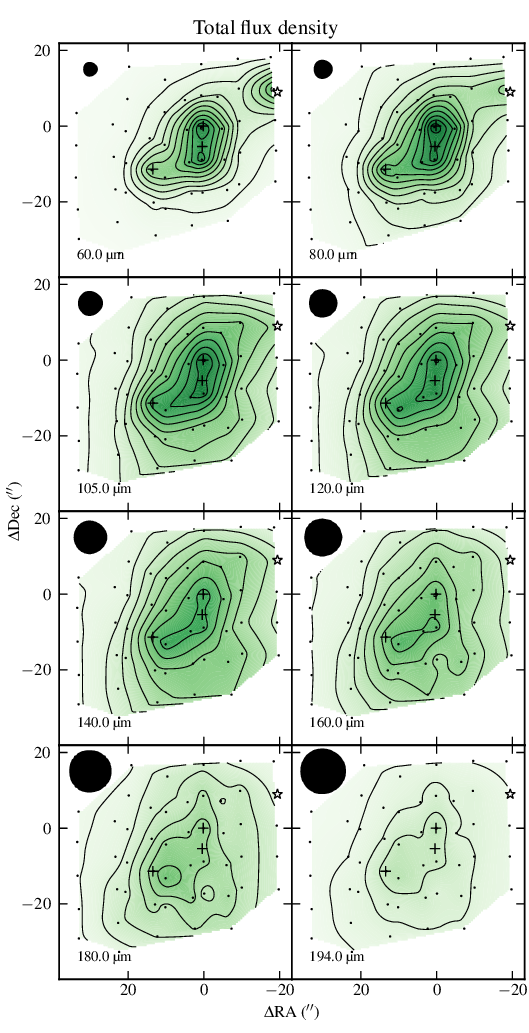} &
    \includegraphics[width=0.48\linewidth]{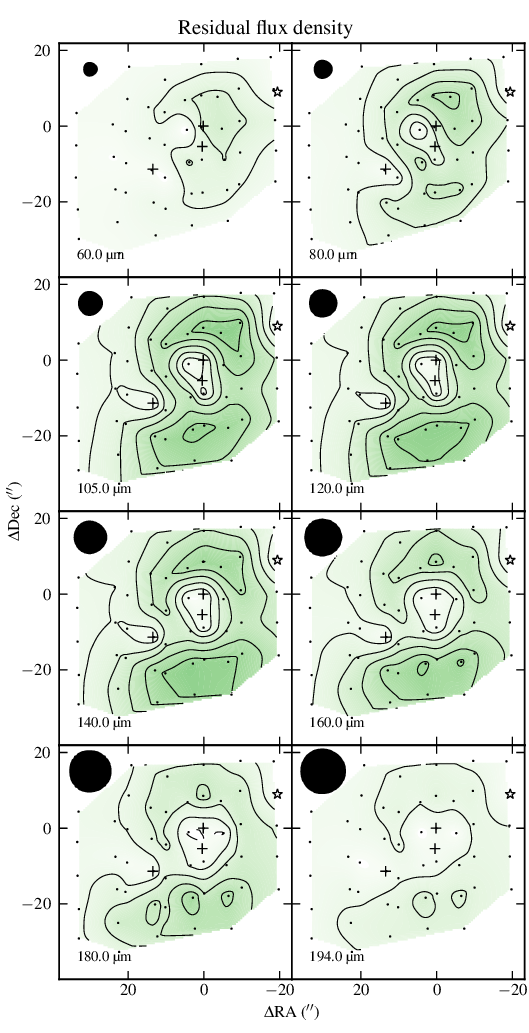} \\
    \end{array}$
    \caption{Continuum maps of the IRS7 region. \textit{Left:} Total flux
density map. \textit{Right:} Residual map. Contour levels are at 10~Jy for all
maps. The \textit{Herschel} PSF for each observation is shown in the top left corner.}\label{fig:contmaps7}
\end{figure*}

\begin{figure*}[!htb]
	\centering
	$\begin{array}{c@{\hspace{0.0cm}}c@{\hspace{0.0cm}}c}
    \includegraphics[width=0.48\linewidth]{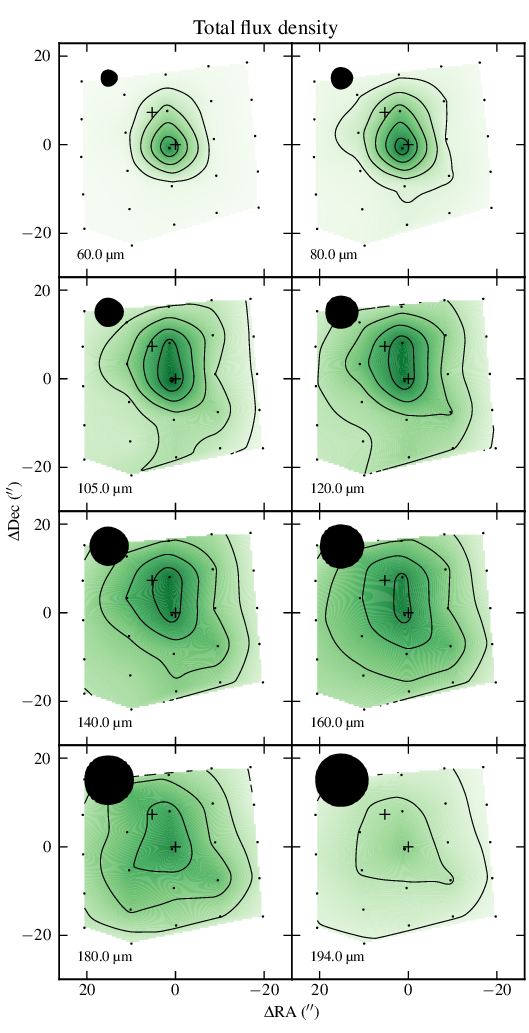} &
    \includegraphics[width=0.48\linewidth]{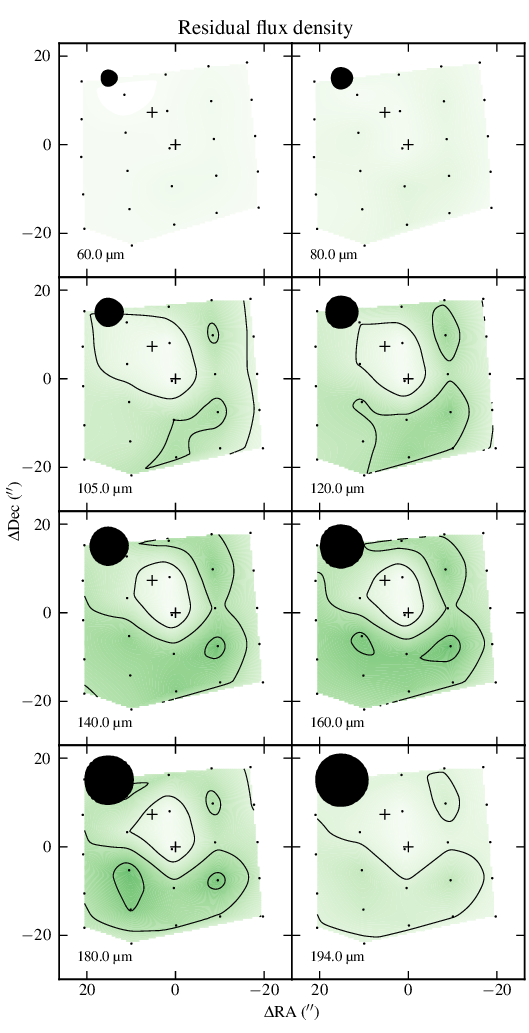} \\
    \end{array}$
    \caption{Continuum maps of the IRS5 region. \textit{Left:} Total flux
density map. \textit{Right:} Residual map. Contour levels are at 3~Jy for all
maps. The \textit{Herschel} PSF for each observation is shown in the top left corner.}\label{fig:contmaps5}
\end{figure*}

The PACS continuum maps and POMAC residuals are found in Figs.~\ref{fig:contmaps7}--\ref{fig:contmaps5}.

\clearpage

\section{Line maps}
\label{app:linemaps}

\begin{figure*}[!htb]
	\centering
	$\begin{array}{c@{\hspace{0.0cm}}c@{\hspace{0.0cm}}c}
    \includegraphics[width=0.48\linewidth]{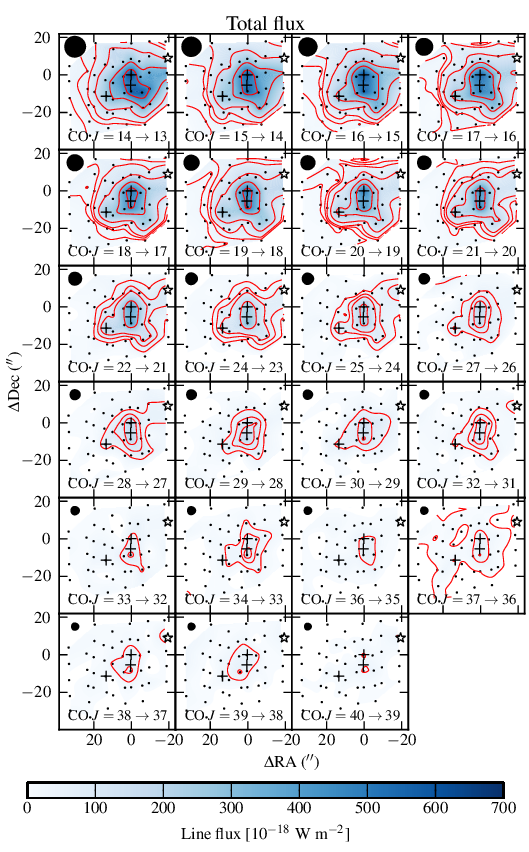} &
    \includegraphics[width=0.48\linewidth]{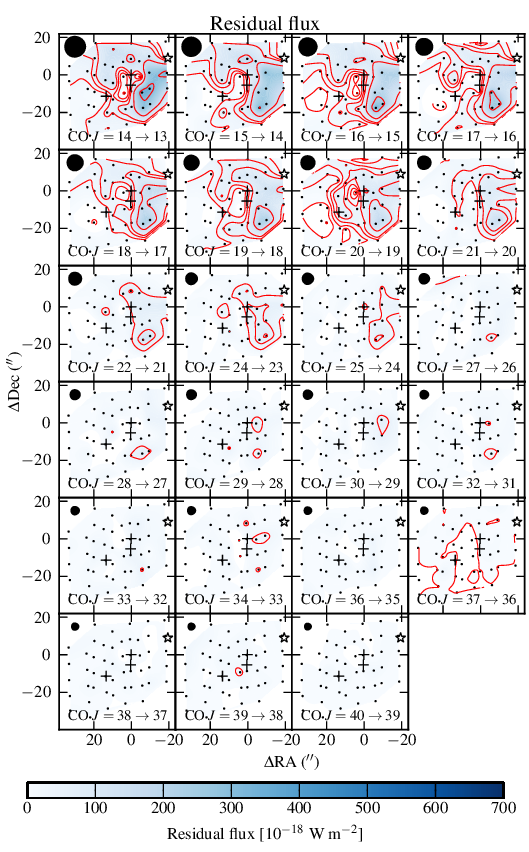} \\
    \end{array}$
    \caption{CO line maps of the IRS7 region. \textit{Left:} Total flux map.
\textit{Right:} Residual map. The red contour levels are at $5\sigma$, $10\sigma$, $15\sigma$, $30\sigma$, $60\sigma$, and $90\sigma$ (dashed contours for negative fluxes). The blue colour maps have the same scale for all maps. The \textit{Herschel} PSF for each observation is shown in the top left corner.
}\label{fig:comaps7}
\end{figure*}

\begin{figure*}[!htb]
	\centering
	$\begin{array}{c@{\hspace{0.0cm}}c@{\hspace{0.0cm}}c}
    \includegraphics[width=0.48\linewidth]{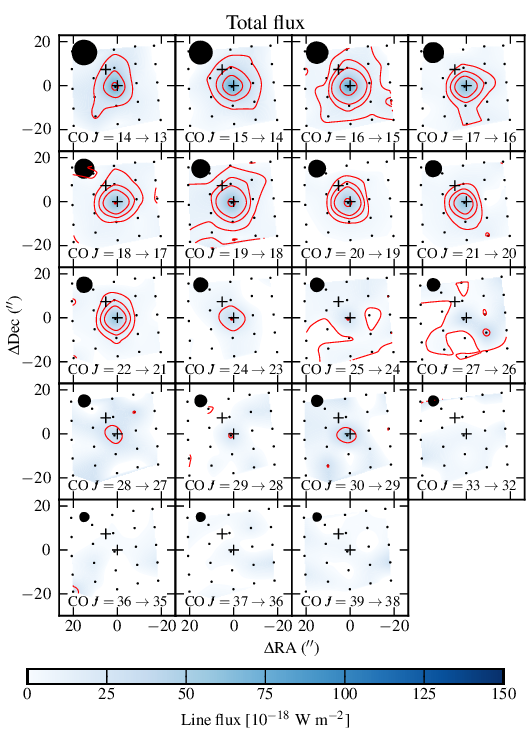} &
    \includegraphics[width=0.48\linewidth]{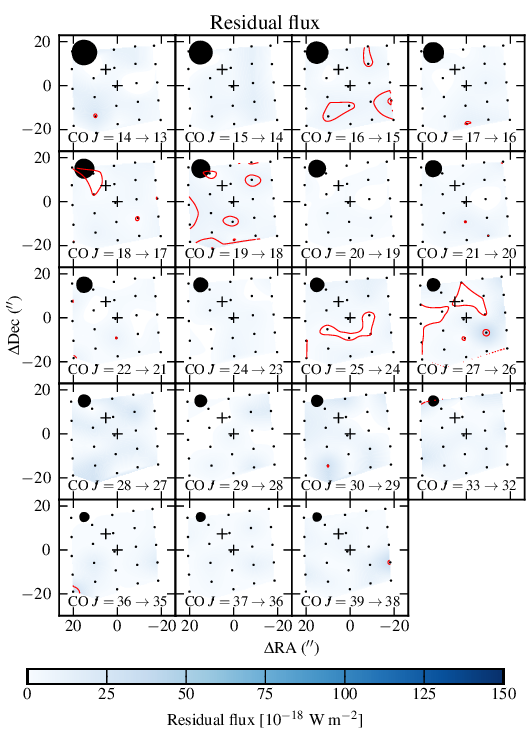} \\
    \end{array}$
    \caption{CO line maps of the IRS5 region. \textit{Left:} Total flux map.
\textit{Right:} Residual map. Contour levels as in Fig.~\ref{fig:comaps7}. The blue colour maps have the same scale for all maps. The \textit{Herschel} PSF for each observation is shown in the top left corner.}\label{fig:comaps5}
\end{figure*}

\begin{figure*}[!htb]
	\centering
	$\begin{array}{c@{\hspace{0.0cm}}c@{\hspace{0.0cm}}c}
    \includegraphics[width=0.48\linewidth]{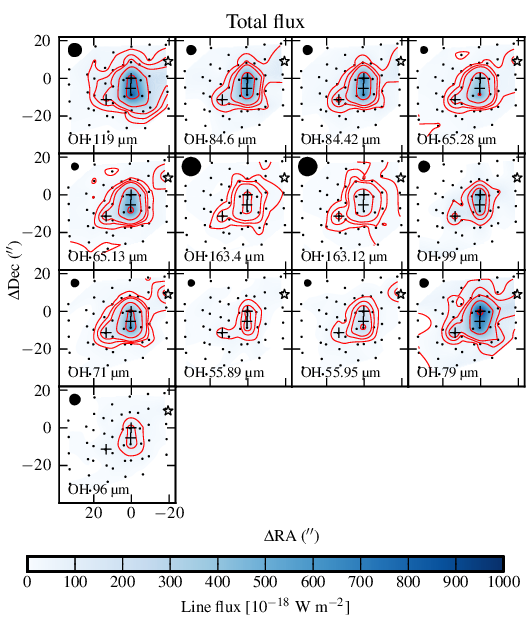} &
    \includegraphics[width=0.48\linewidth]{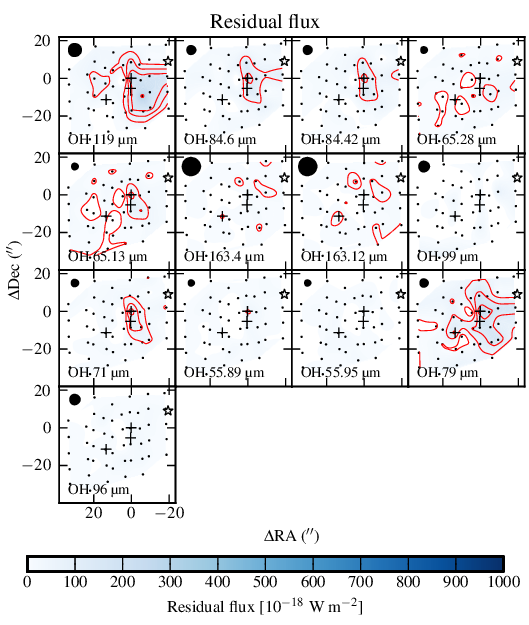} \\
    \end{array}$
    \caption{OH line maps of the IRS7 region. \textit{Left:} Total flux map.
\textit{Right:} Residual map. Contour levels as in Fig.~\ref{fig:comaps7}. The blue colour maps have the same scale for all maps. The \textit{Herschel} PSF for each observation is shown in the top left corner.}\label{fig:ohmaps7}
\end{figure*}

\begin{figure*}[!htb]
	\centering
	$\begin{array}{c@{\hspace{0.0cm}}c@{\hspace{0.0cm}}c}
    \includegraphics[width=0.48\linewidth]{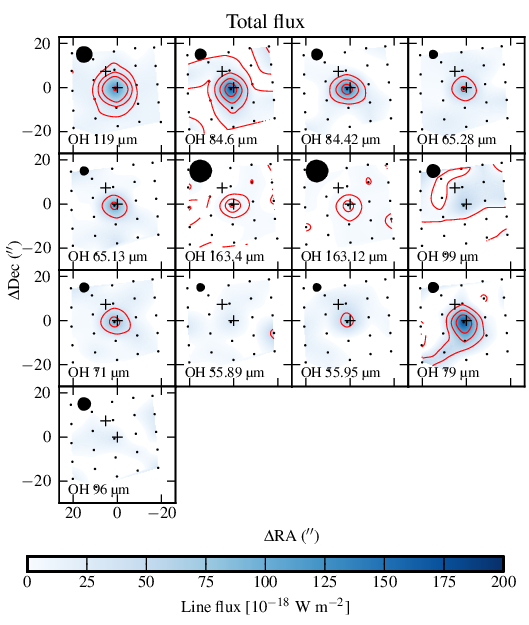} &
    \includegraphics[width=0.48\linewidth]{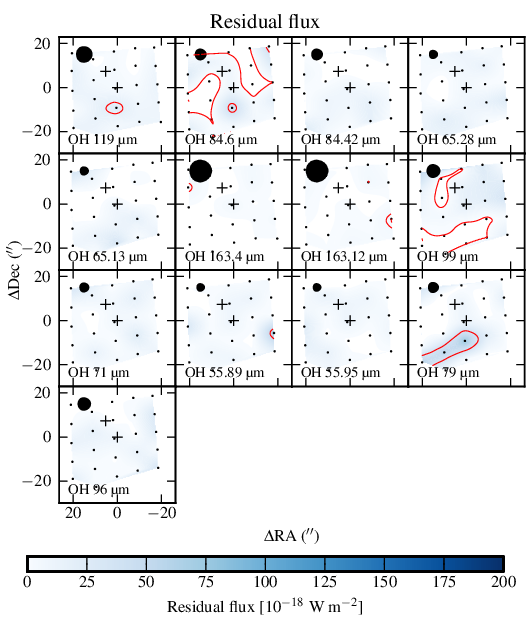} \\
    \end{array}$
    \caption{OH line maps of the IRS5 region. \textit{Left:} Total flux map.
\textit{Right:} Residual map. Contour levels as in Fig.~\ref{fig:comaps7}. The blue colour maps have the same scale for all maps. The \textit{Herschel} PSF for each observation is shown in the top left corner.}\label{fig:ohmaps5}
\end{figure*}

\begin{figure*}[!htb]
	\centering
	$\begin{array}{c@{\hspace{0.0cm}}c@{\hspace{0.0cm}}c}
    \includegraphics[width=0.48\linewidth]{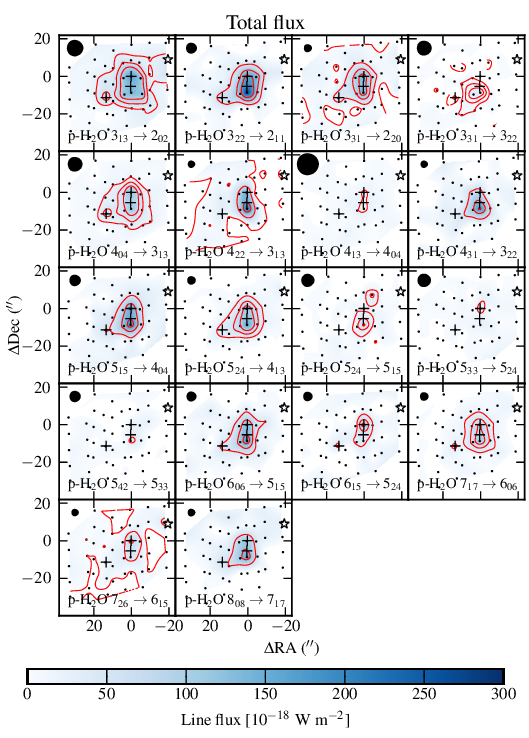} &
    \includegraphics[width=0.48\linewidth]{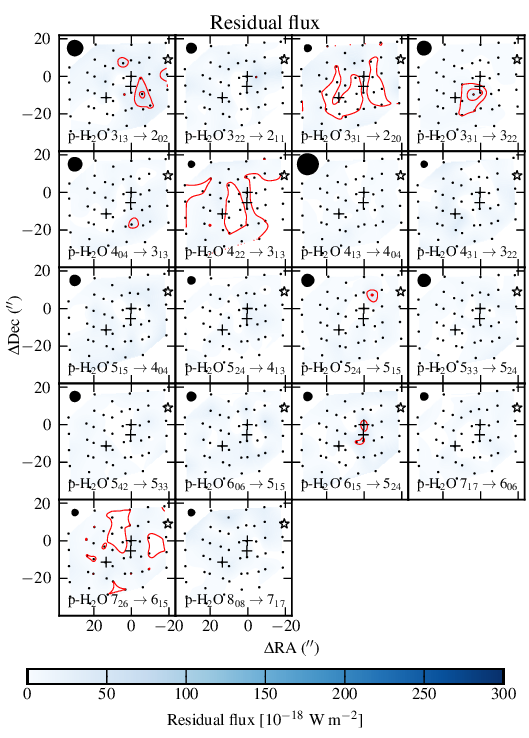} \\
    \end{array}$
    \caption{p-H$_2$O line maps of the IRS7 region. \textit{Left:} Total flux
map. \textit{Right:} Residual map. Contour levels as in Fig.~\ref{fig:comaps7}.The blue colour maps have the same scale for all maps. The \textit{Herschel} PSF for each observation is shown in the top left corner.}\label{fig:ph2omaps7}
\end{figure*}

\begin{figure*}[!htb]
	\centering
	$\begin{array}{c@{\hspace{0.0cm}}c@{\hspace{0.0cm}}c}
    \includegraphics[width=0.48\linewidth]{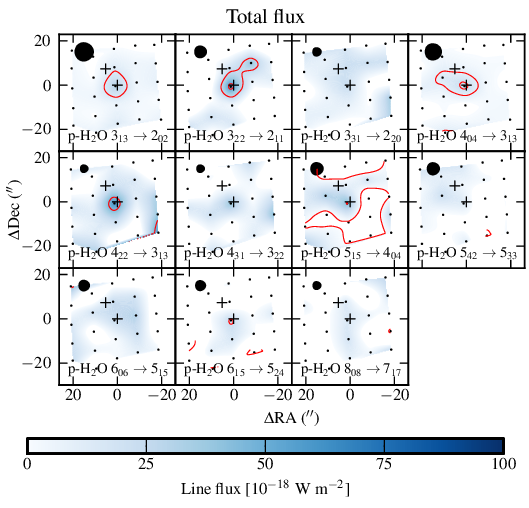} &
    \includegraphics[width=0.48\linewidth]{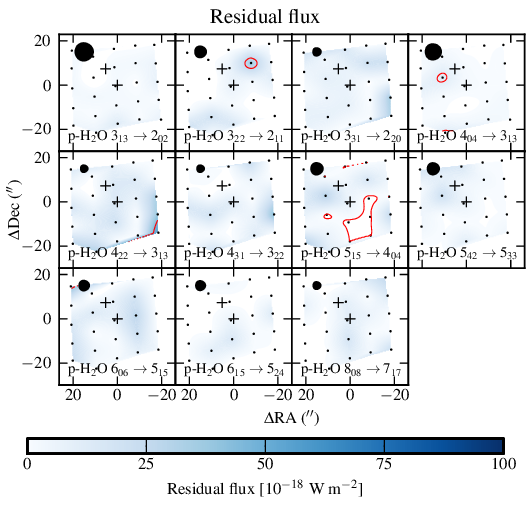} \\
    \end{array}$
    \caption{p-H$_2$O line maps of the IRS5 region. \textit{Left:} Total flux
map. \textit{Right:} Residual map. Contour levels as in Fig.~\ref{fig:comaps7}. The blue colour maps have the same scale for all maps. The \textit{Herschel} PSF for each observation is shown in the top left corner.}\label{fig:ph2omaps5}
\end{figure*}

\begin{figure*}[!htb]
	\centering
	$\begin{array}{c@{\hspace{0.0cm}}c@{\hspace{0.0cm}}c}
    \includegraphics[width=0.48\linewidth]{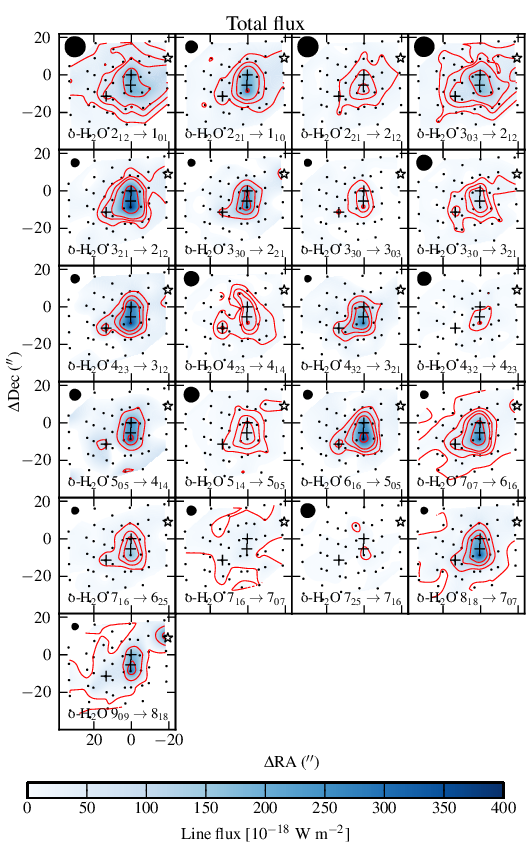} &
    \includegraphics[width=0.48\linewidth]{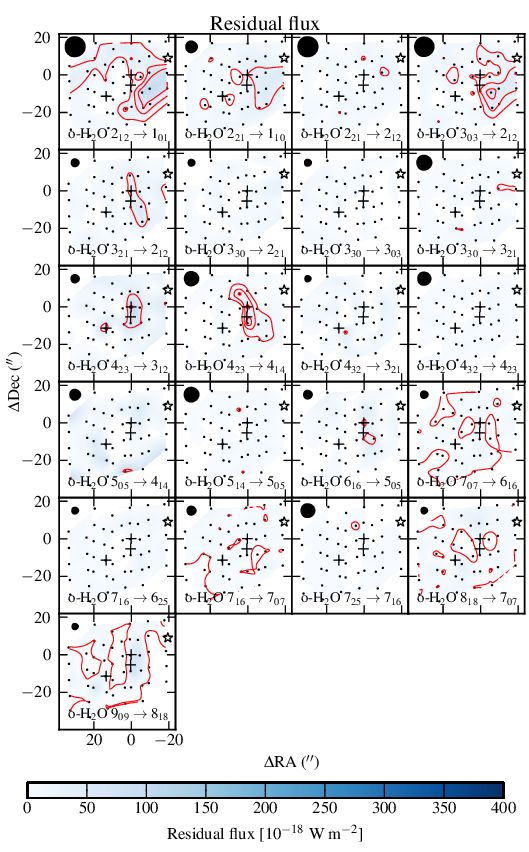} \\
    \end{array}$
    \caption{o-H$_2$O line maps of the IRS7 region. \textit{Left:} Total flux
map. \textit{Right:} Residual map. Contour levels as in Fig.~\ref{fig:comaps7}. The blue colour maps have the same scale for all maps. The \textit{Herschel} PSF for each observation is shown in the top left corner.}\label{fig:oh2omaps7}
\end{figure*}

\begin{figure*}[!htb]
	\centering
	$\begin{array}{c@{\hspace{0.0cm}}c@{\hspace{0.0cm}}c}
    \includegraphics[width=0.48\linewidth]{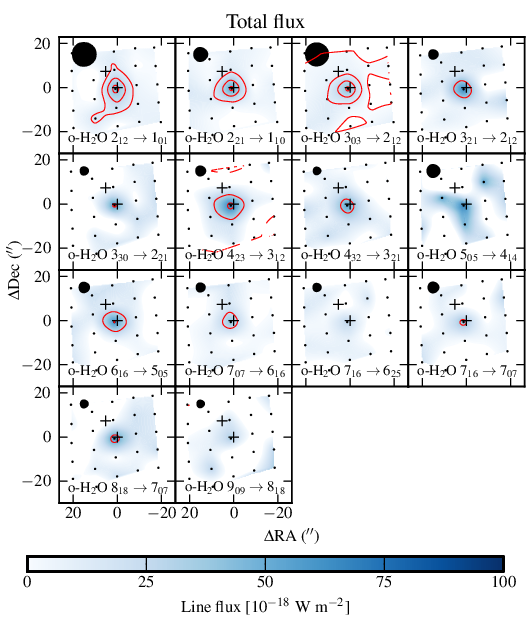} &
    \includegraphics[width=0.48\linewidth]{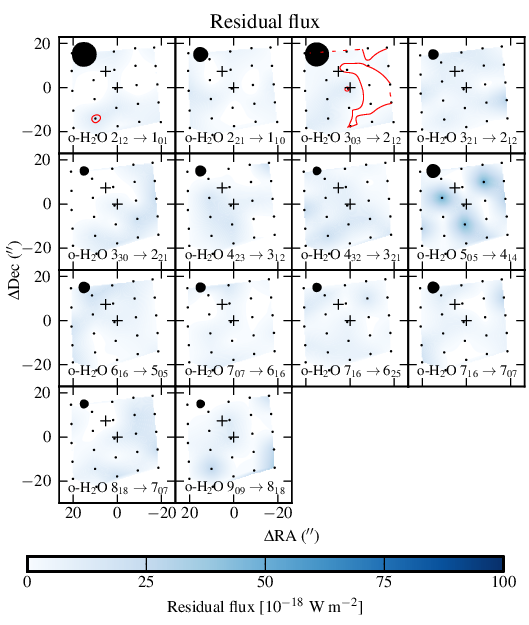} \\
    \end{array}$
    \caption{o-H$_2$O line maps of the IRS5 region. \textit{Left:} Total flux
map. \textit{Right:} Residual map. Contour levels as in Fig.~\ref{fig:comaps7}. The blue colour maps have the same scale for all maps. The \textit{Herschel} PSF for each observation is shown in the top left corner.}\label{fig:oh2omaps5}
\end{figure*}

\begin{figure*}[!htb]
	\centering
	$\begin{array}{c@{\hspace{0.0cm}}c@{\hspace{0.0cm}}c}
    \includegraphics[width=0.48\linewidth]{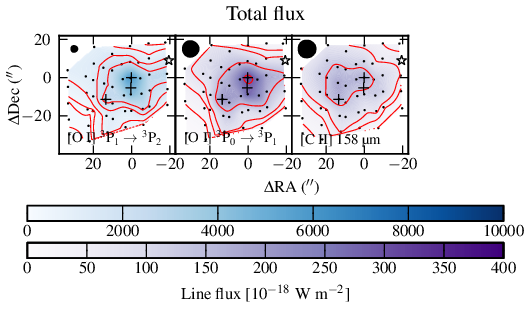} &
    \includegraphics[width=0.48\linewidth]{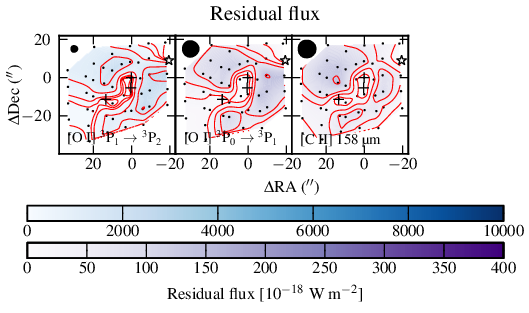} \\
    \end{array}$
    \caption{Atomic line maps of the IRS7 region. \textit{Left:} Total flux
map. \textit{Right:} Residual map. Contour levels as in Fig.~\ref{fig:comaps7}. The blue colour scale is used for the [\ion{O}{i}] 63\micron\ map and the purple colour scale for the [\ion{O}{i}] 146\micron\ and [\ion{C}{ii}] 158\micron\ maps. The \textit{Herschel} PSF for each observation is shown in the top left corner.}\label{fig:atomicmaps7}
\end{figure*}

\begin{figure*}[!htb]
	\centering
	$\begin{array}{c@{\hspace{0.0cm}}c@{\hspace{0.0cm}}c}
    \includegraphics[width=0.48\linewidth]{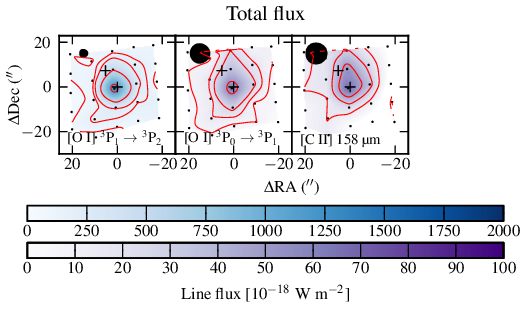} &
    \includegraphics[width=0.48\linewidth]{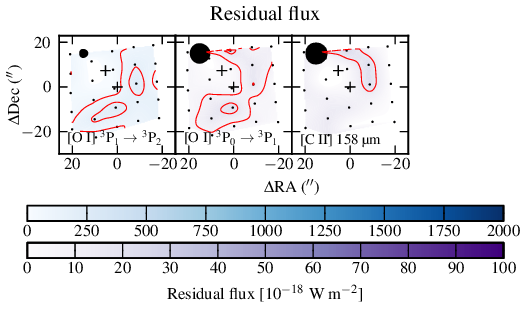} \\
    \end{array}$
    \caption{Atomic line maps of the IRS5 region. \textit{Left:} Total flux
map. \textit{Right:} Residual map. Contour levels as in Fig.~\ref{fig:comaps7}. The blue colour scale is used for the [\ion{O}{i}] 63\micron\ map and the purple colour scale for the [\ion{O}{i}] 146\micron\ and [\ion{C}{ii}] 158\micron\ maps. The \textit{Herschel} PSF for each observation is shown in the top left corner.}\label{fig:atomicmaps5}
\end{figure*}

The line flux maps and POMAC residual maps of the CO lines are found in Figs.~\ref{fig:comaps7}--\ref{fig:comaps5}, the OH maps in Figs.~\ref{fig:ohmaps7}--\ref{fig:ohmaps5}, the H$_2$O maps in Figs.~\ref{fig:ph2omaps7}--\ref{fig:oh2omaps5}, and the [\ion{O}{i}] and [\ion{C}{ii}] maps in Figs.~\ref{fig:atomicmaps7}--\ref{fig:atomicmaps5}.

\clearpage

\section{The POMAC deconvolution algorithm}
\label{app:pomac}
POMAC (Poor Man's CLEAN) is a modification of the CLEAN algorithm
\citep{hogbom74}, adapted to deconvolve data with a small number of data
points, using previous knowledge about point-source positions as a restricting
assumption. The user needs to define the positions of possible point sources
(and point sources can be added or removed after running the script,
iteratively refining the used set of point sources). The algorithm iteratively
subtracts emission corresponding to the pre-defined point sources, finally
leaving a residual map showing any emission not attributed to these point sources
(extended emission or emission from previously unknown point sources).

\begin{figure}[!htb]
	\centering
    \includegraphics[width=0.48\linewidth]{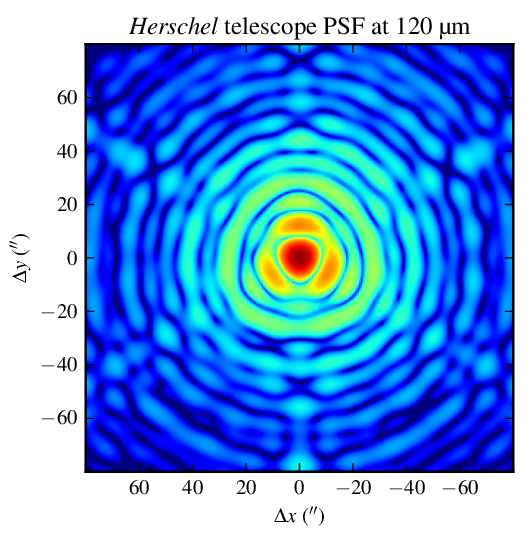}
    \caption{The \textit{Herschel} simulated point-spread function (PSF) at 120\micron\ plotted with logarithmic scaling to better show the side lobes.}
    \label{fig:pacsbeam}
\end{figure}

More specifically, the script creates an $n\times n$ grid (we use $n=400$, giving a
pixel size of $\sim0\farcs2$ for the IRS7 map) for each of the point
sources, covering the whole field-of-view of the PACS footprint(s). For each of
the grids, a unitary point source is convolved with a simulated telescope PSF.
We use the simulated Herschel PSF, which is available for 60, 70, 80, 90, 100,
120, 140, 160, 180, and 200\micron\footnote{Zemax modelled point spread functions, which are found to agree remarkably well with verification observations:\\
\url{http://herschel.esac.esa.int/twiki/pub/Public/PacsCalibrationWeb/PACSPSF_PICC-ME-TN-029_v2.0.pdf}} (see Fig.~\ref{fig:pacsbeam}). We linearly interpolate these for the
intermediate wavelengths. The
script iterates over each grid point to determine which PACS spaxel(s)
the signal in this grid point will fall into (zero or one spaxels per PACS
footprint). The result of this is one PACS instrument PSF (dirty beam) for each
of the point sources in the field. We assume PACS spaxel sizes of
9\farcs4$\times$9\farcs4, centred at the coordinates given by the
telescope data, oriented parallel to the PACS grid. This is, however, a simplified model of the true spaxel footprint, but the simplification will not affect the results more than the calibration accuracy of the instrument (20\%).

% pacsbeam.py

The line flux measured in each of the spaxels are put in a residual flux
matrix. Next, the script determines which spaxel is most nearby each of the
point sources. The script then analyses the residual map flux in each of these
spaxels, and based on the point sources' distances to their nearest spaxels the brightest point source is determined. A user-defined CLEAN gain (we
use 0.01) is multiplied to the calculated flux and added to the CLEAN flux of
this point source. The convolution of the PACS footprint Gaussian and a point
source with the CLEANed flux is then subtracted from the residual map. This is
iterated until one of several possible stop criteria is met:

\begin{enumerate}
\item The strongest point-source residual corresponds to negative flux.
\item The residual flux in any one spaxel near a point-source position drops
below $-3\sigma$.
\item The residual fluxes in any two spaxels drop below $-3\sigma$.
\item The residual flux drops below $1\sigma$ in a spaxel which is associated
with a point source.
\item The average of the residual map drops below 0.
\item The set maximum for CLEAN iterations is reached (this criterium was not
reached in any of our deconvolutions).
\end{enumerate}

After this has been finished, the flux of each point source is found in the
CLEAN flux vector, and the remaining residual map can be investigated to search
for more point sources or extended emission (note, however, that the residual
map is still convolved with the instrument and telescope PSF).

The algorithm was tested on well-centred PACS data of the disc source HD~100546 \citep{sturm10}, which is assumed to behave like a point source in the continuum. The residuals after removing the central point source from the continuum data were found to be less than 5\% of the extracted flux density across the whole spectrum.

It should be noted that in the case of extended emission across the whole field, the algorithm will extract too much flux to the point source, producing a hole in the residual emission. It is, however, impossible to predict the amount of the emission on the point source that should be attributed to the extended emission, and we have thus chosen the most simple approach. In the CrA data, the large-scale extended emission is weak in comparison with the point source emission, so this does not appear to affect the results more than 20\%.

The algorithm works particularly well in the IRS7 field, where we have two
overlapping PACS footprints (creating an almost-Nyquist sampled grid for most
of the PACS bandwidth), but it is also useful for regions with only one PACS
footprint, such as IRS5. The method is useful, not
only to disentangle emission from several point sources, but also to
establish whether a source is a point source or shows extended emission.

\section{SED flux densities}

The flux densities from the literature used in the SED fits in Sect.~\ref{sec:sed_analysis} are listed in Table~\ref{tab:sedfluxes}.

\begin{table*}[!htb]
\centering
\caption[]{Flux densities used in the SED fits in Sect.~\ref{sec:sed_analysis}.}
\label{tab:sedfluxes}
\begin{tabular}{l l l l}
\noalign{\smallskip}
\hline
\hline
\noalign{\smallskip}
Name & Wavelength & Flux density & Reference \\
\hspace{3mm} \textit{Telescope} & [\hbox{\textmu}m] & [Jy] & \\
\noalign{\smallskip}
\hline
\noalign{\smallskip}
IRS7A \\
\hspace{3mm} \textit{Spitzer} & \phantom{00}3.6 & \phantom{0}0.0509 & \citet{peterson11} \\
& \phantom{00}4.5 & \phantom{0}0.239 \\
& \phantom{00}5.8 & \phantom{0}0.546 \\
& \phantom{00}8.0 & \phantom{0}1.05 \\
SMM~1C \\
\hspace{3mm} \textit{JCMT SCUBA}\tablefootmark{a} & 450 & 45 & \citet{nutter05} \\
& 850 & \phantom{0}5.6 \\
IRS7B \\
\hspace{3mm} \textit{Spitzer} & \phantom{00}3.6 & \phantom{0}0.0367 & \citet{peterson11} \\
& \phantom{00}4.5 & \phantom{0}0.257 \\
& \phantom{00}5.8 & \phantom{0}0.697 \\
& \phantom{00}8.0 & \phantom{0}0.957 \\
\hspace{3mm} \textit{JCMT SCUBA}\tablefootmark{a} & 450 & 50 & \citet{nutter05} \\
& 850 & \phantom{0}5.4 \\
R~CrA \\
\hspace{3mm} \textit{SAAO} & \phantom{00}0.36 & \phantom{0}0.0141 & \citet{koen10} \\
& \phantom{00}0.45 & \phantom{0}0.0359 \\
& \phantom{00}0.555 & \phantom{0}0.0647 \\
& \phantom{00}0.67 & \phantom{0}0.116 \\
& \phantom{00}0.87 & \phantom{0}0.180 \\
\hspace{3mm} \textit{2MASS} & \phantom{00}1.25 & \phantom{0}2.68 & \citet{2mass} \\
& \phantom{00}1.65 & 10.7 \\
& \phantom{00}2.2 & 47.9 \\
IRS5A \\
\hspace{3mm} \textit{Spitzer} & \phantom{00}3.6 & \phantom{0}0.389 & \citet{peterson11} \\
& \phantom{00}4.5 & \phantom{0}0.798 \\
& \phantom{00}5.8 & \phantom{0}1.27 \\
& \phantom{00}8.0 & \phantom{0}1.81 \\
IRS5N \\
\hspace{3mm} \textit{Spitzer} & \phantom{00}3.6 & \phantom{0}0.00856 & \citet{peterson11} \\
& \phantom{00}4.5 & \phantom{0}0.0262 \\
& \phantom{00}5.8 & \phantom{0}0.0528 \\
& \phantom{00}8.0 & \phantom{0}0.0809 \\
\hspace{3mm} \textit{JCMT SCUBA}\tablefootmark{a} & 450 & 12\tablefootmark{b}
&\citet{nutter05} \\
& 850 & \phantom{0}1.8\tablefootmark{b} \\
\noalign{\smallskip}
\hline
\end{tabular}
\tablefoot{
	\tablefoottext{a}{The integrated flux density in apertures of $16\arcsec\times16\arcsec$ (SMM~1C), $18\arcsec\times17\arcsec$ (IRS7B), and $24\arcsec\times24\arcsec$ (IRS5N) are given.}
	\tablefoottext{b}{This is the combined value for IRS5A and IRS5N, which are
unresolved. For the SEDs, we assign half of the flux density to each source.}
     	}
\end{table*}

\end{appendix}
\end{document}